\documentclass[a4paper,12pt,parskip=half,english,numbers=noenddot]{scrartcl}

\newcommand{\pMRcOne}{2000}
\newcommand{\pMRcTwo}{1000}
\newcommand{\pFIBa}{15000}
\newcommand{\pFIBb}{3000}
\newcommand{\pFIBc}{1.25}
\newcommand{\pFibLone}{\frac{1}{\sqrt{4.25}} \left[-1; -1; -1.5\right]}
\newcommand{\pFibLtwo}{\frac{1}{\sqrt{4.25}} \left[-1; -1; \phantom{-}1.5\right]}
\newcommand{\pA}{1}
\newcommand{\pB}{1}
\newcommand{\pC}{1}
\newcommand{\pD}{1}
\newcommand{\pE}{1}
\newcommand{\pF}{1}
\newcommand{\pG}{1}
\newcommand{\pH}{1}
\newcommand{\pI}{1}
\newcommand{\pJ}{1}
\newcommand{\pK}{1}
\newcommand{\pL}{1}
\newcommand{\pM}{1}
\newcommand{\pN}{1}
\newcommand{\pO}{1}
\newcommand{\pP}{1}
\newcommand{\pQ}{1}
\newcommand{\pR}{1}
\newcommand{\pS}{1}
\newcommand{\pT}{1}
\newcommand{\pU}{1}
\newcommand{\pV}{1}
\newcommand{\pW}{1}
\newcommand{\pX}{1}
\newcommand{\pY}{1}
\newcommand{\pZ}{1}
\newcommand{\pAA}{1}
\newcommand{\pAB}{1}
\newcommand{\pAC}{1}
\newcommand{\pAD}{1}
\newcommand{\pAE}{1}
\newcommand{\pAF}{1}
\newcommand{\pAG}{1}
\newcommand{\pAH}{1}
\newcommand{\pAI}{1}

\newcommand{\pBA}{1}
\newcommand{\pBB}{1}
\newcommand{\pBC}{1}
\newcommand{\pBD}{1}
\newcommand{\pBE}{1}
\newcommand{\pBF}{1}
\newcommand{\pBG}{1}
\newcommand{\pBH}{1}
\newcommand{\pBI}{1}

\newcommand{\pCA}{1}
\newcommand{\pCB}{1}
\newcommand{\pCC}{1}
\newcommand{\pCD}{1}
\newcommand{\pCE}{1}
\newcommand{\pCF}{1}
\newcommand{\pCG}{1}
\newcommand{\pCH}{1}
\newcommand{\pCI}{1}

\newcommand{\pDA}{1}
\newcommand{\pDB}{1}
\newcommand{\pDC}{1}
\newcommand{\pDD}{1}
\newcommand{\pDE}{1}
\newcommand{\pDF}{1}
\newcommand{\pEA}{1}
\newcommand{\pEB}{1}
\newcommand{\pEC}{1}
\newcommand{\pED}{1}
\newcommand{\pEE}{1}
\newcommand{\pEF}{1}

\usepackage{babel}
\usepackage[T1]{fontenc}
\usepackage{lmodern,textcomp,latexsym}
\usepackage{amsthm,amssymb,amsmath}
\usepackage{graphicx,psfrag,color,rotate}
\usepackage{paralist,threeparttable}
\usepackage[footnotesize,labelfont={bf}]{caption}
\usepackage{alphalph}
\usepackage{longtable,booktabs,multirow}
\usepackage{tabularx,nth}
\usepackage{tikz}
\usepackage{pifont}
\usepackage{enumerate}
\usepackage{ulem,upgreek}
\usepackage{cancel}
\usepackage{hyperref}
\usepackage[boxed]{algorithm2e}

\usepackage{verbatim}

\DeclareCaptionStyle{table_new}{font={footnotesize},justification=raggedright}
\setdefaultenum{1)}{\theenumi.1)}{}{}

\newtheorem{remark}{Remark}

\renewcommand{\vec}{\boldsymbol}
\renewcommand{\d}[1][]{\ensuremath{\,\mathrm{d}#1}}

\renewcommand{\arraystretch}{1.5}

\newcommand{\op}[1]{\ensuremath{\operatorname{#1}}}

\newcommand{\transp}[1]{\ensuremath{^{\scriptstyle \mathrm{T}\mathrm{#1}}}}
\newcommand{\change}[1]{\ensuremath{^{\scriptstyle \mathrm{C}\mathrm{#1}}}}
\newcommand{\h}{\ensuremath{^{\scriptstyle \mathrm{h}}}}
\newcommand{\e}{\ensuremath{^{\scriptstyle \mathrm{e}}}}
\newcommand{\eX}[1]{\ensuremath{^{\scriptstyle \mathrm{e}{#1}}}}
\newcommand{\eh}{\ensuremath{^{\scriptstyle \mathrm{eh}}}}
\newcommand{\unit}[1]{\mathrm{#1}}

\makeatletter
\renewcommand*{\env@matrix}[1][*\c@MaxMatrixCols c]{%
  \hskip -\arraycolsep
  \let\@ifnextchar\new@ifnextchar
  \array{#1}}
\makeatother

\graphicspath{{./pictures/}}

\DeclareOldFontCommand{\it}{\normalfont\itshape}{\mathit}

\newcommand{\tp}{\ensuremath{^{\scriptstyle \mathrm{T}}}}

\renewcommand{\d}{\operatorname{d}\!}

\newcommand{\acos}{\operatorname{acos}}
\newcommand{\ma}[1]{\bar{\vec{#1}}}
\newcommand{\mi}[1]{\vec{#1}}
\newcommand{\trima}[1]{\bar{\mathfrak{#1}}}
\newcommand{\quarma}[1]{\bar{\mathbb{#1}}}
\newcommand{\quarmi}[1]{\mathbb{#1}}
\newcommand{\trimi}[1]{\mathfrak{#1}}
\newcommand{\fluc}[1]{\tilde{\vec{#1}}}
\newcommand{\trifluc}[1]{\tilde{\mathfrak{#1}}}
\newcommand{\RVE}{\mathcal{RVE}}
\newcommand{\intRVE}{\int\limits_{\mathcal{RVE}}}
\newcommand{\dintRVE}{\int\limits_{\partial\mathcal{RVE}}}
\newcommand{\intRVEe}{\int\limits_{\mathcal{R}\e}}
\newcommand{\intRVEh}{\int\limits_{\RVE\h}}
\newcommand{\fmi}[1]{\mathsf{#1}}
\newcommand{\smi}[1]{\mathtt{#1}}

\makeatletter
\DeclareRobustCommand\bigop[1]{%
  \mathop{\vphantom{\sum}\mathpalette\bigop@{#1}}\slimits@
}
\newcommand{\bigop@}[2]{%
  \vcenter{%
    \sbox\z@{$#1\sum$}%
    \hbox{\resizebox{\ifx#1\displaystyle.9\fi\dimexpr\ht\z@+\dp\z@}{!}{$\m@th#2$}}%
  }%
}
\makeatother
\newcommand{\bigA}{\DOTSB\bigop{\mathbf{\mathsf{A}}}}

\renewcommand{\vec}[1]{\pmb{#1}}

\newcommand{\ten}[1]{\pmb{#1}}

\newcommand{\MK}[1]{{\color{black} #1}}

\usepackage{tikz}
\usetikzlibrary{decorations.pathmorphing} 
\usetikzlibrary{matrix} 
\usetikzlibrary{arrows, arrows.meta} 
\usetikzlibrary{calc} 
\usetikzlibrary{shapes}
\usetikzlibrary{shapes.geometric}

\usepgfmodule{nonlineartransformations}
\usepgflibrary{curvilinear}

\usepackage{tikz-3dplot} 

\usepackage{bm}

\tikzstyle{block} = [draw,rectangle,thick,minimum height=2em,minimum width=2em]
\tikzstyle{sum} = [draw,circle,inner sep=0mm,minimum size=2mm]
\tikzstyle{connector} = [->,thick]
\tikzstyle{line} = [thick]
\tikzstyle{branch} = [circle,inner sep=0pt,minimum size=1mm,fill=black,draw=black]
\tikzstyle{guide} = []
\tikzstyle{snakeline} = [connector, decorate, decoration={pre length=0.2cm,
	post length=0.2cm, snake, amplitude=.4mm,
	segment length=2mm},thick, magenta, ->]

\begin{document}

\begin{center}
\large{\textbf{\MK{Computational} homogenization of higher-order continua.}}

{\large Felix Schmidt$^{a}$,  Melanie Kr{\"u}ger$^{a}$, Marc-Andr\'{e} Keip$^{b}$,  Christian Hesch$^{a}$\footnote{Corresponding author. E-mail address: christian.hesch@uni-siegen.de}}

{\small
\(^a\) Chair of Computational Mechanics, University of Siegen, Germany\\
\(^b\) Institute of Applied Mechanics, University of Stuttgart, Germany}

\end{center}

\vspace*{-0.1cm}

\textbf{Abstract.} We introduce a novel computational framework for the multiscale simulation of higher-order continua that allows for the consideration of first-, second- and third-order effects at both micro- and macro-level. In line with classical two-scale approaches, we describe the microstructure via representative volume elements (RVE) that are attached at each integration point of the macroscopic problem. To take account of the extended continuity requirements of independent fields at micro- and macro-level, we discretize both scales via isogeometric analysis (IGA). As a result, we obtain an IGA\textsuperscript{2}-method that is conceptually similar to the well-known FE\textsuperscript{2}-method. We demonstrate the functionality and accuracy of this novel multiscale method by means of a series of multiscale simulations involving different kinds of higher-order continua.


\section{Introduction}
Whether or not morphological features of a material are visible depends on the observed length scale. While a material may appear perfectly homogeneous at one scale, it may be heterogeneous at another. A typical example for such a \MK{material} is a composite, whose phases are distinguishable only at a small length scale and whose heterogeneous properties are linked to homogeneous properties at a larger scale. In general, the involved length scales are considered \textit{separated} if \MK{their contrast} is sufficiently high.%
\footnote{%
More strictly speaking, separation of scales is present when the wavelengths of physical fields at the higher scale are very much larger than the dimensions of heterogeneities at the lower scale
(\cite{sanchez-palencia_homogenization_1983}).
}
In such cases, it is reasonable to describe the \MK{homogenized} behavior with classical, first-order theories. In contrast to that, when the scales are not clearly separated, the description of the
\MK{homogenized} behavior needs to be based on generalized, higher-order theories.

Generalized theories for materials are nowadays well-established. They trace back to the seminal work of \cite{cosserat_theorie_1909}, who investigated the emergence and significance of couple stresses for the modeling of the size-dependent response of materials more than a hundred years ago. In their theory, the Cosserat brothers linked \textit{couple stresses} to the gradient of a microscopic rotation field and classical \textit{force stresses} to the gradient of the macroscopic translation field (i.e., the displacements). In that context, the microscopic rotation field is understood independent from the macroscopic rotation field. An extended theory based on the consideration of both macroscopic translational and macroscopic rotational degrees of freedom was later developed by \cite{toupin_elastic_1962,toupin_theories_1964,mindlin_effects_1962}. We refer to \cite{altenbach_cosserat_2013} and \cite{ehlers_cosserat_2020} for rigorous expositions of the Cosserats' couple-stress theory as well as to \cite{eringen_mechanics_1968,eringen_microcontinuum_2012,hadjesfandiari_couple_2011} for further developments and generalizations.

Next to classical couple-stress theories, there exist a number of further approaches to the modeling of size effects in materials. An important branch is given by so-called \textit{strain-gradient theories}, which for linear elastic solids have first been proposed by \cite{mindlin_micro-structure_1964}. Associated theories are based on the incorporation of higher-order gradients of the displacement field into the material description. We refer to \cite{mindlin_first_1968} for further specifications based on the first gradient of strain and to \cite{mindlin_second_1965} for an extension involving the second gradient of strain. The interested reader is further referred to \cite{germain_method_1973} and \cite{kirchner_unifying_2005} for underlying virtual-work and variational principles, respectively, to \cite{steinmann2013} for extensions towards fully nonlinear settings, and to \cite{zervos_finite_2008,askes_gradient_2011,fischer_isogeometric_2011} for possible numerical implementations. A general overview of gradient-extended continua is available through the monographs of \cite{maugin_mechanics_2010,altenbach_mechanics_2011,bertram_mechanics_2020}. Next to gradient-theories for elastic materials, there exists a rich theory on gradient-extended models for dissipative solids. In these cases, the gradient extensions are classicaly linked to internal variables like the plastic strains (\cite{aifantis_physics_1987,fleck_strain_1994}) and the damage field (\cite{peerlings_gradient_1996}). We refer to \cite{miehe_multi-field_2011} for associated variational treatments.

The above mentioned formulations have in common that they incorporate the notion of microstructure (and its size) in a phenomenological way. In contrast to that, microstructural information about morphology and material properties can be accounted for in an explicit manner by means of \textit{homogenization methods}. As in case of phenomenological material modeling, size effects may be incorporated in related schemes, depending on the existence of scale separation. If the considered scales are clearly separated, \textit{classical} or \textit{first-order} homogenization schemes are applicable; if they are not, \textit{generalized} or \textit{higher-order} schemes become necessary.

In the context of first-order homogenization schemes, we refer to
\cite{hill_elastic_1952,hashin_variational_1963,hill_self-consistent_1965,mori_average_1973,willis_bounds_1977}
for fundamental analytical approaches and to
\cite{smit_prediction_1998,miehe_computational_1999-1,feyel_fe2_2000,kouznetsova_approach_2001}
for seminal contributions to two-scale finite-element (FE) simulations. In the context of higher-order and generalized continua, analytical approaches have been explored by
\cite{diener_bounds_1984,gambin_higher-order_1989,boutin_microstructural_1996,drugan_micromechanics-based_1996},
see also the overview by
\cite{forest_homogenization_2002}
as well as the more recent contributions of
\cite{hutter_homogenization_2017},
\MK{\cite{maurice_second_2019,ganghoffer_determination_2019}},
\cite{ganghoffer_variational_2021},
\MK{\cite{alavi_construction_2021}.}
Associated computational homogenization schemes have been developed in the framework of couple-stress and micromorphic theories by
\cite{bouyge_micromechanically_2001,feyel_multilevel_2003,janicke_two-scale_2009,rokos_micromorphic_2019}
and in the framework of macroscopic strain-gradient approaches by
\cite{kouznetsova2002,kouznetsova_multi-scale_2004,bacigalupo_second-order_2010,yvonnet_computational_2020}.
We refer to
\cite{forest_cosserat_1998,forest_generalized_2011}
for seminal treatments.

The present work is devoted to the \textit{multiscale computational homogenization of gradient-extended continua} and unites ingredients of the works of
\cite{kouznetsova2002}
with respect to the gradient extensions at the homogenized scale, of
\cite{miehe_computational_1999}
with regard to the algorithmic linearization of the macroscopic field equations, and of
\cite{cirak_subdivision_2000,hughes2005}
with regard to spatial discretizations. In contrast to the contribution of
\cite{kouznetsova2002},
which combines a Cauchy continuum at the lower length scale with a gradient-extended continuum at the larger length scale, we will take into account \textit{gradient-extended continua at both scales}. This endeavor poses additional challenges not only for the theoretical treatment, but also for the numerical implementation.

\MK{%
From a theoretical perspective, we are dealing with overall three spatial scales given by (i) a \textit{macroscopic scale}, at which the homogenized, gradient-extended behavior will be obtained through computational homogenization of (ii) a \textit{mesoscopic scale}, at which we assume the presence of representative volume elements ($\cal RVE$), which are themselves characterized by size dependent material response at each mesoscopic material point and thus inherently linked to (iii) a \textit{microscopic scale}, at which we assume the existence of a microstructure that we capture with phenomenological, gradient-extended material models. The latter could be motivated, for example, through the presence of microscopic fibers with spatial extensions and distributions that could still be distinguished from further morphological entities like holes, inclusions, etc.\ at the level of the $\cal RVE$. We refer the interested reader to
\cite{schulte_isogeometric_2020,khristenko2021}
for associated analytical, numerical and experimental details.}

\MK{From a numerical perspective, challenges arise because} the gradient extensions come along with the requirement of $C^1$-continuous approximations of independent fields at both scales. Such a requirement can be captured in an elegant way by employing \textit{isogeometric analysis} (IGA) in the sense of \cite{cirak_subdivision_2000,hughes2005}. A further algorithmic feature of the proposed implementation is due to the linearization of the macroscopic boundary value problem. Here, we employ the approach advocated by \cite{miehe_computational_1999}, which was originally developed in the context of first-order homogenization. As we will see, the associated gradient extensions result in settings that remind of the linearized structures appearing in the coupled homogenization schemes considered by \cite{keip2012,keip2014,keip_multiscale_2016}.

\MK{%
As the present work proposes a computational multiscale method based on numerical discretizations involving isogeometric analysis at two scales, we denote it as IGA\textsuperscript{2}-method in analogy to the well-known FE\textsuperscript{2}-methods mentioned above. We refer to \cite{schroeder2014} for a review of FE\textsuperscript{2}-methods and to \cite{matous_review_2017}
for a general overview of computational multiscale techniques. As already mentioned, the motivation behind using IGA instead of classical finite elements is due to the straightforward and elegant implementation of $C^1$-continuous independent fields. In case of classical FE methods, the contruction and implementation of higher-order element continuities is usually cumbersome. It could, for example, be realized by the use of Hermite shape functions, which however come with a complex algebraic structure and a high number of degrees of freedom, in particular in three spatial dimensions. Alternative FE approximations are given by mixed and non-conforming methods. While mixed methods can be implemented with standard $C^0$-type shape functions, they need to satisfy the inf-sup condition
(\cite{brezzi_mixed_1991}).
Non-conforming finite elements indeed allow for a more or less straightforward numerical implementation at a reasonable amount of degrees of freedom
(\cite{teichtmeister_phase_2017}),
still their finite-element function space is not a subspace of the solution space
(\cite{crouzeix_conforming_1973}).
In contrast to that, IGA-based schemes do not suffer from such limitations, but allow for an elegant implementation of higher-order continuities.%
}
\footnote{%
Alternative schemes with even $C^\infty$-continuous interpolations at the microscopic level have been proposed by \cite{moulinec_numerical_1998} and were recently implemented in the framework of so-called FE-FFT methods. As the name suggests, associated schemes combine macroscopic solvers based on finite elements with microscopic solvers based on spectral methods (Fast Fourier Transforms; FFT), see \cite{spahn_multiscale_2014,kochmann_two-scale_2016,gokuzum_algorithmically_2018}.
}
\MK{
As IGA-based multiscale methods have thus far been limited to homogeneous macroscopic problems (\cite{alberdi_framework_2018,wang_optimal_2018}), we believe that the here proposed IGA\textsuperscript{2}-method provides a useful and innovative framework for the modeling of higher-order continua across length scales.%
}

\MK{%
The outline of the paper is as follows. In Section 2 we discuss fundamental concepts of the multiscale modeling of higher-gradient continua. In that consequence, suitable boundary conditions based on an energetically consistent scale transition are derived. In Section 3 we discuss the numerical implementation of the proposed scheme. Here, we put an emphasis on the consistent linearization of the macroscopic field equations and the IGA-based discretization of representative volume elements (RVE). In Section 4 we present a number of benchmark tests to demonstrate the performance and accuracy of the proposed multiscale technique. We close the paper with a summary and a conclusion in Section 5.
}

\section{Preliminaries and problem description}\label{sec:preliminaries}
In this section, we present the basic concepts for the homogenization of second- and third-gradient media for the macro- and microcontinuum.  Moreover, suitable boundary conditions with respect to energetic criteria for the scale transition are provided. As higher-order tensor notations and operations on them are required, a brief summary is given in Appendix \ref{app:notation}. 

\subsection{Macroscopic boundary value problem}
\begin{figure}[htb]
	\centering
	\begin{minipage}[top]{0.6\textwidth}
	\psfrag{x}{$\ma{x}$}
	\psfrag{X}{$\ma{X}$}
	\psfrag{x1}[][]{$\vec{e}_1$}
	\psfrag{x2}{$\vec{e}_2$}
	\psfrag{x3}{$\vec{e}_3$}
	\psfrag{p}{$p$}
	\psfrag{P}{$P$}
	\psfrag{B0}{$\bar{\Omega}_0$}
	\psfrag{Bt}{$\bar{\Omega}$}
	\psfrag{dB0}{$\partial\bar{\Omega}_0$}
	\psfrag{dBt}{$\partial\bar{\Omega}$}
		\includegraphics[width=\textwidth]{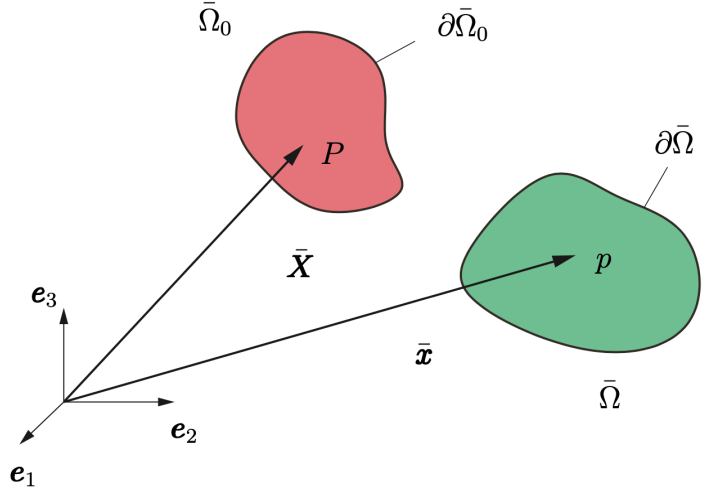}
	\end{minipage}
	\caption{Reference and current configuration.}
	\label{fig:Ref_Act}
\end{figure}

We start with a short summary of the second-gradient macroscopic continuum. Therefore, we introduce a reference configuration $\bar{\Omega}_0\subset \mathbb{R}^3$ with boundary $\partial\bar{\Omega}_0$ and outward unit normal $\ma{N}$ and a current configuration $\bar{\Omega}\subset \mathbb{R}^3$,  with outward unit normal $\ma{n}$ and  boundary $\partial\bar{\Omega}$, \MK{with subsets $\bar{\Gamma}^\varphi$ and $\bar{\Gamma}^\sigma$, and properties $\bar{\Gamma}^\varphi\cap\bar{\Gamma}^\sigma=\emptyset$ and $\bar{\Gamma}^\varphi\cup\bar{\Gamma}^\sigma=\delta\bar{\Omega}$. } The deformation mapping $\ma{\varphi}:\bar{\Omega}_0\rightarrow\mathbb{R}^3$ relates the reference and current configuration to each other, $\bar{\Omega}=\ma{\varphi}(\bar{\Omega}_0)$. Furthermore, the vector to an arbitrary material point $P$ is labelled by $\ma{X}\in\bar{\Omega}_0$. In the current configuration, the location of the corresponding point $p$ is given by $\ma{x}=\ma{\varphi}(\ma{X})$, see Figure \ref{fig:Ref_Act}.

\begin{figure}[htb]
	\centering
	\begin{minipage}[top]{0.8\textwidth}
	\psfrag{barX}[][]{$\ma{X}\in\bar{\Omega}_0$}
	\psfrag{X}{$\mi{X}$}
	\psfrag{x1}{$\vec{e}_1$}
	\psfrag{x2}{$\vec{e}_2$}
	\psfrag{x3}{$\vec{e}_3$}
	\psfrag{Bu}{$\bar{\Gamma}^\varphi$}
	\psfrag{Br}{$\bar{\Gamma}^{\nabla\sigma}$}
	\psfrag{Bs}{$\bar{\Gamma}^\sigma$}
	\psfrag{micro}{$\RVE$}
	\psfrag{di}{$\vec{d}_I$}
	\psfrag{x}{$\mi{x}$}
	\psfrag{Rd}[][]{\begin{minipage}{4cm}\begin{center}\begin{footnotesize}
 $\RVE$ driven by\\ $\ma{F},\,\trima{F}$	\end{footnotesize}\end{center}\end{minipage}}
	\psfrag{cons}[][]{\begin{minipage}{4cm}\begin{center}\begin{footnotesize}
constitutive quantities\\ $\ma{P},\,\trima{P},\Delta\ma{P},\,\Delta\trima{P}$	\end{footnotesize}\end{center}\end{minipage}}
	\includegraphics[width=\textwidth]{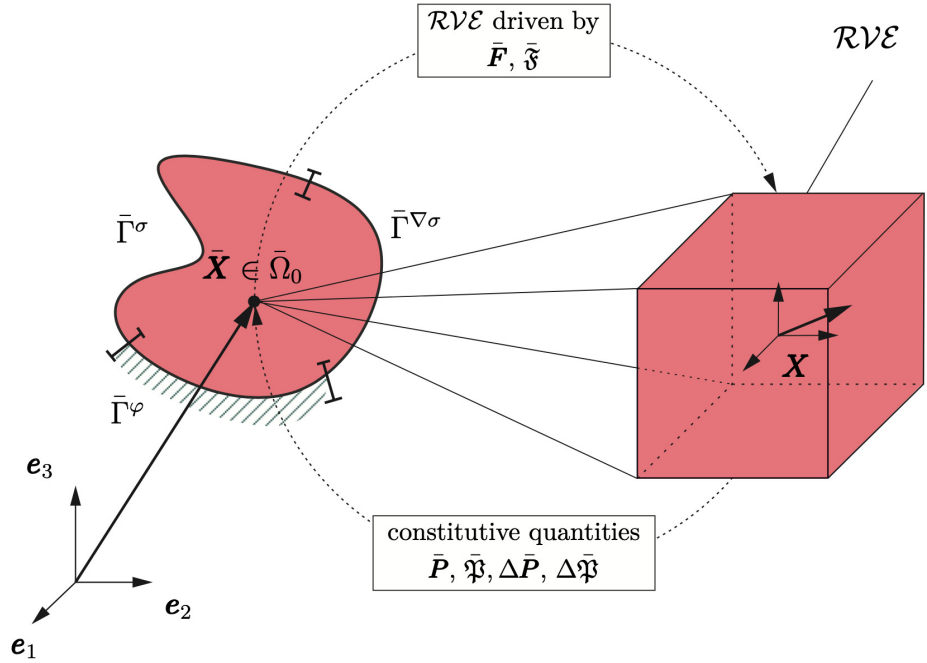}
	\end{minipage}
	\caption{Meso-macro transition of the mechanical boundary value problem, left: boundary decomposition of the macroscopic continuum in Dirichlet boundaries $\bar{\Gamma}^\varphi$ and Neumann boundaries $\bar{\Gamma}^{\sigma}$, $\bar{\Gamma}^{\nabla\sigma}$ of the traction force and the hyperstress traction force, right: $\RVE$ as defined for every macroscopic point.}
	\label{fig:Macro_RVE}
\end{figure}

The first order deformation measure $\ma{F}:\bar{\Omega}_0\rightarrow\mathbb{R}^{3\times3}$ and the second order deformation measure $\trima{F}:\bar{\Omega}_0\rightarrow\mathbb{R}^{3\times3\times3}$ are given by the first and second gradient of the mapping $\ma{\varphi}(\ma{X})$ as
\begin{equation}
\ma{F}=\bar\nabla\ma{\varphi}\qquad\text{and}\qquad\trima{F}=\bar\nabla^2\ma{\varphi}\,.
\end{equation}
Here, $\bar\nabla$ refers to the gradient with respect to $\bar{\vec{X}}$, see (\ref{macro_grad}) in Appendix \ref{app:notation}.  Following \cite{kouznetsovaPHD2002,steinmann2013,dittmann2020a},  we postulate the virtual work of the internal contributions as  
\begin{equation}\label{virt_int_work}
\delta\bar{\Pi}^{\mathrm{int}}=\int\limits_{\bar{\Omega}_0}\left(\ma{P}:\delta\ma{F}+\trima{P}\,\vdots\,\delta\trima{F}\right)\,\d V\,,
\end{equation}
where $\ma{P}$ denotes the macroscopic two-point first Piola-Kirchhoff stress tensor and $\trima{P}$ the macroscopic two-point third-order stress tensor, conjugate to $\trima{F}$.  Moreover,
\begin{equation}
\delta\ma{F}=\bar\nabla\delta\ma{\varphi}\qquad\text{and}\qquad\delta\trima{F}=\bar\nabla^2\delta\ma{\varphi}\,,
\end{equation}
where the space of virtual or admissible test functions is given by
\begin{equation}
\mathcal{V} = \{\delta\ma{\varphi}\in \mathcal{H}^2(\bar{\Omega})\quad |\quad \delta\ma{\varphi} = \vec{0},\,\bar{\nabla}\delta\ma{\varphi}\,\ma{N} = \vec{0} \quad\text{on}\quad\bar{\Gamma}^{\varphi}\}
\end{equation}
with boundary $\bar{\Gamma}^{\varphi}$, see Figure \ref{fig:Macro_RVE}.  \MK{Applying integration by parts twice in \eqref{virt_int_work} yields
\begin{equation}
\delta\bar{\Pi}^{int}(\vec{\varphi}) = \int\limits_{\bar{\Omega}_0}\bar{\nabla}\cdot(\bar{\nabla}\cdot\bar{\mathfrak{P}}-\bar{\vec{P}})\cdot\delta\bar{\vec{\varphi}} \d V +
\int\limits_{\partial\bar{\Omega}_0}\delta\bar{\vec{\varphi}}\cdot(\bar{\vec{P}}-\bar{\nabla}\cdot\bar{\mathfrak{P}})\,\bar{\vec{N}} + \bar{\nabla}\delta\bar{\vec{\varphi}}:(\bar{\mathfrak{P}}\cdot\bar{\vec{N}})\d A.
\end{equation}
Introducing the orthogonal decomposition $\bar{\nabla}_{\bot}\cdot(\bullet) = \bar{\nabla}(\bullet):(\bar{\vec{N}}\otimes\bar{\vec{N}})$ and $\bar{\nabla}_{\|}\cdot(\bullet) = \bar{\nabla}(\bullet):(\ten{I}-\bar{\vec{N}}\otimes\bar{\vec{N}})$, we obtain after some further technical steps
\begin{equation}\label{eq:gradBoundary}
\begin{aligned}
\delta\bar{\Pi}^{int}(\vec{\varphi}) =& \,\int\limits_{\bar{\Omega}_0}\bar{\nabla}\cdot(\bar{\nabla}\cdot\bar{\mathfrak{P}}-\bar{\vec{P}})\cdot\delta\bar{\vec{\varphi}} \d V +
\int\limits_{\partial\bar{\Omega}_0}\delta\bar{\vec{\varphi}}\cdot(\bar{\vec{P}}-\bar{\nabla}\cdot\bar{\mathfrak{P}})\,\bar{\vec{N}} \d A \\
&-\int\limits_{\partial\bar{\Omega}_0}\left[\delta\bar{\vec{\varphi}}\cdot(K\,(\bar{\mathfrak{P}}\,\bar{\vec{N}})\,\bar{\vec{N}}+\bar{\nabla}_{\|}\cdot(\bar{\mathfrak{P}}\,\bar{\vec{N}}))  - \bar{\nabla}_{\bot}\delta\bar{\vec{\varphi}}:\left(\bar{\mathfrak{P}}\,\bar{\vec{N}}\right)\right]\d A \\ 
&+\int\limits_{\partial^2\bar{\Omega}_0}\delta\bar{\vec{\varphi}}\cdot(\bar{\mathfrak{P}}:(\hat{\bar{\vec{N}}}\otimes\bar{\vec{N}}))\d S,
\end{aligned}
\end{equation}
for a sufficiently smooth $\bar{\Omega}_0$, where $\hat{\bar{\vec{N}}}$ is the normal to $\partial^2\bar{\Omega}_0$ and the tangent to $\partial\bar{\Omega}_0$.  Note that $\partial^2\bar{\Omega}_0$ is defined by the union of the boundary curves of the boundary surface patches and thus, $\hat{\bar{\vec{N}}}$ can be defined differently from both adjacent surfaces, see Javili et al. \cite{steinmann2013} and the citations therein for details. Moreover, $K = -\bar{\nabla}_{\|}\cdot\bar{\vec{N}}$ is the curvature of the surface.  

Omitting line forces for the ease of exposition, the external contributions to the virtual work are given by
\begin{equation}
\delta\bar{\Pi}^{\mathrm{ext}}= \int\limits_{\bar{\Omega}_0} \ma{B}_{\mathrm{ext}}\cdot\delta\ma{\varphi}\,\d V +
\int\limits_{\bar{\Gamma}^{\sigma}} \ma{T}_{\mathrm{ext}}\cdot\delta\ma{\varphi}\, \d A + \int\limits_{\bar{\Gamma}^{\nabla\sigma}}\ma{M}_{\mathrm{ext}}:\bar{\nabla}_{\bot}\delta\ma{\varphi}\,\d A
\end{equation}
with the common body force per unit volume $\ma{B}_{\mathrm{ext}}$, the traction forces $\ma{T}_{\mathrm{ext}}$ on boundary $\bar{\Gamma}^{\sigma}$ and the hyperstress traction force $\ma{M}_{\mathrm{ext}}$ on boundary $\bar{\Gamma}^{\nabla\sigma}$, see once again Figure \ref{fig:Macro_RVE}. }

Thus, the principle of virtual work reads
\begin{equation}\label{eq:pdva}
\delta\bar{\Pi}^{\mathrm{int}} - \delta\bar{\Pi}^{\mathrm{ext}} = 0\,,\qquad\forall\quad\delta\ma{\varphi}\in\mathcal{V}
\end{equation}
and the internal contributions can be related by applying partial integration and the Gaussian integral theorem to the external contributions, see Javili et al.\ \cite{steinmann2013}:
\begin{equation}\label{int_ext_cont}
\begin{aligned}
\ma{T}_{\mathrm{ext}}&=(\ma{P}-\bar{\nabla}\cdot\trima{P})\,\ma{N}\,,\\
\ma{M}_{\mathrm{ext}}&=\trima{P}\,\ma{N}\,.
\end{aligned}
\end{equation}
Note, that the last equation can be decomposed in tangential and normal components, see Madeo et al. \cite{neff2016} for details. Taking the balance equation
\begin{equation}\label{eq:macro_strong}
\bar{\nabla}\cdot(\ma{P}-\bar{\nabla}\cdot\trima{P}) + \ma{B}_{\mathrm{ext}}=\vec{0}\\
\end{equation}
into account,  completes the set of equations for the strong form of the second-gradient boundary value problem.  In the following, we omit volumetric body forces as gravity forces, thereby $\ma{B}_{\mathrm{ext}}=\vec{0}$.

\subsection{Mesoscopic boundary value problem}\label{sec:micro}
In every material point $\bar{P}$ of the macroscopic domain, we assume the existence of a representative volume element $\RVE$ \MK{on a mesoscale, sufficiently separated from the macroscale and sufficiently large to be \textit{representative}}, containing the information on the inhomogeneous mesoscopic continuum, see Figure \ref{fig:Macro_RVE}. To be specific, we postulate a second-gradient material in the $\RVE$ analogous to \eqref{eq:pdva} on the macroscale for two reasons: First, this general approach for the homogenization from a second-gradient micro-continuum towards a second-gradient macro-continuum allows us to demonstrate that the formulation proposed by Kouznetsova et al.\ \cite{kouznetsova2002} is a special case of the methodology presented next. Second, we can now generalize this concept for general higher-order materials.

We start with the mapping for the microscopic relative position of the material points $\mi{x}=\mi{\varphi}(\mi{X})$:
\begin{equation}\label{eq:mi_kinematic}
\mi{\varphi}(\mi{X})= \ma{F}\,\mi{X}+\frac{1}{2}\,\trima{F}:(\mi{X}\otimes\mi{X})+\fluc{w}\,.
\end{equation}
Here, $\fluc{w}$ describes the unknown microcroscopic fluctuation field, which includes all higher-order terms of the Taylor series expansion, see Kouznetsova et al.\ \cite{kouznetsova2002}.  
In analogy to the macroscopic quantities, we obtain the microscopic first-order deformation measure $\mi{F}=\nabla\mi{\varphi}$ and the second-order deformation measure $\trimi{F}=\nabla^2\mi{\varphi}$: 
\begin{equation}\label{eq:deform}
\mi{F}= \ma{F}+\trima{F}\,\mi{X}+\fluc{F}\qquad\text{and}\qquad\trimi{F}= \trima{F}+\trifluc{F}\,,
\end{equation}
where $\fluc{F}:=\nabla\fluc{w}$ and $\trifluc{F}:=\nabla^2\fluc{w}$.  The averaged microscopic deformations over the volume of the $\RVE$ can be connected to the macroscopic counterparts $\ma{F}$ and $\trima{F}$ via
\begin{equation}\label{eq:mi_ma_F_triF}
\frac{1}{V}\,\intRVE\mi{F}\,\d V=\ma{F}
\qquad\text{and}\qquad\frac{1}{V}\,\intRVE\trimi{F}\,\d V=\trima{F}\,,
\end{equation}
see Appendix \ref{app:kinematic}\footnote{All appendices are written most generally with regard to a third-gradient medium. For the proposed second-gradient material, the corresponding terms of the third gradient can be removed easily.} for further information. The local balance equation of the microscopic second-gradient continuum is given analogously to (\ref{eq:macro_strong}) by:
\begin{equation}\label{eq:micro_strong}
\nabla\cdot\left[\mi{P} - \nabla\cdot\trimi{P}\right]=\vec{0}\,,
\end{equation}
where $\mi{P}:=\partial_{\mi{F}}\Psi(\mi{F},\trimi{F})$ and $\trimi{P}:=\partial_{\trimi{F}}\Psi(\mi{F},\trimi{F})$ are defined in terms of a Helmholtz energy function $\Psi$.  

The macro-homogeneity condition is given by an energetic criterion that states that the virtual work applied to the system in the material point $\bar{P}$ is equal to the virtual work in the $\RVE$, hence we assume
\begin{equation}\label{eq:Hill_Mandel}
\frac{1}{V}\intRVE\bigl(\mi{P}:\nabla\delta\mi{\varphi}+ \mathfrak{P}\,\vdots\,\nabla^2\delta\mi{\varphi}\bigr)\d V =\ma{P}:\delta\ma{F} +\trima{P}\,\vdots\,\delta\trima{F}\,.
\end{equation}
\MK{Note that this excludes Neumann conditions on the $\RVE$, which would add an effective contribution to the virtual work on the mesoscale.} The left-hand side of the energetic criterion can be rewritten as 
\begin{equation}\label{eq:hillMandel}
\frac{1}{V}\,\intRVE\mi{P}\,\d V:\delta\ma{F}+\frac{1}{V}\,\intRVE\bigl(\mi{P}\otimes\mi{X}+\trimi{P}\bigr)\,\d V\,\vdots\,\delta\trima{F} =\ma{P}:\delta\ma{F} +\trima{P}\,\vdots\,\delta\trima{F}\,,
\end{equation}
see Appendix \ref{app:Deriv_stresses}. Comparing the left- and right-hand sides of the last equation, yields 
\begin{equation}\label{eq:mi_ma_stresses}
\ma{P}=\frac{1}{V}\intRVE\mi{P}\,\d V\qquad\text{and}\qquad
\trima{P}=\underbrace{\frac{1}{V}\intRVE\mi{P}\otimes\mi{X}\, \d V}_{\trima{P}^{\mi{P}}}+\underbrace{\frac{1}{V}\intRVE\trimi{P}\, \d V}_{\trima{P}^{\trimi{P}}}\,.  
\end{equation}
Here, the macroscopic third-order stress tensor $\trima{P}$ is split into $\trima{P}^{\mi{P}}$, which is given by the volume average of the first moment of the microscopic stresses $\mi{P}$, and $\trima{P}^{\trimi{P}}$, which is a volume average of the microscopic third-order stress tensor $\trimi{P}$. Note that if a first-gradient material within the $\RVE$ is assumed, the macroscopic hyperstress $\trima{P}^{\trimi{P}}$ vanishes and we obtain the formulation provided by Kouznetsova et al.\ \cite{kouznetsova2002}. To obtain information about the boundary conditions, \eqref{eq:Hill_Mandel} can be rewritten as:
\begin{equation}\label{eq:deriv_boundary1}
\frac{1}{V}\intRVE\left(\left[\ma{P}-\mi{P}\right]:[\delta\ma{F}+\delta\trima{F}\,\vec{X}-\delta\mi{F}]+\left[\trima{P}^{\trimi{P}}-\trimi{P}\right]\,\vdots\,\left[\delta\trima{F}-\delta\trimi{F}\right]\right)\,\d V=0\,,
\end{equation}
see Appendix \ref{app:rew_Hill_Mandel}. Obviously, the simplest assumption for all points of the \MK{mesoscale}, that fulfils the last equation is given by postulating the constraints $\ma{P}:=\mi{P}$ or $\delta\ma{F}+\delta\trima{F}\,\vec{X}:=\delta\mi{F}$ and additionally $\trima{P}^\trimi{P}:=\trimi{P}$ or $\delta\trima{F}:=\delta\trimi{F}$, compare Schröder \cite{schroeder2014} in the context of first-order theories. An alternative expression of (\ref{eq:deriv_boundary1}) yields:
\begin{equation}\label{eq:deriv_boundary2}
\begin{aligned}
&\frac{1}{V}\dintRVE\left(\left[\trima{P}^\trimi{P}-\trimi{P}\right]\,\mi{N}\right):\left[\delta\ma{F}+\delta\trima{F}\,\mi{X}-\delta\mi{F}\right]\,\d A\\
+\,&\frac{1}{V}\dintRVE\left(\left[\ma{P}-\left(\mi{P}-\nabla\cdot\trimi{P}\right)\right]\,\mi{N}\right)\cdot\left[\delta\ma{F}\,\mi{X}+\frac{1}{2}\,\delta\trima{F}:(\mi{X}\otimes\mi{X})-\delta\mi{\varphi}\right]\,\d A=0\,,\end{aligned} 
\end{equation}
see Appendix \ref{app:bc_Hill_Mandel} for further information. Thus, regarding a deformation-driven approach, suitable Dirichlet boundary conditions on the boundary $\partial\RVE$ are 
\begin{equation}\label{eq:dirichlet}
\begin{aligned}
\ma{F}\,\mi{X}+\frac{1}{2}\,\trima{F}:(\mi{X}\otimes\mi{X})-\mi{\varphi}&=\vec{0}\,,\\
\ma{F}+\trima{F}\,\mi{X}-\mi{F}&=\vec{0}\,,
\end{aligned}
\end{equation}
satisfying (\ref{eq:deriv_boundary2}). Note that due to $\trima{F}$ the boundaries are quadratic functions. 

\begin{figure}[htb]
	\centering
	\begin{minipage}[top]{0.6\textwidth}
	\psfrag{a}{$\partial\RVE$}
	\psfrag{b}{$\partial\RVE$}
	\psfrag{phi}{$\mi{\varphi}(\mi{X})$}
	\psfrag{x}{$\mi{x}$}
	\psfrag{X}{$\mi{X}$}
	\psfrag{np}{$\mi{N}^+$}
	\psfrag{nm}{$\mi{N}^-$}
	\psfrag{+}{$+$}
	\psfrag{-}{$-$}
	\includegraphics[width=\textwidth]{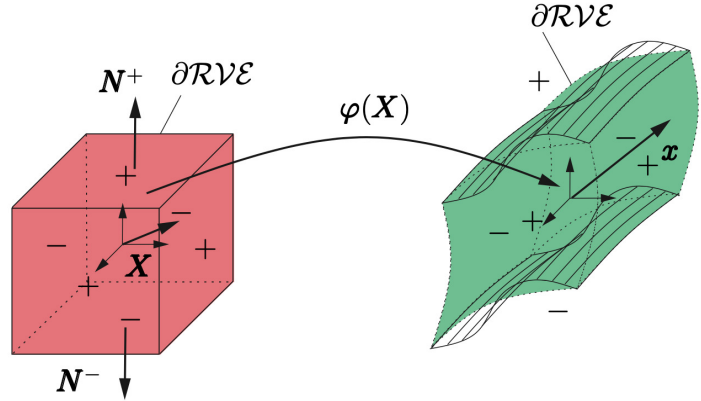}
	\end{minipage}
	\caption{Mesoscopic boundary value problem, periodic boundary conditions on $\partial\RVE$, here only displayed for top and bottom for better understanding.}
	\label{fig:periodic}
\end{figure}

For a stress driven approach, (\ref{eq:deriv_boundary2}) yields possible Neumann boundary conditions, however, that would render an inherently complex implementation for large deformations, see Kouznetsova \cite{kouznetsovaPHD2002}. A comparison of the Dirichlet boundary conditions with the mappings \eqref{eq:mi_kinematic} and \eqref{eq:deform}$_1$ provides the following relationship for these conditions, $\fluc{w}=\vec{0}$ and $\nabla\fluc{w}=\vec{0}$ on the boundary. Furthermore, the microscopic stress tractions are $\mi{T}_{\mathrm{ext}}=\left(\mi{P}-\nabla\cdot\trimi{P}\right)\,\mi{N}$ and the hyperstress tractions are given by $\mi{M}_{\mathrm{ext}}=\trimi{P}\,\mi{N}$, periodic boundary conditions as shown in Figure \ref{fig:periodic} require
\begin{equation}\label{eq:periodic_BC}
\begin{aligned}
\fluc{w}(\mi{X}^+)&=\fluc{w}(\mi{X}^-)\,,\qquad&\qquad\mi{T}_{\mathrm{ext}}(\mi{X}^+)&=-\mi{T}_{\mathrm{ext}}(\mi{X}^-)\,,\\
\nabla\fluc{w}(\mi{X}^+)&=\nabla\fluc{w}(\mi{X}^-)\,,&\mi{M}_{\mathrm{ext}}(\mi{X}^+)&=-\mi{M}_{\mathrm{ext}}(\mi{X}^-)\,,
\end{aligned}
\end{equation}
satisfying the energetic criterion (\ref{eq:Hill_Mandel}).  \MK{Here,  $\mi{X}^+$ and $\mi{X}^-$ refer to opposite surfaces, see Figure \ref{fig:periodic} for details. Note, that the tangential part of the constraint  $\nabla\fluc{w}=\vec{0} $ is already fulfilled by the condition $\fluc{w}=\vec{0}$. Therefore, we can either restrict the gradient term to the normal component or, alternatively, make use of a least-square minimization approach within the context of Mortar domain decomposition methods. We refer to \cite{SCHU201991} for details on the theoretical background and to \cite{dittmann2018c,hesch2019b} for the implementation.}

Note that the periodicity is given in terms of the fluctuation $\fluc{w}$,  i.e.\ with regard to \eqref{eq:mi_kinematic} follows immediately that the geometrical boundaries for a second-order problem are not periodic within the $\RVE$ in contrast to a first-order problem. To be specific, the boundary deformation emanating from $\ma{F}$ is periodic whereas the deformation emanating from $\trima{F}$ is not due to the quadratic formulation in $\mi{X}$. The latter term does not drop out if \eqref{eq:periodic_BC}, left, is formulated in the total deformation $\mi{\varphi}(\mi{X})$. 


\begin{remark}\label{rem:third_gradient}
Third-gradient medium: The proposed formulation at hand can be extended in a straightforward manner towards a macroscopic third-gradient medium with hyperstress $\quarma{P}$ and the conjugate deformation measure $\quarma{F} = \bar{\nabla}^3\ma{\varphi}$. The corresponding application of the energetic criterion reads
\begin{equation}\label{eq:Hill_Mandel_third}
\frac{1}{V}\intRVE\bigl(\mi{P}:\nabla\delta\mi{\varphi}+ \trimi{P}\,\vdots\,\nabla^2\delta\mi{\varphi}\bigr)\,\d V =\ma{P}:\delta\ma{F} +\trima{P}\,\vdots\,\delta\trima{F} +\quarma{P}::\delta\quarma{F} \,.
\end{equation}
The mapping of the microscopic position reads
\begin{equation}
\mi{\varphi}(\mi{X})= \ma{F}\,\mi{X}+\frac{1}{2}\,\trima{F}:(\mi{X}\otimes\mi{X})+\frac{1}{6}\,\quarma{F}\,\vdots\,(\mi{X}\otimes\mi{X}\otimes\mi{X})+\fluc{w}\,.
\end{equation}
Insertion yields the relations
\begin{equation}\label{eq:mi_ma_stresses_third}
\begin{aligned}
\ma{P} &= \frac{1}{V}\intRVE\mi{P}\,\d V,\\
\trima{P} &= \MK{\underbrace{\frac{1}{V}\intRVE\mi{P}\otimes\mi{X}\, \d V}_{\trima{P}^{\mi{P}}}+\underbrace{\frac{1}{V}\intRVE\trimi{P}\, \d V}_{\trima{P}^{\trimi{P}}}},\\
\quarma{P} &= \underbrace{\frac{1}{V}\intRVE\frac{1}{2}\,\mi{P}\otimes \mi{X}\otimes\mi{X}\,\d V}_{\quarma{P}^{\mi{P}}} + \underbrace{\frac{1}{V}\intRVE\trimi{P}\otimes\mi{X}\,\d V}_{\quarma{P}^{\trimi{P}}}\,,
\end{aligned}
\end{equation}
where we have again made use of $\intRVE\mi{X}\d V = \vec{0}$, see Appendices \ref{app:kinematic} to \ref{app:Deriv_boundary}. This yields the set of Dirichlet boundary conditions
\begin{equation}
\begin{aligned}
\ma{F}\,\mi{X}+\frac{1}{2}\,\trima{F}:(\mi{X}\otimes\mi{X}) + \frac{1}{6}\,\quarma{F}\,\vdots\,(\mi{X}\otimes\mi{X}\otimes\mi{X})-\mi{\varphi} &= \vec{0}\,,\\
\ma{F}+\trima{F}\,\mi{X}+\frac{1}{2}\,\quarma{F}\,:\,(\mi{X}\otimes\mi{X})-\mi{F} &= \vec{0}\,,
\end{aligned}
\end{equation}
where we omit again further discussion on possible (periodic) Neumann conditions.  With this at hand, a first-gradient medium within the $\RVE$ can be established by removing all terms related to $\mathfrak{P}$. An extension towards a third-gradient medium within the $\RVE$ seems plausible, but up to now constitutive equations for this need further investigations. Moreover, we note here, that a typical $\RVE$ is in the range of $\upmu\mathrm{m}$, and thus, inhomogeneities in the first Piola-Kirchhoff stress tensor are weighted with $\upmu\mathrm{m}^2$ in $\quarma{P}$, which is often negligible and the reason, why we do not further take this into account here. For further information on scale separation,  see Schr\"oder \cite{schroeder2014}.
\end{remark}

\section{Consistent linearization and discretization}\label{sec:discretization}
For the computation of the macroscopic boundary value problem with attached \MK{mesoscopic} $\RVE$s, we introduce here the $\text{IGA}^2$-method, analogous to the $\text{FE}^2$-method, see Schr\"oder \cite{schroeder2014} and references therein.  We omit here details on the spline-based discretization of the macroscale within the concept of $\text{IGA}$, as numerous papers have already presented this and instead focus on the \MK{mesoscopic} $\RVE$, assuming that the discrete macroscopic quantities of the deformations ($\ma{F}$, $\trima{F}$) are known at the particular Newton step.  Note that higher-order continua at the macroscale require appropriate continuity of the spline based discretization. 

Thus,  in a first step the macroscopic quantities ($\ma{F}$, $\trima{F}$) are transferred to the \MK{mesoscale} $\RVE$ at every material point, see Figure \ref{fig:Macro_RVE}. After that, the boundary value problem on the \MK{mesoscale} is solved using suitable boundary conditions and the homogenization is performed using volumetric averaged \MK{mesoscopic} quantities as well as the linearization of these quantities. In the last step, the macroscopic boundary value problem is solved and the next Newton iteration starts. 

\subsection{Linearization of macroscopic stresses and hyperstresses}
Since the macroscopic boundary value problem is solved with a Newton-Raphson iteration, we need a consistent linearization of the macroscopic field equations. Therefore, it is necessary to linearize the stresses $\ma{P}$ and $\trima{P}$,  evaluated via the incremental relations:
\begin{equation}\label{eq:incremental_rel}
\Delta\ma{P}:=\frac{\partial \ma{P}}{\partial\ma{F}}:\Delta\ma{F}+\frac{\partial \ma{P}}{\partial\trima{F}}\,\vdots\,\Delta\trima{F}\qquad\text{and}\qquad\Delta\trima{P}:=\frac{\partial \trima{P}}{\partial\ma{F}}:\Delta\ma{F}+\frac{\partial \trima{P}}{\partial\trima{F}}\,\vdots\,\Delta\trima{F}\,.
\end{equation}
However, the macroscopic quantities are given by the averaged \MK{mesoscopic} stresses and hyperstresses, hence $\ma{P}:=\ma{P}(\mi{P}(\mi{F},\trimi{F}))$ and $\trima{P}:=\trima{P}(\mi{P}(\mi{F},\trimi{F}),\trimi{P}(\mi{F},\trimi{F}))$, see (\ref{eq:mi_ma_stresses}). Thus we have to use the chain rule for the partial derivative of the macroscopic stresses with respect to the corresponding deformations and end up after some calculations with:
\begin{equation}\label{eq:deriv_P}
\begin{aligned}
\left[\Delta\ma{P}\right]_{iJ}=\phantom{+\,}&\frac{1}{V}\intRVE\left[\quarmi{C}\right]_{iJsT}\,\d V\,\left[\Delta\ma{F}\right]_{sT}\\
+\,&\frac{1}{V}\intRVE\bigl(\left[\quarmi{C}\right]_{iJsT}\,\left[\mi{X}\right]_{U}+\left[\fmi{D}\right]_{iJsTU}\bigr)\d V\,\left[\Delta\trima{F}\right]_{sTU}\\
+\,&\frac{1}{V}\intRVE\bigl(\left[\quarmi{C}\right]_{iJsT}\,\left[\Delta\fluc{F}\right]_{sT}+\left[\fmi{D}\right]_{iJsTU}\,\left[\Delta\trifluc{F}\right]_{sTU}\bigr)\d V
\end{aligned}
\end{equation}

and
\begin{equation}\label{eq:deriv_tri_P}
\begin{aligned}
\left[\Delta\trima{P}\right]_{iJK}=\phantom{+\,}&\frac{1}{V}\intRVE\bigl(\left[\quarmi{C}\right]_{iJsT}\,\left[\mi{X}\right]_{K}+\left[\fmi{E}\right]_{iJKsT}\bigr)\,\d V\,\left[\Delta\ma{F}\right]_{sT}\\
+\,&\frac{1}{V}\intRVE\bigl(\left[\quarmi{C}\right]_{iJsT}\,\left[\mi{X}\right]_{K}\,\left[\mi{X}\right]_{U}+\left[\fmi{D}\right]_{iJsTU}\,\left[\mi{X}\right]_{K}\bigr.\\
&\hspace*{2cm}+\bigl.\left[\fmi{E}\right]_{iJKsT}\,\left[\mi{X}\right]_{U}+\left[\smi{G}\right]_{iJKsTU}\bigr)\d V\,\left[\Delta\trima{F}\right]_{sTU}\\
+\,&\frac{1}{V}\intRVE\bigl(\left[\quarmi{C}\right]_{iJsT}\,\left[\mi{X}\right]_{K}+\left[\fmi{E}\right]_{iJKsT}\bigr)\,\left[\Delta\fluc{F}\right]_{sT}\,\d V\\
+\,&\frac{1}{V}\intRVE\bigl(\left[\fmi{D}\right]_{iJsTU}\,\left[\mi{X}\right]_{K}+\left[\smi{G}\right]_{iJKsTU}\bigr)\,\left[\Delta\trifluc{F}\right]_{sTU}\d V\,,
\end{aligned}
\end{equation}
where the derivatives of the stresses are defined by:
\begin{equation}\label{eq:deriv_stresses}
\begin{aligned}
\quarmi{C}:=\frac{\partial\mi{P}}{\partial\mi{F}}\quad,\quad\fmi{D}:=\frac{\partial\mi{P}}{\partial\trimi{F}}\quad,\quad\fmi{E}:=\frac{\partial\trimi{P}}{\partial\mi{F}}\quad\text{and}\quad\smi{G}:=\frac{\partial\trimi{P}}{\partial\trimi{F}}\,,
\end{aligned}
\end{equation}
see Appendix \ref{app:linearization} for more details.  It is obvious, that the linearizations of the macroscopic stresses $\ma{P}$ and $\trima{P}$ depend on the sensitivity of the \MK{mesoscopic} fluctuations $\Delta\fluc{F}$ and $\Delta\trifluc{F}$, defined in \eqref{eq:deform}.  The correlation between these sensitivities and the change of the corresponding macroscopic fields $\Delta\ma{F}$ and $\Delta\trima{F}$ can be done in the discrete setting by linearization of the virtual work of the \MK{mesoscopic} boundary value problem in the solution point, as shown next.

\subsection{Linearization of \MK{mesoscopic} boundary value problem}\label{sec:linearization}
The relationship between these sensitivities and the macroscopic fields follows from the \MK{mesoscopic} boundary value problem.  With regard to (\ref{eq:micro_strong}) and assuming that $\delta\fluc{w} = \vec{0}$ holds on the whole boundary,  we obtain
\begin{equation}
G:=\intRVE\left(\mi{P}:\delta\fluc{F}+\trimi{P}\,\vdots\,\delta\trifluc{F}\right)\,\d V\,.
\end{equation}
Solving the problem such that $G = 0$, it follows immediately that  $\Delta G=0$.  Hence, the linearization in the equilibrium state reads 
\begin{equation}\label{eq:micro_equilibrium}
\Delta G:=\intRVE\left(\delta\fluc{F}:\left[\quarmi{C}:\Delta\mi{F}+\fmi{D}\,\vdots\,\Delta\trimi{F}\right]+\delta\trifluc{F}\,\vdots\,\left[\fmi{E}:\Delta\mi{F}+\smi{G}\,\vdots\,\Delta\trimi{F}\right]\right)\,\d V=0\,,
\end{equation}
where
\begin{equation}\label{eq:incremental_def}
\Delta\mi{F}=\Delta\ma{F}+\Delta\trima{F}\,\mi{X}+\Delta\fluc{F}\qquad\text{and}\qquad\Delta\trimi{F}=\Delta\trima{F}+\Delta\trifluc{F}\,.
\end{equation}
This can be evaluated in the discrete setting, as will be shown next. 

\subsection{\MK{Mesoscopic} finite element approximation}
Next, we have to approximate the fluctuation field, the virtual and the incremental fluctuation fields:
\begin{equation}\label{eq:approx_fluc}
\fluc{w}\h=\sum\limits_{A\in\mathcal{I}}R^A\,\fluc{q}^A\quad,\quad\delta\fluc{w}\h=\sum\limits_{A\in\mathcal{I}}R^A\,\delta\fluc{q}^A\quad\text{and}\quad\Delta\fluc{w}\h=\sum\limits_{A\in\mathcal{I}}R^A\,\Delta\fluc{q}^A\,,
\end{equation} 
where $R^A:\RVE\rightarrow\mathbb{R}$ are B-Spline\footnote{B-Splines are used without loss of generality,  NURBS can also be applied if necessary.} based shape functions of order $p$ with associated control points $A\in\mathcal{I}={1,\hdots,m}$ with the overall number of control points $m$. Furthermore, $\left[\fluc{q}^A,\,\delta\fluc{q}^A,\,\Delta\fluc{q}^A\right]\in\mathbb{R}^3$. So, the deformation tensors lead to the approximation
\begin{equation}\label{eq:approx_def}
\fluc{F}\h=\sum\limits_{A\in\mathcal{I}}\fluc{q}^A\otimes \nabla R^A\quad\text{and}\quad \trifluc{F}\h=\sum\limits_{A\in\mathcal{I}}\fluc{q}^A\otimes \nabla^2 R^A\,,
\end{equation} 
which are given analogously for the  virtual ($\delta\fluc{F}\h$, $\delta\trifluc{F}\h$) and the incremental ($\Delta\fluc{F}\h$, $\Delta\trifluc{F}\h$) deformation tensors.  Note that we can also discretize the displacement field $\vec{\varphi}(\mi{X})$ using \eqref{eq:mi_kinematic} as well. 

For the boundary conditions we first introduce Dirichlet conditions as presented in \eqref{eq:dirichlet}.  For the implementation of a first-order mesoscale continuum is straightforward, as we only have to deal with linear conditions in $\mi{X}$. Using open knot vectors, which are interpolatory at the boundaries, the control points of the spline has to be distributed linearly along the boundaries of the $\RVE$.  For higher-order problems, we obtain  quadratic (second-order formulations) and cubic (third-order formulations) boundaries in $\mi{X}$. Therefore, we make use of a least-square optimization for the ease of implementation.  However, the problem itself can be solved exactly, i.e.\  quadratic \MK{or higher order} splines can reproduce a quadratic boundary, c.f.\ \cite{cottrell2009}. Introducing a set of evaluation points $\hat{\vec{q}}_i$ along the boundary and a set of control points $\fluc{q}_j$ for the splines-based discretization of the discrete boundary $\partial\RVE^{\h}$, the least-square problem reads
\begin{equation}\label{eq:least}
\{\fluc{q}_j\} = \underbrace{\text{min}}_{\fluc{q}_i\in \partial\RVE^{\h}}\|\hat{\vec{q}}_i - \sum\limits_{j}R^j(\vec{\xi}_i)\, \fluc{q}^j\|.
\end{equation}

Note, that $\hat{\vec{q}}_i = \ma{F}\,{\vec{q}}_i+\frac{1}{2}\,\trima{F}:({\vec{q}}_i\otimes{\vec{q}}_i)\,$ and $\nabla\hat{\vec{q}}_i = \ma{F}+\trima{F}\,{\vec{q}}_i\,$, with the position of the evaluation point in the reference configuration ${\vec{q}}_i$.  \MK{We refer to the textbook \cite{cottrell2009} and the discussion therein on the enforcement of Dirichlet conditions for further information on the evaluation of the least-square problem.} For second-order boundaries, the least-square problem is expanded by the constraint $\nabla\fluc{w}(\mi{X})= \vec{0}$ on all surfaces to
\begin{equation}\label{eq:least_diri_2ndOrder}
\{\fluc{q}_j\} = \underbrace{\text{min}}_{\fluc{q}_i\in \partial\RVE^{\h}} \begin{Vmatrix} \hat{\vec{q}}_i - \sum\limits_{j}R^j(\vec{\xi}_i)\, \fluc{q}^j \\ \nabla\hat{\vec{q}}_i - \sum\limits_{j}\nabla R^j(\vec{\xi}_i)\, \fluc{q}^j\end{Vmatrix} .
\end{equation}
For periodic boundary conditions,  we have to ensure that \eqref{eq:periodic_BC}, left, is valid. \MK{For general higher-order domain decomposition problems using non-conforming meshes, we refer to our previous developments in \cite{dittmann2018c, hesch2019b}, applied here on conforming meshes. } For the ease of implementation, we note that a least-square optimization using
\begin{equation}\label{eq:least_peri_2ndOrder}
\{\fluc{q}_j^-\} = \underbrace{\text{min}}_{\fluc{q}_i^-\in \partial\RVE^{\h}} \begin{Vmatrix}    \sum\limits_{k}R^k(\vec{\xi}_i^+)\, \fluc{q}^{k} -\sum\limits_{j}R^j(\vec{\xi}_i^-)\, \fluc{q}^{j} + \left(\hat{\vec{q}}_i^+ - \hat{\vec{q}}_i^- \right) \\ \sum\limits_{k}\nabla R^k(\vec{\xi}_i^+)\, \fluc{q}^{k} -\sum\limits_{j}\nabla R^j(\vec{\xi}_i^-)\, \fluc{q}^{j} + \left(\nabla\hat{\vec{q}}_i^+ - \nabla\hat{\vec{q}}_i^- \right) \end{Vmatrix}\, ,
\end{equation}
can also be applied, leaving a nodal dependency in the form $\fluc{q}_j^- := \fluc{q}_j^-(\fluc{q}_j^+)$ for the set of opposing evaluation points $\{\vec{q}_i^+,\vec{q}_i^-\}$.

Next, we can establish a relationship between the mesoscopic sensitivities and the change of corresponding macroscopic fields. For this, we discretize the last two sections in reverse order and insert the approximations in a first step in the equilibrium state of the mesoscopic boundary value problem (\ref{eq:micro_equilibrium})
\begin{equation}\label{eq:discr_micro_equil}
\begin{aligned}
\Delta G\h:=\phantom{+}&\intRVEh\delta\fluc{F}\h:\left[\quarmi{C}\h:\left(\Delta\ma{F}+\Delta\trima{F}\,\mi{X}\h+\Delta\fluc{F}\h\right)+\fmi{D}\h\,\vdots\,\left(\Delta\trima{F}+\Delta\trifluc{F}\h\right)\right]\,\d V\\
+&\intRVEh\delta\trifluc{F}\h\,\vdots\,\left[\fmi{E}\h:\left(\Delta\ma{F}+\Delta\trima{F}\,\mi{X}\h+\Delta\fluc{F}\h\right)+\smi{G}\h\,\vdots\,\left(\Delta\trima{F}+\Delta\trifluc{F}\h\right)\right]\,\d V=0\,,
\end{aligned}
\end{equation}
where the discrete derivatives of the stresses are defined by
\begin{equation}\label{eq:discr_stresses}
\begin{aligned}
\quarmi{C}\h:=\quarmi{C}\left(\mi{F}\h,\,\trimi{F}\h\right)\,,\,\fmi{D}\h:=\fmi{D}\left(\mi{F}\h,\,\trimi{F}\h\right)\,,\,\fmi{E}\h:=\fmi{E}\left(\mi{F}\h,\,\trimi{F}\h\right)\,\text{and}\,\smi{G}\h:=\smi{G}\left(\mi{F}\h,\,\trimi{F}\h\right)\,.
\end{aligned}
\end{equation}
After some calculations, see Appendix \ref{app:approx_BVP} for further information, we arrive at the discrete correlation between the mesoscopic sensitivities and the change of corresponding macroscopic fields:
\begin{equation}\label{eq:correl_fluc_macro}
\left[\Delta\fluc{q}\right]^{B}_{s}=-\left(\left[\mi{K}\right]^{AB}_{ls}\right)^{-1}\,\left(\left[\trimi{L}\right]^{A}_{lrT}\,\left[\Delta\ma{F}\right]_{rT}+\left[\quarmi{M}\right]^{A}_{lrTU}\,\left[\Delta\trima{F}\right]_{rTU}\right)\,.
\end{equation}
Here, $\mi{K}$, $\trimi{L}$ and $\quarmi{M}$ are the stiffness matrices of the mesoscopic boundary value problem.

In a second step, we discretize the macroscopic stresses (\ref{eq:deriv_P}) and (\ref{eq:deriv_tri_P}), where we use the correlation of the mesoscopic  sensitivities to the macroscopic quantities (\ref{eq:correl_fluc_macro}) and end up in:
\begin{equation}\label{eq:disc_lin_P}
\begin{aligned}
\left[\Delta\ma{P}\right]\h_{iJ}=\phantom{+\,}&\biggl\{\left[\quarmi{V}^{\quarmi{C}}\right]\h_{iJrT}-\left[\trimi{N}\,\right]^{B}_{iJs}\,\left(\left[\mi{K}\right]^{AB}_{ls}\right)^{-1}\,\left[\trimi{L}\right]^{A}_{lrT}\biggr\}\,\left[\Delta\ma{F}\right]_{rT}\\
+\,&\biggl\{\left[\fmi{V}^{\quarmi{C}\fmi{D}}\right]\h_{iJrTU}-\left[\trimi{N}\right]^{B}_{iJs}\,\left(\left[\mi{K}\right]^{AB}_{ls}\right)^{-1}\,\left[\quarmi{M}\right]^{A}_{lrTU}\biggr\}\,\left[\Delta\trima{F}\right]_{rTU}\,,\\
\end{aligned}
\end{equation}
for the linearization of the stresses and:
\begin{equation}\label{eq:disc_lin_triP}
\begin{aligned}
\left[\Delta\trima{P}\right]\h_{iJK}=\phantom{+\,}&\bigl\{\left[\fmi{V}^{\quarmi{C}\fmi{E}}\right]\h_{iJKrT}-\left[\quarmi{N}\right]^{B}_{iJKs}\,\left(\left[\mi{K}\right]^{AB}_{ls}\right)^{-1}\,\left[\trimi{L}\right]^{A}_{lrT}\bigr\}
\,\left[\Delta\ma{F}\right]_{rT}\\
 +\,&\bigl\{\left[\smi{V}^{\quarmi{C}\fmi{D}\fmi{E}\smi{G}}\right]\h_{iJKrTU}-\left[\quarmi{N}\right]^{B}_{iJKs}\,\left(\left[\mi{K}\right]^{AB}_{ls}\right)^{-1}\,\left[\quarmi{M}\right]^{A}_{lrTU}\bigr\}\left[\Delta\trima{F}\right]_{rTU}\,,\\
\end{aligned}
\end{equation}
for the linearization of the hyperstresses, see Appendix \ref{app:discr_lin_stresses} for further information on the volume averaged tensors $\quarmi{V}^{\quarmi{C}}$, $\fmi{V}^{\quarmi{C}\fmi{D}}$, $\fmi{V}^{\quarmi{C}\fmi{E}}$ and $\smi{V}^{\quarmi{C}\fmi{D}\fmi{E}\smi{G}}$ as well as $\trimi{N}$ and $\quarmi{N}$. This description of the linearization of the macroscopic stresses and hyperstresses is free of the \MK{mesoscopic} fluctuations $\fluc{w}$  and the discretized version $\fluc{q}^A$, respectively.  


\section{Numerical experiments}\label{sec:numerics}
In this section we investigate the performance and accuracy of the homogenization technique for different materials. We start with some benchmark tests for a Mooney-Rivlin material and a second-gradient material for fiber-reinforced polymers for the $\RVE$. Afterwards, we use this second-gradient material for investigations using the well-known Cook's membrane.

\subsection{Benchmark test: Mooney-Rivlin material}\label{sec:Mooney}

\renewcommand{\pA}{\phantom{-}0.897}
\renewcommand{\pB}{\phantom{-}0.500}
\renewcommand{\pC}{-0.400}
\renewcommand{\pD}{-0.070}
\renewcommand{\pE}{\phantom{-}1.001}
\renewcommand{\pF}{-0.100}
\renewcommand{\pG}{\phantom{-}0.082}
\renewcommand{\pH}{\phantom{-}0.020}
\renewcommand{\pI}{\phantom{-}0.997}

\begin{figure}[ht!]
	\centering
	\begin{minipage}[top]{0.67\textwidth}
	\psfrag{d}[][]{$\partial\RVE:\,\tilde{\mi{w}}=\mi{0}\,\unit{mm}$}
	\psfrag{X}{$\mi{X}$}
	\psfrag{-}[][]{$-$}
	\psfrag{+}[][]{$+$}
	\psfrag{p}[][]{pcs: $\fluc{w}(\mi{X}^+)=\fluc{w}(\mi{X}^-)$}
	\psfrag{c}[l][l]{cn: $\tilde{\mi{w}}=\mi{0}\,\unit{mm}$}
		\includegraphics[width=\textwidth]{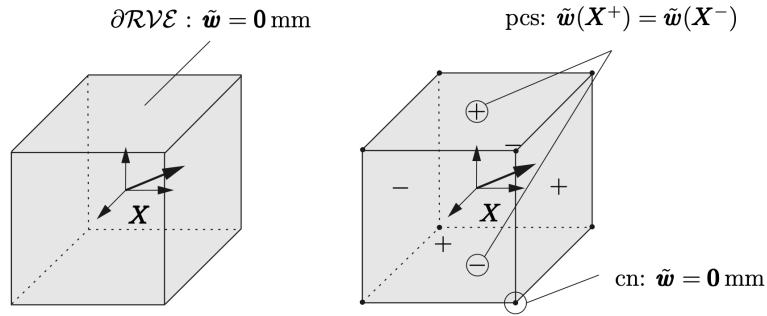}
	\end{minipage}
	\caption{\textbf{Mooney-Rivlin material.} Left: $\RVE$ (edge length $0.1$ mm) with Dirichlet boundaries $\tilde{\mi{w}}=\mi{0}\,\unit{mm}$ on $\partial\RVE$. Right: $\RVE$ with periodic boundary conditions $\fluc{w}(\mi{X}^+)=\fluc{w}(\mi{X}^-)$ for the periodically contiguous surfaces (pcs) top-bottom, right-left, front-back and eight constrained  corner nodes (cn) with $\tilde{\mi{w}}=\mi{0}\,\unit{mm}$.}
	\label{fig:MR_BC}
\end{figure}

As a first proof of concept, we investigate $\RVE$ using a homogeneous Mooney-Rivlin material, see \cite{khristenko2021}. The edge length of the $\RVE$ cube is $0.1$ mm and the coordinate system is fixed in the center of the cube, see Figure \ref{fig:MR_BC}. 
The first-order constitutive relation is given by
\begin{equation}\label{MR_material}
\Psi(J, I_1,I_2) = c\,(J-1)^2 - d\, \op{ln}(J) + c_1\,(I_1-3) + c_2\,(I_2-3)\,.
\end{equation}
Here, $J = \op{det}(\vec{F})$, $I_1 = \op{tr}(\vec{F}\tp\,\vec{F}) = \vec{F}:\vec{F}$ and $I_2 = \op{tr}(\op{cof}(\vec{F}\tp\,\vec{F}))$. Moreover, $c = 1/3\,(c_1 + c_2)$, $d = 2\,(c_1 + 2\,c_2)$, {$c_1 = \pMRcOne\, \unit{MPa}$ and $c_2 = \pMRcTwo\,\unit{MPa}$}. To test the $\RVE$, we define the macroscopic deformation tensor:
\begin{equation}\label{macro_F}
\ma{F} :=
\begin{bmatrix}
{\pA} &{\pB} &{\pC} \\
{\pD} &{\pE} &{\pF} \\
{\pG} &{\pH} &{\pI} 
\end{bmatrix}\,,
\end{equation}
and assume the macroscopic second gradient to be $\trima{F}:=\trimi{0}$. With this information, we solve the microscopic boundary value problem, where we apply in a first step Dirichlet boundaries on $\partial\RVE$ and in a second step periodic boundaries, see Figure \ref{fig:MR_BC} for details. 

\renewcommand{\pA}{$6132.72515830 \, \unit{MPa}$}
\renewcommand{\pB}{$6132.72515830 \, \unit{MPa}$}

\begin{figure}[ht!]
\begin{minipage}{0.49\textwidth}
\includegraphics[width=\textwidth]{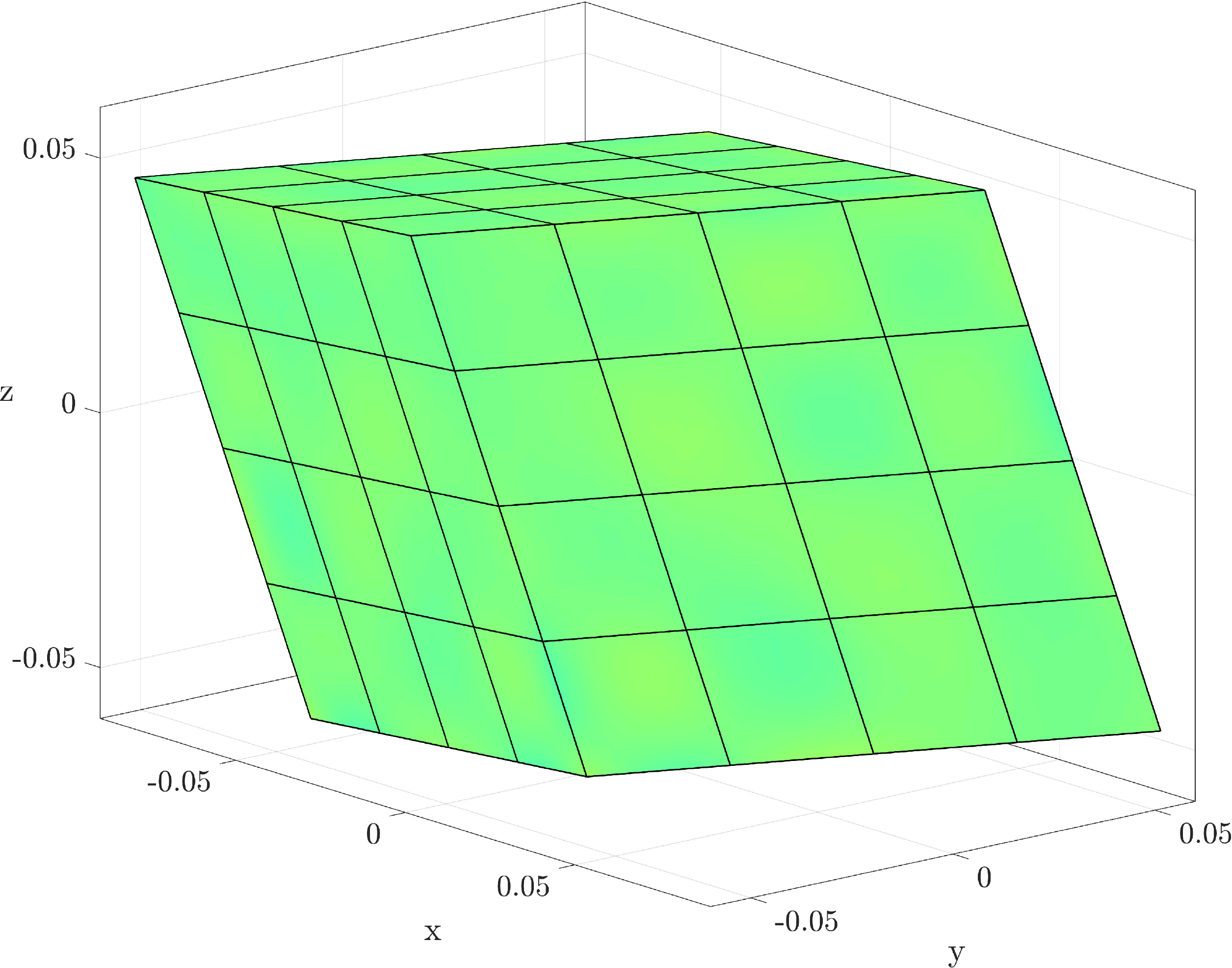}
\end{minipage}\hfill
\begin{minipage}{0.49\textwidth}
\includegraphics[width=\textwidth]{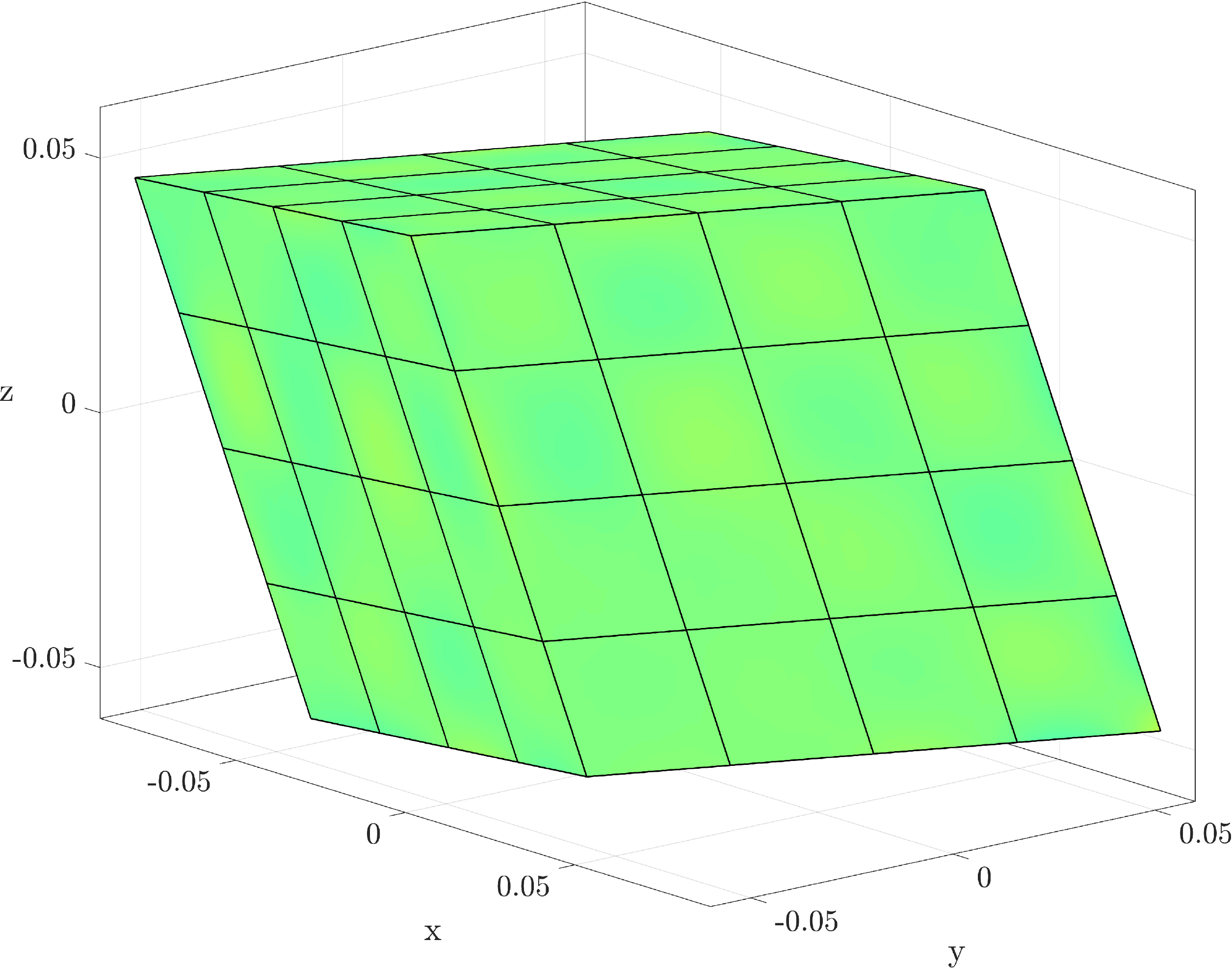}
\end{minipage}\\
\begin{minipage}{0.49\textwidth}
\includegraphics[width=\textwidth]{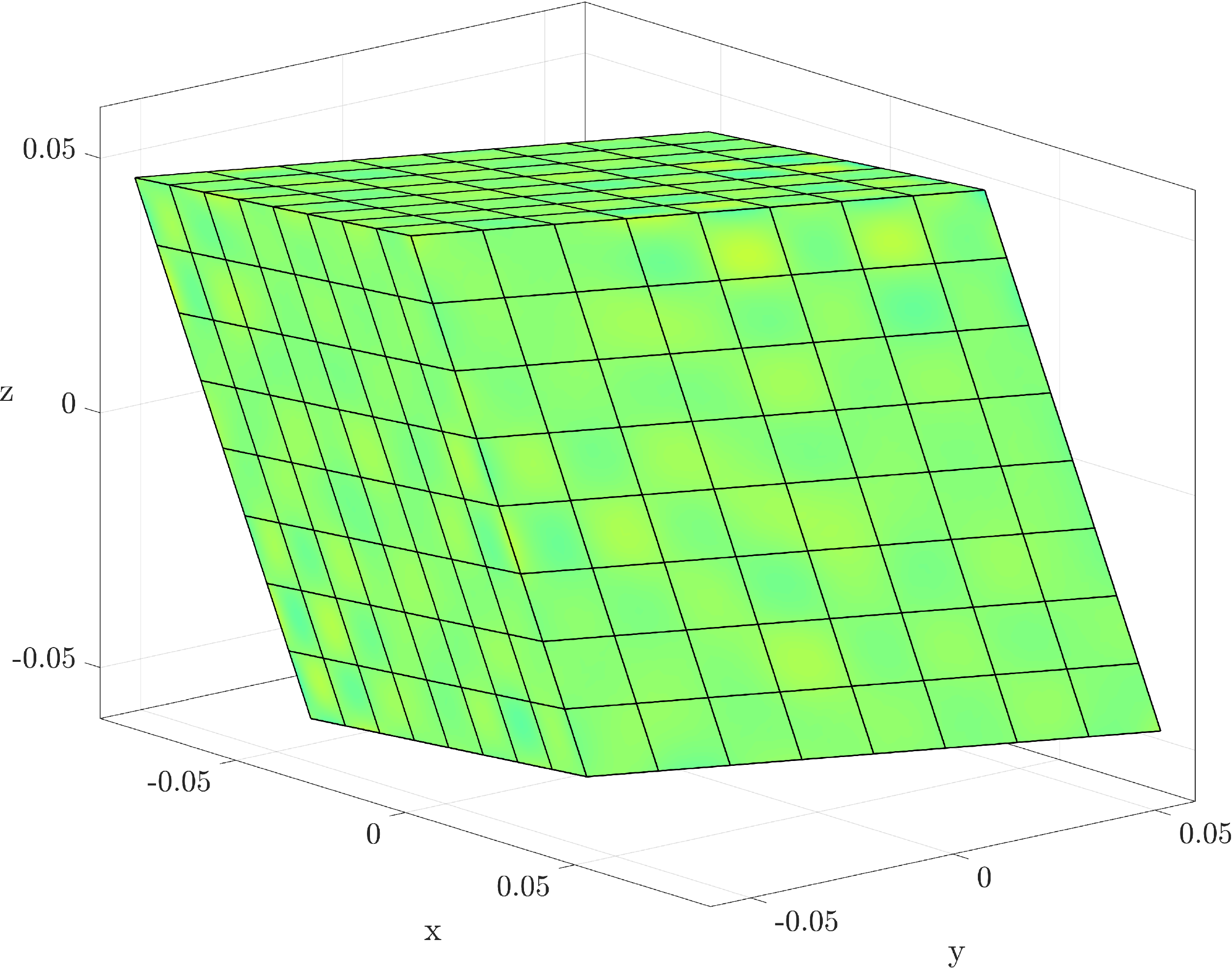}
\end{minipage}\hfill
\begin{minipage}{0.49\textwidth}
\includegraphics[width=\textwidth]{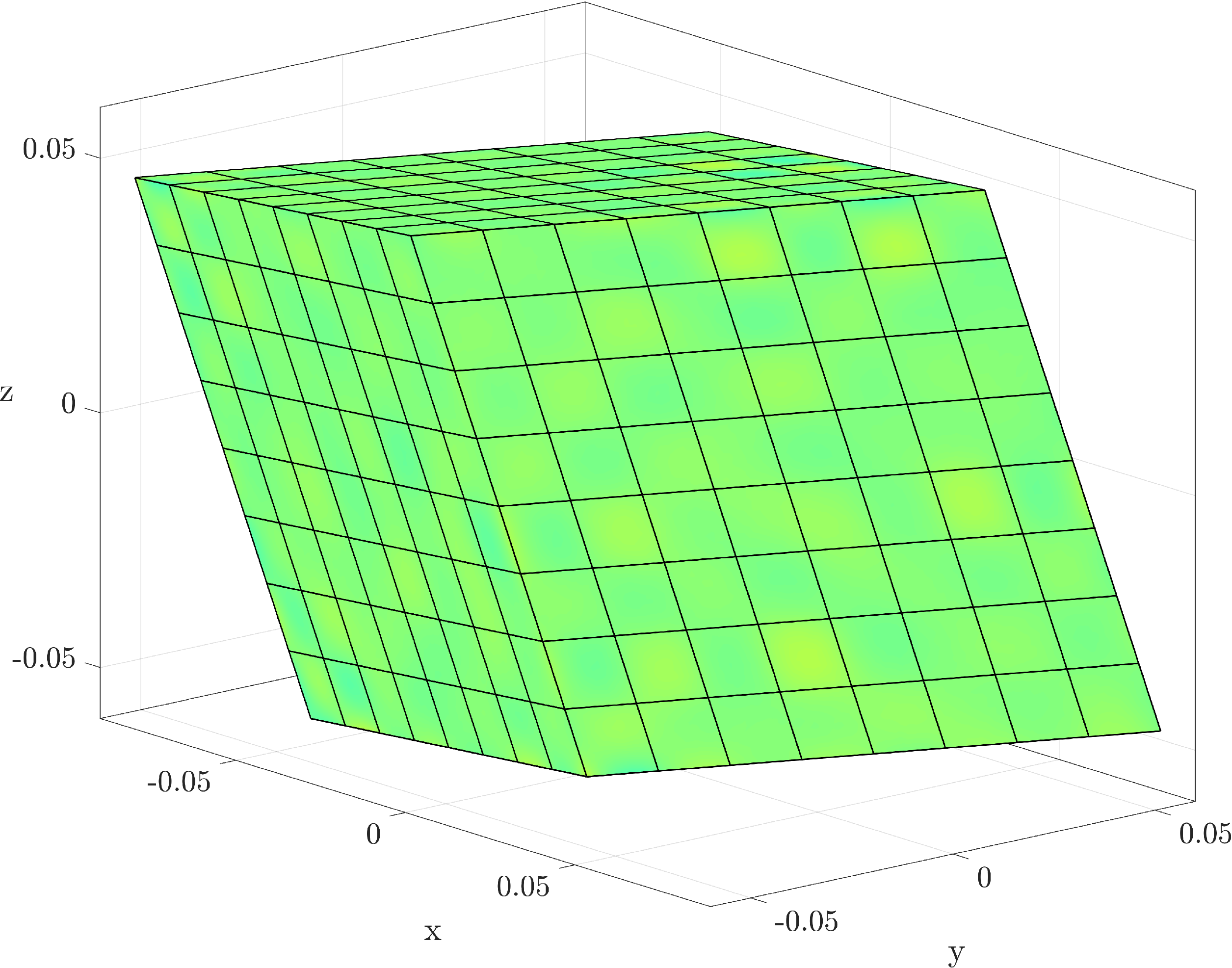}
\end{minipage}\\
\begin{minipage}{0.49\textwidth}
\includegraphics[width=\textwidth]{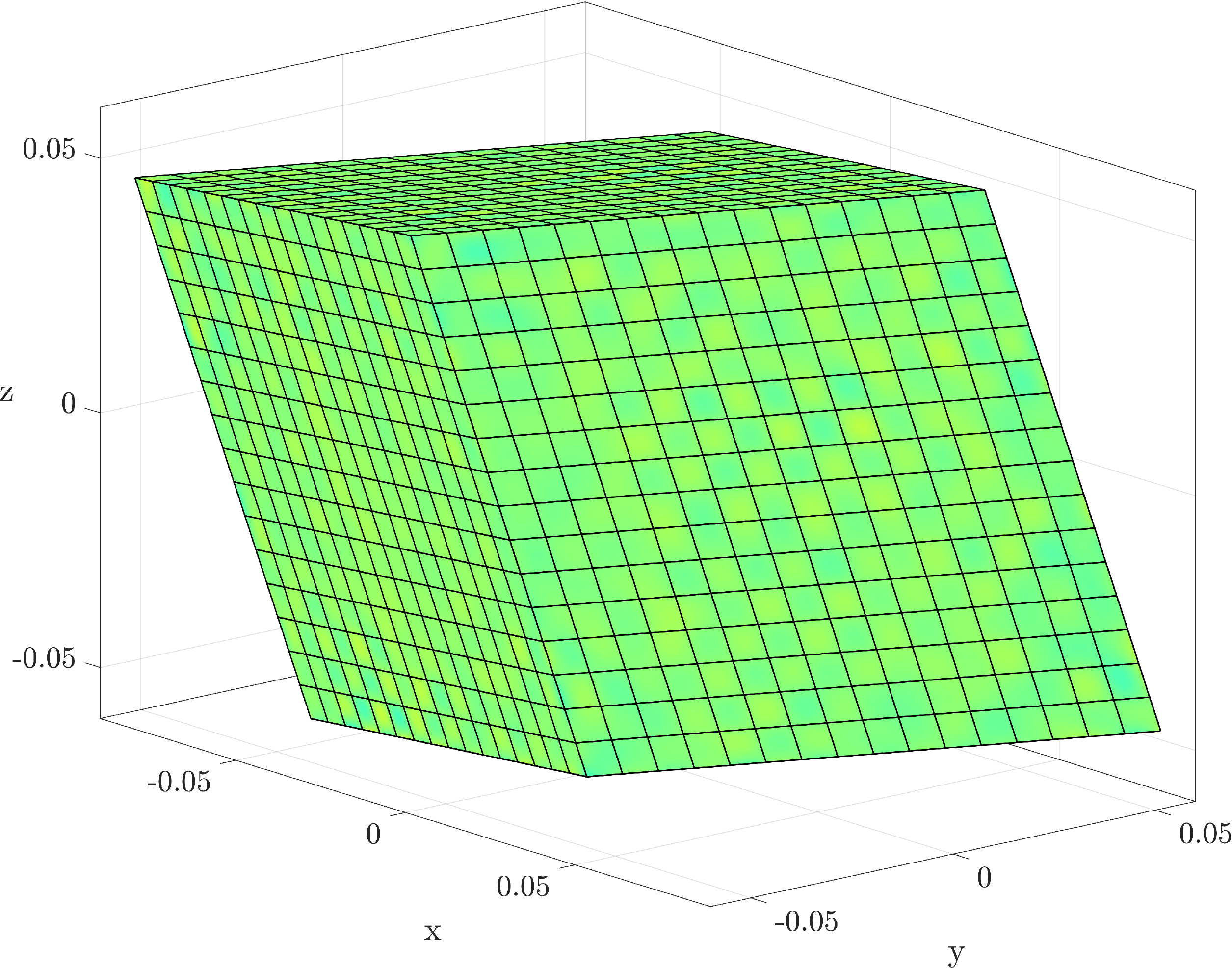}
\end{minipage}\hfill
\begin{minipage}{0.49\textwidth}
\includegraphics[width=\textwidth]{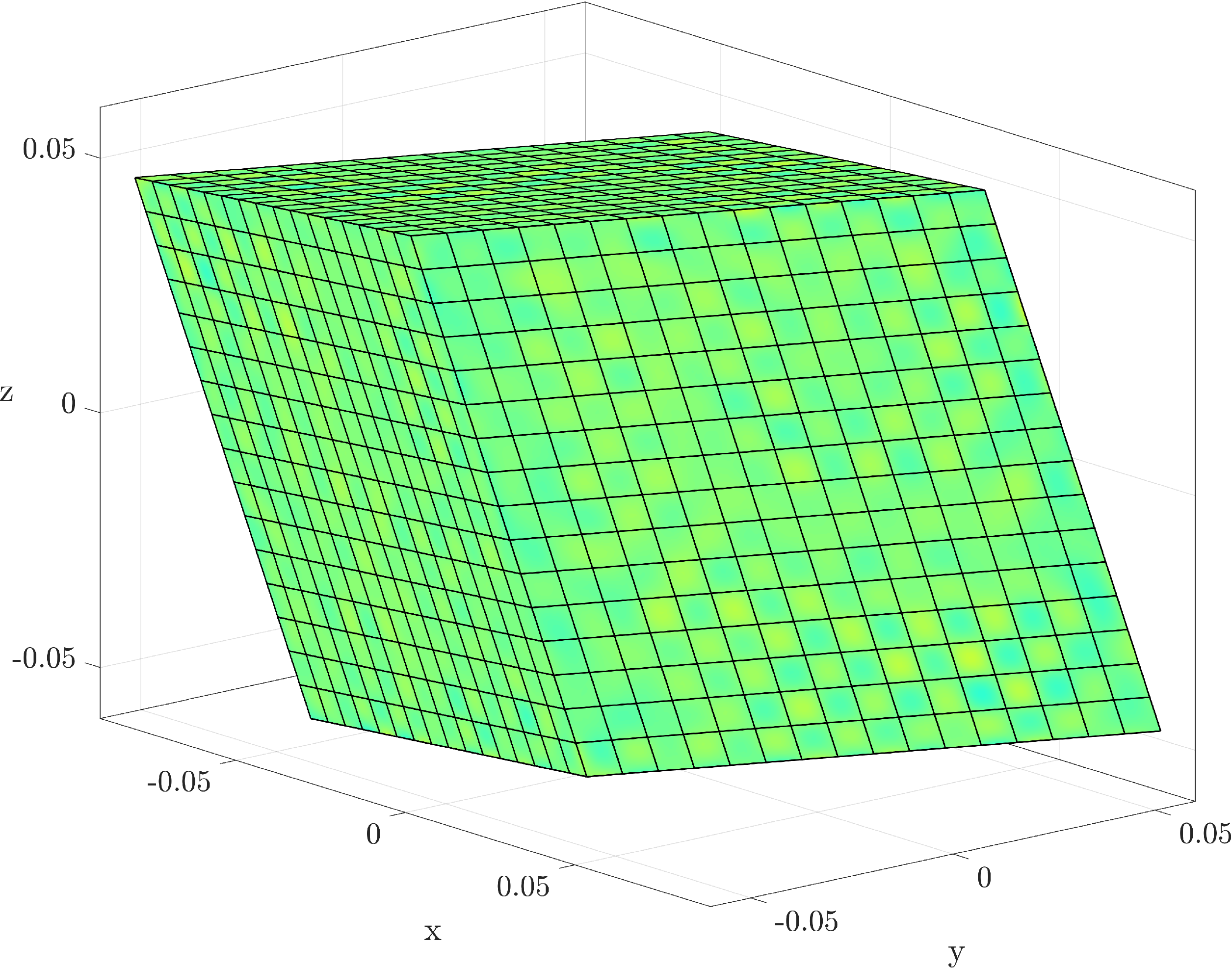}
\end{minipage}\\
\centering
\psfrag{0}[c][c]{\small\pA}
\psfrag{1}{}
\psfrag{2}[c][c]{\small\pB}
\includegraphics[width=0.5\textwidth]{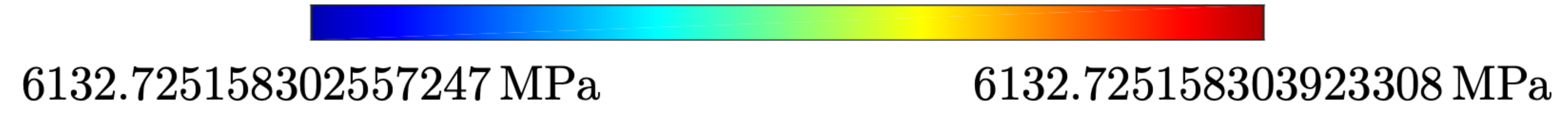}
\caption{\textbf{Mooney-Rivlin material.} Von Mises stresses - left to right: $\RVE$ with Dirichlet and periodic boundaries, top to bottom: $4$, $8$ and $16$ elements in each direction.}
\label{fig:Mooney_stress}
\end{figure}

In Figure \ref{fig:Mooney_stress}, the von Mises stresses are plotted for the $\RVE$ with Dirichlet and periodic boundary conditions. In particular, we increase the number of elements in each direction of the cube from $4$, $8$ to $16$ elements using B-splines of order $p=2$.  Since the Mooney-Rivlin material is of first order with linear constraints on the boundary, we obtain a homogeneous distribution of the stress field. 

\renewcommand{\pA}{$6.18\mathrm{E}{-}16$} 
\renewcommand{\pB}{$3.38\mathrm{E}{-}16$} 
\renewcommand{\pC}{$5.03\mathrm{E}{-}16$} 
\renewcommand{\pD}{$2.23\mathrm{E}{-}16$} 
\renewcommand{\pE}{$4.36\mathrm{E}{-}16$} 

\renewcommand{\pF}{$3.71\mathrm{E}{-}16$} 
\renewcommand{\pG}{$2.24\mathrm{E}{-}16$} 
\renewcommand{\pH}{$3.49\mathrm{E}{-}16$} 
\renewcommand{\pI}{$1.48\mathrm{E}{-}16$} 
\renewcommand{\pJ}{$4.00\mathrm{E}{-}16$} 

\renewcommand{\pK}{$1.00\mathrm{E}{-}16$} 
\renewcommand{\pL}{$1.43\mathrm{E}{-}15$} 
\renewcommand{\pM}{$1.62\mathrm{E}{-}15$} 
\renewcommand{\pN}{$4.45\mathrm{E}{-}16$} 
\renewcommand{\pO}{$1.27\mathrm{E}{-}15$} 

\renewcommand{\pP}{$1.00\mathrm{E}{-}16$} 
\renewcommand{\pQ}{$1.43\mathrm{E}{-}15$} 
\renewcommand{\pR}{$1.62\mathrm{E}{-}15$} 
\renewcommand{\pS}{$4.45\mathrm{E}{-}16$} 
\renewcommand{\pT}{$1.26\mathrm{E}{-}15$} 

\renewcommand{\pU}{$8.29\mathrm{E}{-}15$} 
\renewcommand{\pV}{$7.65\mathrm{E}{-}15$} 
\renewcommand{\pW}{$1.19\mathrm{E}{-}14$} 
\renewcommand{\pX}{$4.68\mathrm{E}{-}15$} 
\renewcommand{\pY}{$1.27\mathrm{E}{-}14$} 

\renewcommand{\pZ}{$8.29\mathrm{E}{-}15$} 
\renewcommand{\pAA}{$7.65\mathrm{E}{-}15$} 
\renewcommand{\pAB}{$1.20\mathrm{E}{-}14$} 
\renewcommand{\pAC}{$4.68\mathrm{E}{-}15$} 
\renewcommand{\pAD}{$1.27\mathrm{E}{-}14$} 

\renewcommand{\pAE}{$1.48\mathrm{E}{-}16$}
\renewcommand{\pAF}{$1.27\mathrm{E}{-}14$}

\begin{table}[ht!]
\begin{center}
\small
\begin{tabular}{lcccccc}
\toprule
Elements&\multicolumn{2}{c}{$4\times4\times4$}&\multicolumn{2}{c}{$8\times8\times8$}&\multicolumn{2}{c}{$16\times16\times16$}\\ 
Boundary&Dirichlet& Periodic&Dirichlet& Periodic&Dirichlet& Periodic \\
\midrule
$E_{\mathrm{max}}(\bar{\Psi})$ 					 & \pA & \pF & \pK & \pP & \pU & \pZ \\ 
$E_{\mathrm{max}}(\ma{P})$         			 & \pB & \pG & \pL & \pQ & \pV & \pAA \\ 
$E_{\mathrm{norm}}(\ma{P})$        			 & \pC & \pH & \pM & \pR & \pW & \pAB \\ 
$E_{\mathrm{max}}(\partial_{\ma{F}}\ma{P})$  & \pD & \pI & \pN & \pS & \pX & \pAC \\ 
$E_{\mathrm{norm}}(\partial_{\ma{F}}\ma{P})$ & \pE & \pJ & \pO & \pT & \pY & \pAD \\ 
\bottomrule
\end{tabular}
\end{center}
\caption{\textbf{Mooney-Rivlin material.} Relative maximum error of the energies $E_{\mathrm{max}}(\bar{\Psi})$ (1st row). Relative maximum error $E_{\mathrm{max}}(\bullet)$ and relative error in the norm $E_{\mathrm{norm}}(\bullet)$ for the stresses and tangent (2nd - 5th row). Here, for $4$, $8$ and $16$ elements in each direction and the Dirichlet and periodic boundaries, respectively.}
\label{table:Mooney_error}
\end{table}

Since we use an energetic criterion within the homogenization, we compare the maximum error $E_{\mathrm{max}}(\Psi)$ of the (analytically evaluated) strain energy $\Psi_{\mathrm{ana}} := \Psi(\ma{F})$ with the averaged strain energy $\bar{\Psi}_{\RVE}:=\frac{1}{V}\int_{\RVE}\Psi(\mi{F}\h)\d V$ of the $\RVE$, see Table \ref{table:Mooney_error}.  In particular, we make use of the following error definitions for the relative maximal error $E_{\mathrm{max}}$ and the relative error of the norm $E_{\mathrm{norm}}$
\begin{equation}\label{eq:errors}
E_{\mathrm{max}}(\bullet)=\frac{\mathrm{max}(\mathrm{abs}((\bullet)_{\mathrm{ana}}-(\bullet)_{\RVE}))}{||(\bullet)_{\mathrm{ana}}||}\,,\qquad  E_{\mathrm{norm}}(\bullet)=\frac{||(\bullet)_{\mathrm{ana}}-(\bullet)_{\RVE}||}{||(\bullet)_{\mathrm{ana}}||}\,.
\end{equation}
Moreover, we make use of the same error definition for the stresses $\partial_{\mi{F}} \Psi(\ma{F})$ and for the tangent $\partial^2_{\mi{F}} \Psi(\ma{F})$.  Note that the relative maximum errors $E_{\mathrm{max}}(\bullet)$ and the relative errors in the norm $E_{\mathrm{norm}}(\bullet)$ for the energy, stresses and tangent are in the range of {\pAE} to {\pAF}.

\subsection{Benchmark test: Second-gradient material}\label{sec:second}
In this second example, we apply the proposed concept for second-order gradient materials.  In \cite{schulte_isogeometric_2020} the whole deformation has been prescribed such that a constant curvature generates a homogeneous hyperstress field. Here, we prescribe again the boundary of the RVE and evaluate the balance equations to obtain the aimed hyperstress field. To be precise,  we make use of fiber-reinforced polymers ($\mathrm{frp}$) as proposed in \cite{dittmann2020a, menzel2017a} with a composed stored energy function of the form
\begin{equation}
\Psi_{\mathrm{frp}} := \zeta\,\Psi_{\mathrm{mat}}+\frac{1-\zeta}{2}\,\Psi_{\mathrm{fib}}\,,
\end{equation}
where $\zeta\in[0,\,1]$ is the volume fraction of the matrix material. $\Psi_{\mathrm{mat}}$ denotes the stored energy function of the matrix material and $\Psi_{\mathrm{fib}}$ denotes the stored energy function of the fibers, both given as follows
\begin{equation}\label{eq:secondGrad}
\begin{aligned}
\Psi_{\mathrm{mat}}&:=\Psi(J, I_1,I_2)\,,\\
\Psi_{\mathrm{fib}}&:=a_\mathrm{F}\,\tan^2{\varphi}+\frac{1}{2}\,\sum\limits_{\alpha}\left[b_\mathrm{F}\,\left(\lambda^\alpha-1\right)^2+c_\mathrm{F}\,\mi{\kappa}^\alpha\cdot\left(\mi{F}\,\mi{F}\transp\,\mi{\kappa}^\alpha\right)\right]\,,
\end{aligned}
\end{equation}
where we make use of the Mooney-Rivlin material given in (\ref{MR_material}) for the matrix material $\Psi_{\mathrm{mat}}$. The stiffness parameter $a$, $b$ and $c$ are related to the shear, stretch and curvature of the fiber material. 

\begin{table}[ht]
\begin{center}
\small
\begin{tabular}{llrl}
\toprule
parameter of matrix material&$c_1\qquad$&$\pMRcOne$ &$\unit{MPa}$\\
parameter of matrix material&$c_2$&$\pMRcTwo$  &$\unit{MPa}$\\
volume fraction of matrix material &$\zeta$&$0.5$&$-$\\
shear parameter of fiber material& $a_\mathrm{F}$ &$\pFIBa$ &$\unit{MPa}$\\
stretch parameter of fiber material& $b_\mathrm{F}$ & $\pFIBb$&$\unit{MPa}$\\
curvature parameter of fiber material& $c_\mathrm{F}$ & $\pFIBc$&$\unit{N}$\\
orientation of fiber $1$ &$\mi{L}^1$& ${\pFibLtwo}$ &$-$\\
orientation of fiber $2$ &$\mi{L}^2$& ${\pFibLone}$ &$-$\\
initial angle of fibers& $\beta$ &$\acos\left(\mi{L}^1\cdot\mi{L}^2\right)$ &$\unit{rad}$\\
\bottomrule
\end{tabular}
\end{center}
\caption{\textbf{Second-gradient material.} Material setting of the fiber-reinforced polymer. }
\label{table:second}
\end{table}

Using bidirectional fibers with $\alpha=[1,2]$, for the normalized fiber orientation $\mi{L}^{\alpha}$ in the reference configuration and the initial angle $\beta$ between both directions, the spatial field of the fiber directions reads $\mi{l}^{\alpha}=\mi{F}\,\mi{L}^{\alpha}$. The stretch of the fibers $\lambda^\alpha$ can now be expressed as
\begin{equation}
\lambda^\alpha=||\mi{l}^{\alpha}||=||\mi{F}\,\mi{L}^{\alpha}||\,,
\end{equation}
whereas the spatial angle reads
\begin{equation}
\varphi=\acos\left(\tilde{\mi{l}}^1\cdot\tilde{\mi{l}}^2\right)-\beta\,.
\end{equation}
Hence, we can write for the deformed fiber configuration $\mi{l}^\alpha=\lambda^\alpha\,\tilde{\mi{l}}^\alpha$. The curvature measure for the fiber initially aligned in $\mi{L}^\alpha$-direction is introduced as follows
\begin{equation}
\mi{\kappa}^\alpha = \frac{1}{\left(\lambda^\alpha\right)^2}\,\left(\mi{I}-\tilde{\mi{l}}^\alpha\otimes\tilde{\mi{l}}^\alpha\right)\,\trimi{F}:\left(\mi{L}^{\alpha}\otimes\mi{L}^{\alpha}\right)\,,
\end{equation}


\renewcommand{\pA}{\phantom{-}0.897}
\renewcommand{\pB}{\phantom{-}0.500}
\renewcommand{\pC}{-0.400}
\renewcommand{\pD}{-0.070}
\renewcommand{\pE}{\phantom{-}1.001}
\renewcommand{\pF}{-0.100}
\renewcommand{\pG}{\phantom{-}0.082}
\renewcommand{\pH}{\phantom{-}0.020}
\renewcommand{\pI}{\phantom{-}0.997}

\renewcommand{\pAA}{-0.033}
\renewcommand{\pAB}{\phantom{-}0.015}
\renewcommand{\pAC}{-0.020}
\renewcommand{\pAD}{\phantom{-}0.015}
\renewcommand{\pAE}{\phantom{-}0.013}
\renewcommand{\pAF}{\phantom{-}0.043}
\renewcommand{\pAG}{-0.020}
\renewcommand{\pAH}{\phantom{-}0.043}
\renewcommand{\pAI}{\phantom{-}0.029}
\renewcommand{\pBA}{\phantom{-}0.015}
\renewcommand{\pBB}{-0.005}
\renewcommand{\pBC}{\phantom{-}0.024}
\renewcommand{\pBD}{-0.005}
\renewcommand{\pBE}{\phantom{-}0.028}
\renewcommand{\pBF}{\phantom{-}0.028}
\renewcommand{\pBG}{\phantom{-}0.024}
\renewcommand{\pBH}{\phantom{-}0.028}
\renewcommand{\pBI}{\phantom{-}0.014}
\renewcommand{\pCA}{\phantom{-}0.023}
\renewcommand{\pCB}{\phantom{-}0.005}
\renewcommand{\pCC}{-0.031}
\renewcommand{\pCD}{\phantom{-}0.005}
\renewcommand{\pCE}{-0.042}
\renewcommand{\pCF}{-0.001}
\renewcommand{\pCG}{-0.031}
\renewcommand{\pCH}{-0.001}
\renewcommand{\pCI}{-0.012}

\begin{figure}[ht]
	\centering
	\begin{minipage}[top]{0.89\textwidth}
	\psfrag{d}[][]{$\partial\RVE:\tilde{\mi{w}}=\mi{0}\,\unit{mm},\; \nabla\tilde{\mi{w}}=\mi{0}\,\unit{mm}$}
	\psfrag{X}{$\mi{X}$}
	\psfrag{-}[][]{$-$}
	\psfrag{+}[][]{$+$}
	\psfrag{l1}[c][c]{$\mi{L}^1$}
    \psfrag{l2}[c][c]{$\mi{L}^2$}
    \psfrag{b}[][]{$\beta$}
    \psfrag{p}[][]{pcs: $\begin{matrix}\fluc{w}(\mi{X}^+)=\fluc{w}(\mi{X}^-),\\ \nabla\fluc{w}(\mi{X}^+)=\nabla\fluc{w}(\mi{X}^-)\end{matrix}$}
	\psfrag{corn}[][]{cn:$\begin{matrix}\tilde{\mi{w}}=\mi{0}\,\unit{mm},\\ \nabla\tilde{\mi{w}}=\mi{0}\,\unit{mm}\end{matrix}$}
		\includegraphics[width=\textwidth]{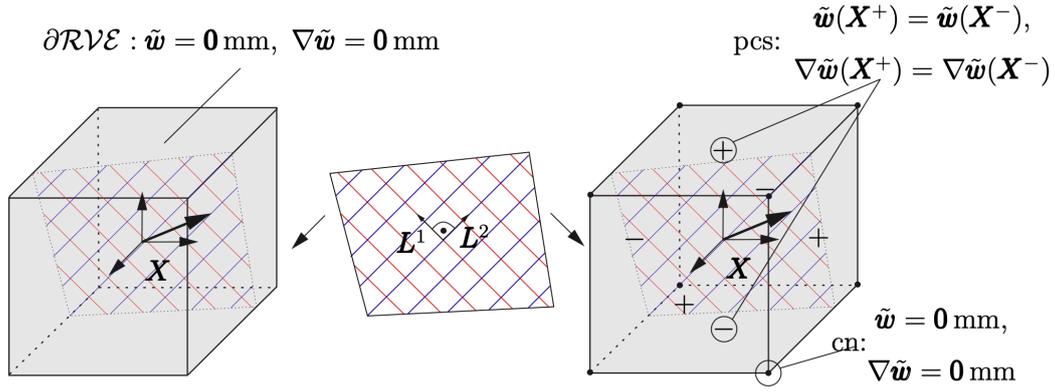}
	\end{minipage}
	\caption{\textbf{Homogeneous second-gradient material.} Left: $\RVE$ in the reference configuration with edge length $0.1$ mm and Dirichlet boundaries on $\partial\RVE$. Middle: schematic representation of the long fibers in the polymer with direction $\mi{L}^\alpha$. Right: $\RVE$ with periodic boundary conditions, for the periodically contiguous surfaces (pcs) top-bottom, right-left, front-back and constrained corner nodes (cn).}
	\label{fig:second_BC}
\end{figure}

The macroscopic values of $\ma{F}$ and $\trima{F}$ are again predefined 
\begin{equation}\label{macro_F_triF}
\begin{aligned}
\ma{F}&:=
\begin{bmatrix}
{\pA} &{\pB} &{\pC} \\
{\pD} &{\pE} &{\pF} \\
{\pG} &{\pH} &{\pI} 
\end{bmatrix}\,,&\,\,
\trima{F}(1,:)&&:=
\begin{bmatrix}
{\pAA} &{\pAB} &{\pAC} \\
{\pAD} &{\pAE} &{\pAF} \\
{\pAG} &{\pAH} &{\pAI} 
\end{bmatrix}\,, \\[2mm]
\trima{F}(2,:)&:=
\begin{bmatrix}
{\pBA} &{\pBB} &{\pBC} \\
{\pBD} &{\pBE} &{\pBF} \\
{\pBG} &{\pBH} &{\pBI} 
\end{bmatrix}\,, &\trima{F}(3,:)&&:=
\begin{bmatrix}
{\pCA} &{\pCB} &{\pCC} \\
{\pCD} &{\pCE} &{\pCF} \\
{\pCG} &{\pCH} &{\pCI} 
\end{bmatrix}\,.
\end{aligned}
\end{equation}
In a first step, Dirichlet boundaries are applied on the boundary $\partial\RVE$, see Figure \ref{fig:second_BC}, left.  Therefore, the boundaries are deformed satisfying the quadratic configuration provided in \eqref{eq:dirichlet}, constraining $\tilde{\mi{w}}=\mi{0}$ and $\nabla\tilde{\mi{w}}=\mi{0}$.

In a second step, we have applied periodic boundaries on $\partial\RVE$, requiring a higher-order coupling of all opposing surfaces. Moreover,  the predefined macroscopic deformation due to  $\ma{F}$ and $\trima{F}$ has to be satisfied on all eight corner nodes,  see Figure \ref{fig:second_BC}.  The edge length of the $\RVE$ cube is $l=0.1$ mm and the coordinate system is placed in the center of the cube, see Figure \ref{fig:second_BC}. 

In Figure \ref{fig:SG16_stress}, the von Mises stress and the norm of the second-order stress $\trimi{P}$ are plotted for $16$ elements in every direction. Again, we compare the solution of the constitutive relation at the \MK{mesoscale} as defined in \eqref{eq:secondGrad} with the analytical solution of \eqref{eq:secondGrad} applied on the macroscale, see Table \ref{tab:hom_sec_comparison} for additional details. It can be seen, that the two shown ways of the enforcement of the energetic criterion result in different stress distributions, especially regarding the second-order contributions.

\renewcommand{\pA}{$37555 \, \unit{MPa}$}
\renewcommand{\pB}{$38128 \, \unit{MPa}$}
\renewcommand{\pC}{$0.038 \, \unit{MPa \, mm^{-1}}$}
\renewcommand{\pD}{$0.115 \, \unit{MPa \, mm^{-1}}$}
\begin{figure}[ht!]
\begin{minipage}{0.49\textwidth}
\includegraphics[width=\textwidth]{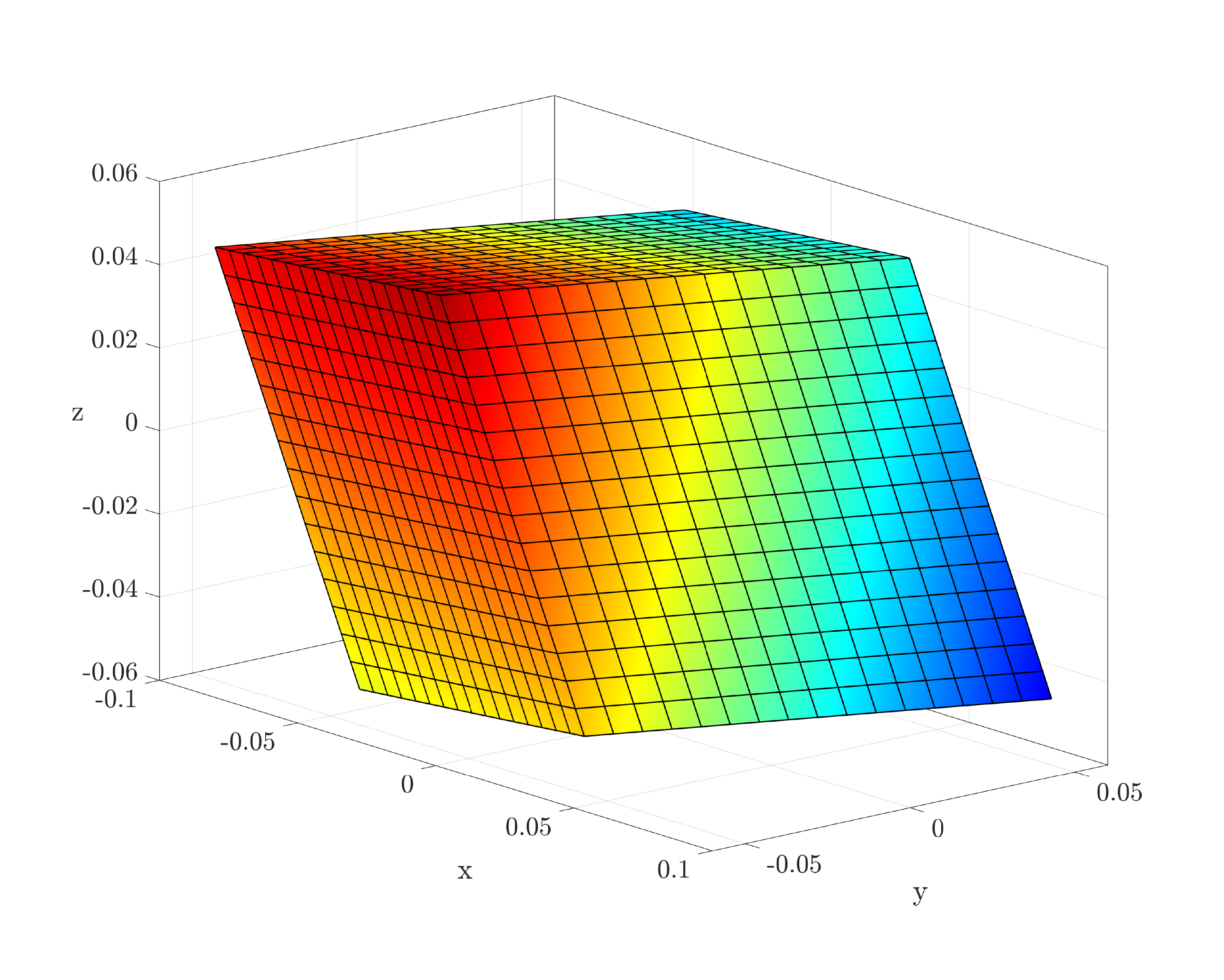}
\end{minipage}
\begin{minipage}{0.49\textwidth}
\includegraphics[width=\textwidth]{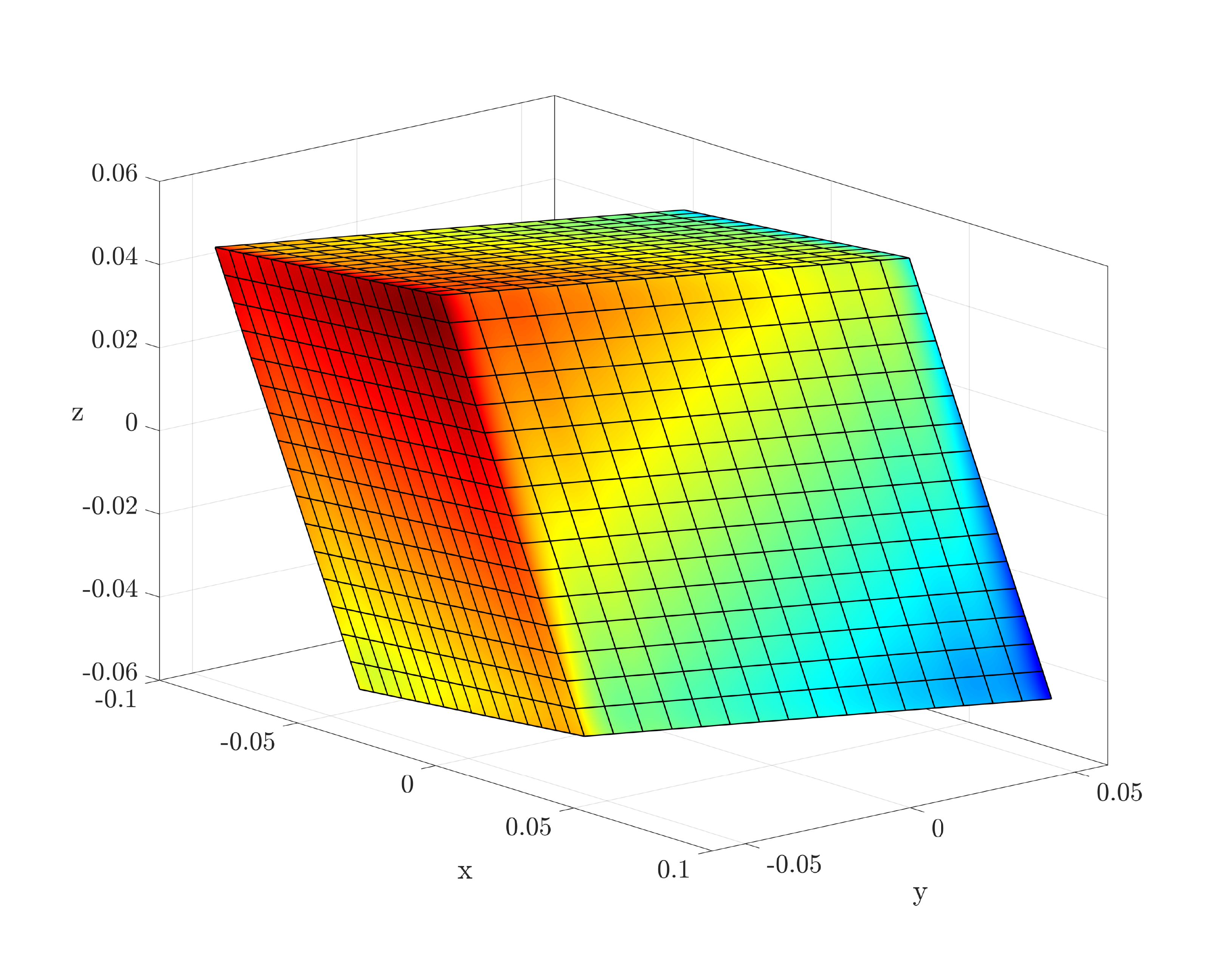}
\end{minipage}
\centering
\psfrag{0}[c][c]{\small\pA}
\psfrag{1}{}
\psfrag{2}[c][c]{\small\pB}
\includegraphics[width=0.5\textwidth]{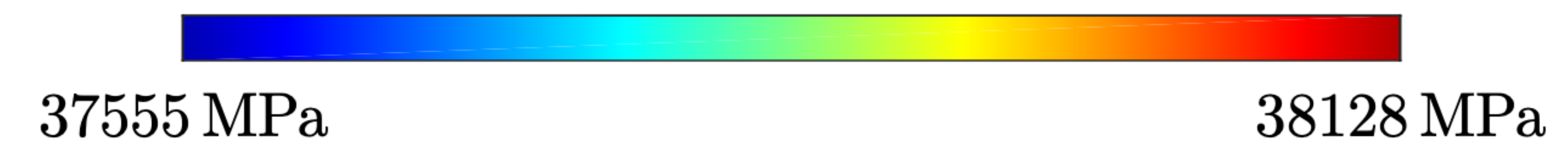}\\\vspace*{1cm}
\begin{minipage}{0.49\textwidth}
\includegraphics[width=\textwidth]{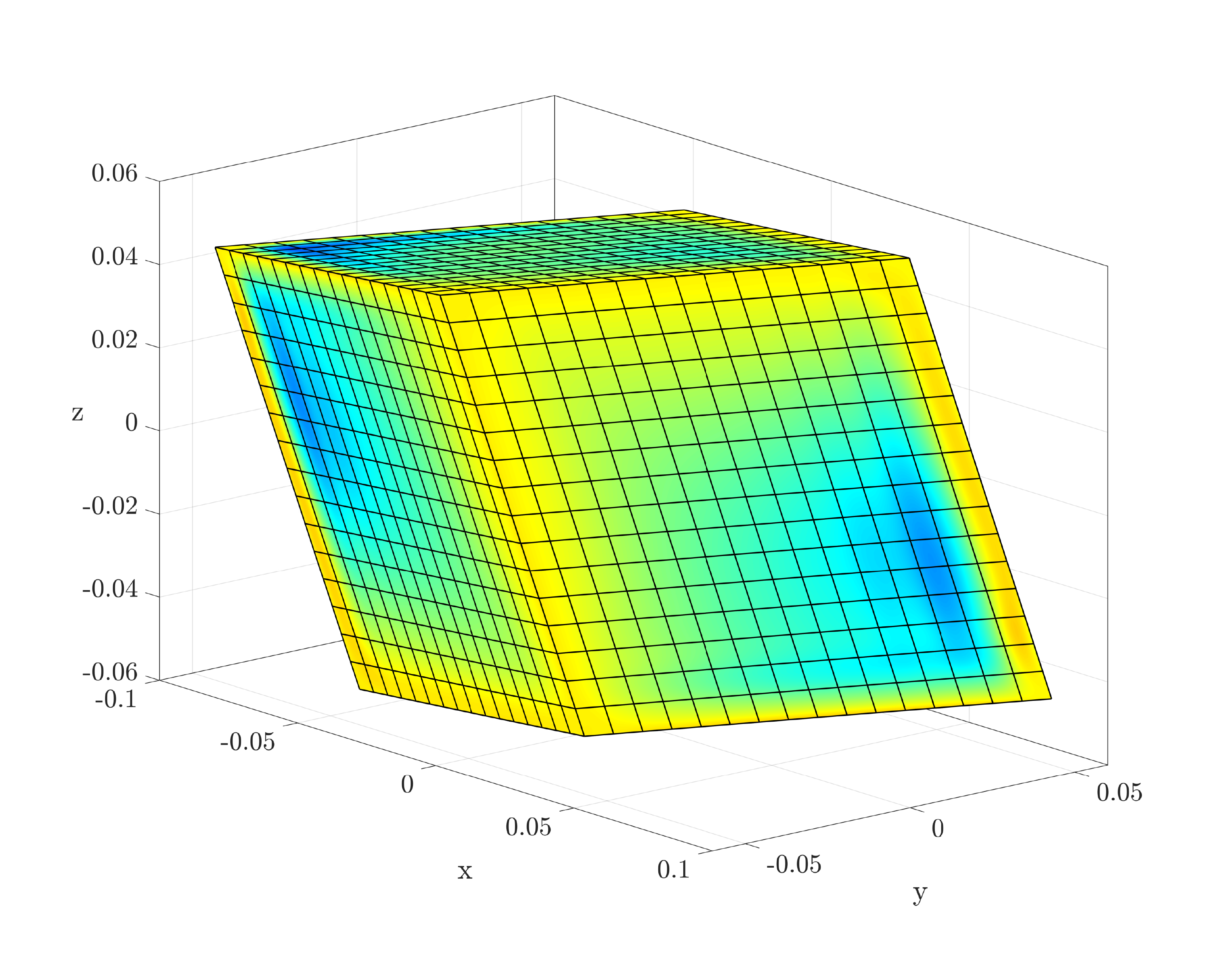}
\end{minipage}
\begin{minipage}{0.49\textwidth}
\includegraphics[width=\textwidth]{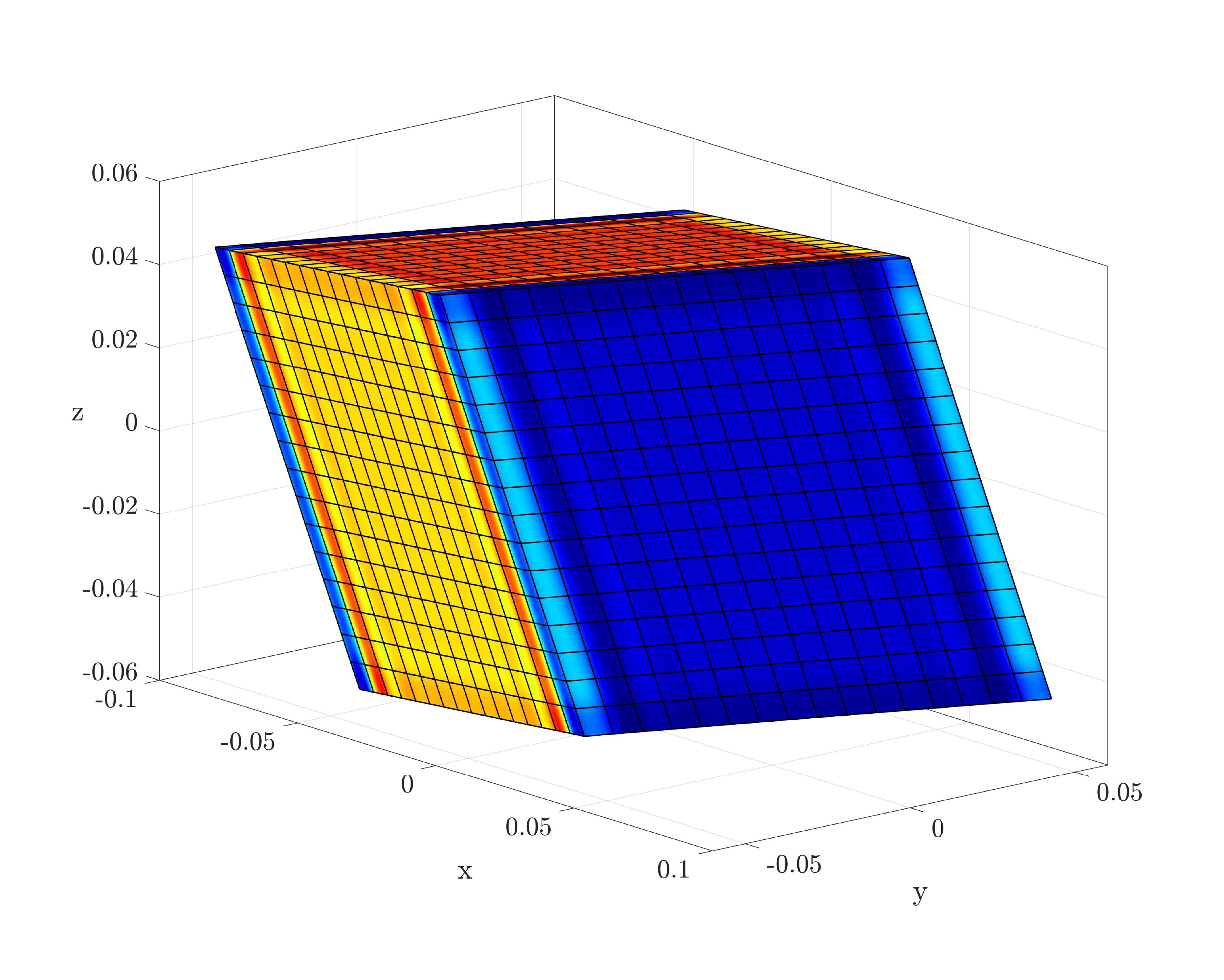}
\end{minipage}
\centering
\psfrag{0}[c][c]{\small\pC}
\psfrag{1}{}
\psfrag{2}[c][c]{\small\pD}
\includegraphics[width=0.5\textwidth]{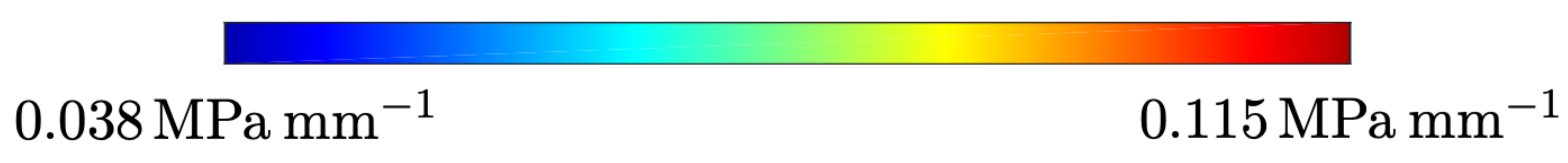}
\caption{\textbf{Homogeneous second-gradient material.} Stresses for $\RVE$ with $16$ elements in each direction -
left to right: Dirichlet and periodic boundaries, top to bottom: von Mises stress and $|| \trimi{P} ||$.}
\label{fig:SG16_stress}
\end{figure}

In addition,  to demonstrate the accuracy of the formulation, we aim at a pure second-gradient material.  Since this anisotropic second-gradient contribution is not well defined \MK{(it is singular without first-gradient contributions),} we have to stabilize the formulation using small first-gradient contributions.  To be specific, we reduced the constitutive parameters successively up to a factor of $1\mathrm{E}{-}08$.  In each direction $16$ elements using quadratic B-splines ($p=2$) for the analysis have been applied \MK{with Dirichlet boundaries,  see \eqref{eq:dirichlet}}.  The maximum absolute error of the averaged values of  $\mi{F}\h$, $\trimi{F}\h$ and $\trimi{P}\h$ for a second-gradient material is shown with regard to the (analytically evaluated) values on the macroscale.  The remaining error $E_{\mathrm{max}}(\trimi{P}\h)$ depends directly on the remaining first-gradient stiffness contributions. Thus, the second-gradient contributions converge to the correct analytical value as expected for a second-gradient material for a constant $\trimi{F}\h$ deformation, as shown in Table \ref{tab:hom_sec_comparison}.

\renewcommand{\pAA}{$1.58\mathrm{E}{-}14$}
\renewcommand{\pAB}{$4.59\mathrm{E}{-}13$}
\renewcommand{\pAC}{$1.74\mathrm{E}{+}01$}
\renewcommand{\pAD}{$4.32\mathrm{E}{-}02$}
\renewcommand{\pAE}{$1.28\mathrm{E}{+}00$}
\renewcommand{\pAF}{$100\%$}
\renewcommand{\pBA}{$1.57\mathrm{E}{-}14$}
\renewcommand{\pBB}{$4.59\mathrm{E}{-}13$}
\renewcommand{\pBC}{$1.81\mathrm{E}{-}01$}
\renewcommand{\pBD}{$4.32\mathrm{E}{-}02$}
\renewcommand{\pBE}{$1.36\mathrm{E}{-}02$}
\renewcommand{\pBF}{$21.6\%$}
\renewcommand{\pCA}{$1.59\mathrm{E}{-}14$}
\renewcommand{\pCB}{$4.60\mathrm{E}{-}13$}
\renewcommand{\pCC}{$1.80\mathrm{E}{-}03$}
\renewcommand{\pCD}{$4.32\mathrm{E}{-}02$}
\renewcommand{\pCE}{$1.39\mathrm{E}{-}04$}
\renewcommand{\pCF}{$0.23\%$}
\renewcommand{\pDA}{$1.59\mathrm{E}{-}14$}
\renewcommand{\pDB}{$4.60\mathrm{E}{-}13$}
\renewcommand{\pDC}{$2.58\mathrm{E}{-}05$}
\renewcommand{\pDD}{$4.32\mathrm{E}{-}02$}
\renewcommand{\pDE}{$1.87\mathrm{E}{-}06$}
\renewcommand{\pDF}{$0.00367\%$}
\renewcommand{\pEA}{$1.56\mathrm{E}{-}14$}
\renewcommand{\pEB}{$4.59\mathrm{E}{-}13$}
\renewcommand{\pEC}{$1.84\mathrm{E}{-}05$}
\renewcommand{\pED}{$4.32\mathrm{E}{-}02$}
\renewcommand{\pEE}{$1.01\mathrm{E}{-}06$}
\renewcommand{\pEF}{$0.00217\%$}
\begin{table}[ht!]
\begin{center}
\small
\MK{
\begin{tabular}{lcccccc}
\toprule
Scaling & $1\mathrm{E}{-}0$ & $1\mathrm{E}{-}2$ & $1\mathrm{E}{-}4$&  $1\mathrm{E}{-}6$ &$1\mathrm{E}{-}8$ \\ 
\midrule
$E_{\mathrm{max}}(\mi{F}\h)$ 					& \pAA & \pBA & \pCA & \pDA & \pEA  \\ 
$E_{\mathrm{max}}(\trimi{F}\h)$ 				& \pAB & \pBB & \pCB & \pDB & \pEB  \\ 
$E_{\mathrm{max}}(\trimi{P}\h)$ 				& \pAC & \pBC & \pCC & \pDC & \pEC  \\ 
${\|\trima{P}^{\mi{P}}\|}/{\|\trima{P}\|}$ 	& \pAF & \pBF & \pCF & \pDF & \pEF  \\ 
\bottomrule
\end{tabular}}
\end{center}
\caption{\textbf{Homogeneous second-gradient material.} Relative absolute maximum error of  $\mi{F}\h$, $\trimi{F}\h$ and $\trimi{P}\h$ \MK{along with the norm of $\trima{P}^{\mi{P}}$ in relation to the total norm of $\trima{P}$} for a second-gradient material with minimal first-gradient contributions, scaled by the parameter as given in the row ''Scaling''.}
\label{tab:hom_sec_comparison}
\end{table}

\subsection{Second-gradient material with inclusions}\label{sec:void}
\begin{figure}[ht]
	\centering
	\vspace{1cm}
	\begin{minipage}[top]{0.89\textwidth}
	\psfrag{d}[][]{$\partial\RVE:\tilde{\mi{w}}=\mi{0}\,\unit{mm},\; \nabla\tilde{\mi{w}}=\mi{0}\,\unit{mm}$}
	\psfrag{X}{$\mi{X}$}
	\psfrag{-}[][]{$-$}
	\psfrag{+}[][]{$+$}
	\psfrag{l1}[c][c]{$\mi{L}^1$}
    \psfrag{l2}[c][c]{$\mi{L}^2$}
    \psfrag{b}[][]{$\beta$}
    \psfrag{p}[][]{pcs: $\begin{matrix}\fluc{w}(\mi{X}^+)=\fluc{w}(\mi{X}^-),\\ \nabla\fluc{w}(\mi{X}^+)=\nabla\fluc{w}(\mi{X}^-)\end{matrix}$}
	\psfrag{corn}[][]{cn:$\begin{matrix}\tilde{\mi{w}}=\mi{0}\,\unit{mm},\\ \nabla\tilde{\mi{w}}=\mi{0}\,\unit{mm}\end{matrix}$}
		\includegraphics[width=\textwidth]{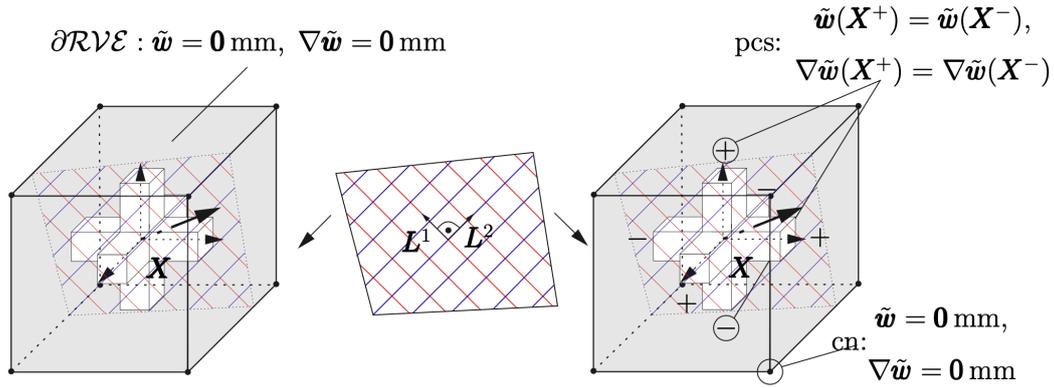}
	\end{minipage}
	\caption{\textbf{Second-gradient material with a void.} Left: $\RVE$ (edge length $0.1$ mm) with a 3D cross void and Dirichlet boundaries  on $\partial\RVE$, see (\ref{eq:dirichlet}), middle: schematic representation of the long fibers in the \MK{polymer} with direction $\mi{L}^\alpha$,  right: $\RVE$ with a 3D cross void and periodic boundary conditions, see (\ref{eq:periodic_BC}), for the periodically contiguous surfaces (pcs) top-bottom, right-left, front-back except the eight corner nodes (cn), where Dirichlet boundaries are used, see (\ref{eq:dirichlet}).}
	\label{fig:RVE_void}
\end{figure}

Next,  the previously introduced second-gradient material is applied on a geometrically inhomogeneous $\RVE$ with a 3D cross inclusion in the center.  These types of inhomogeneities are used e.g. to reduce weight in 3D printed materials. In this example, we consider the inclusions as a void by setting the material parameters to approximately zero\footnote{Setting the parameters strictly to zero may lead to numerical instabilities.}.  Again, Dirichlet and periodic boundary conditions are applied,  see Figure \ref{fig:RVE_void}. The edge length of the $\RVE$ cube is again $l=0.1$ mm with the coordinate system placed in the center. The 3D cross consists of two different edge lengths. The short edges are of the length $l/6$ and the long edges are of the length $l/4$,  see Figure \ref{fig:RVE_void} for details.

In Figure \ref{fig:void_stress}, the von Mises stresses of the matrix and the fibers are plotted, cutting the $\RVE$ in half. Here,  $24$ quadratic B-splines elements in each direction of the $\RVE$ are used.  Note that we observe  the expected anisotropic stress distribution.

\renewcommand{\pA}{$6341 \, \unit{MPa}$}
\renewcommand{\pB}{$116516 \, \unit{MPa}$}

\renewcommand{\pC}{$0.000 \, \unit{MPa \, mm^{-1}}$}
\renewcommand{\pD}{$141.727 \, \unit{MPa \, mm^{-1}}$}
\begin{figure}[ht!]
\begin{minipage}{0.49\textwidth}
\includegraphics[width=\textwidth]{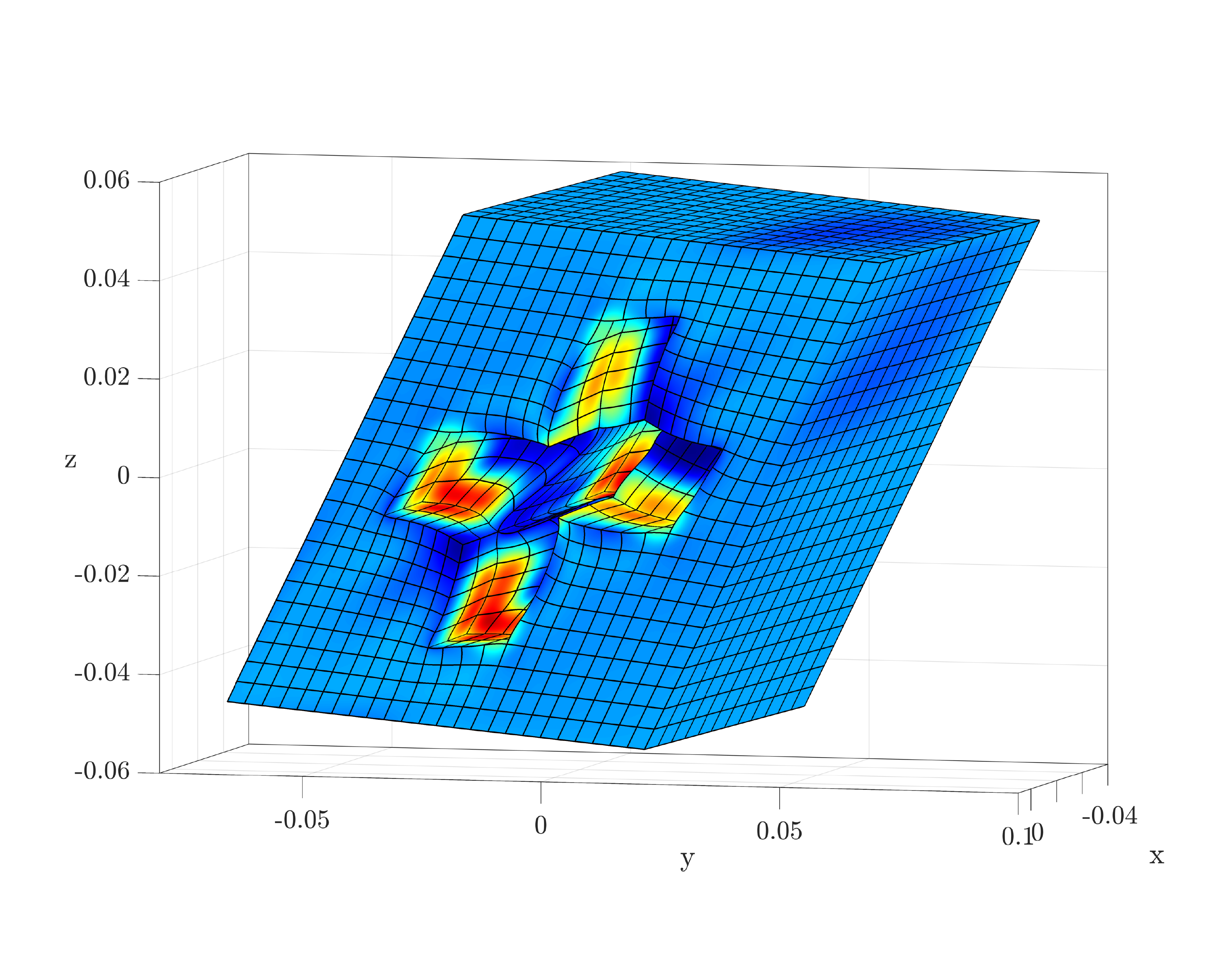}
\end{minipage}
\begin{minipage}{0.49\textwidth}
\includegraphics[width=\textwidth]{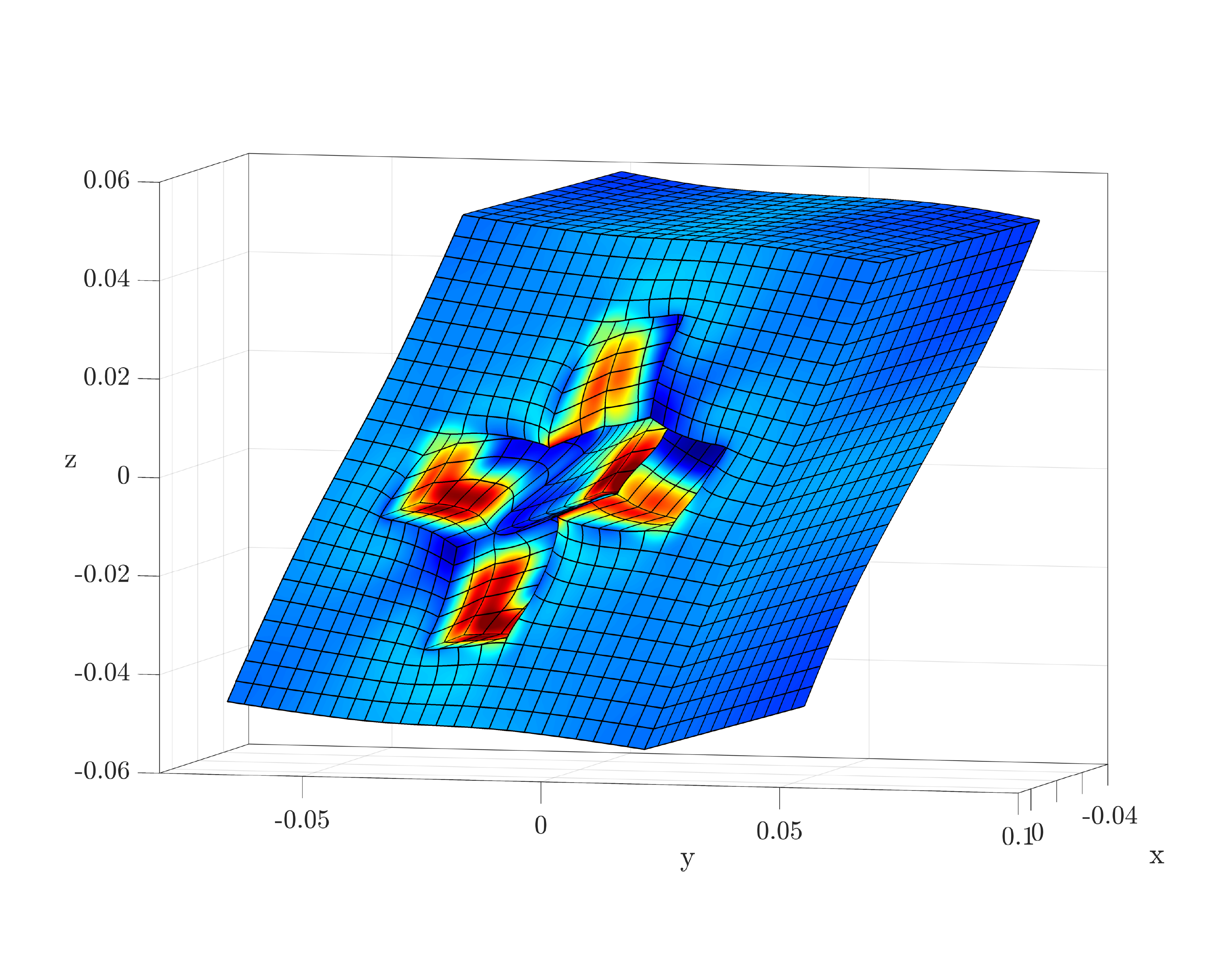}
\end{minipage}
\centering
\psfrag{0}[c][c]{\small\pA}
\psfrag{1}{}
\psfrag{2}[c][c]{\small\pB}
\includegraphics[width=0.5\textwidth]{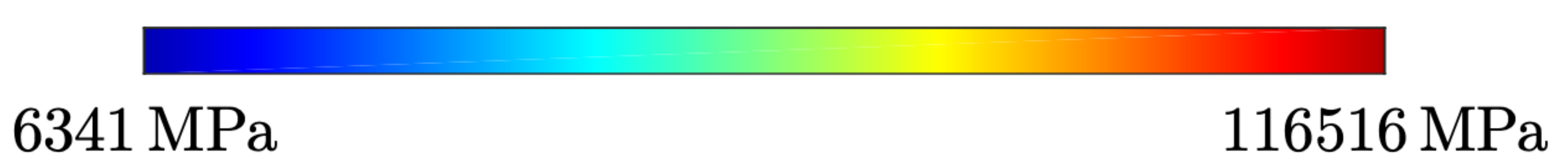}\\
\vspace*{.4cm}
\begin{minipage}{0.49\textwidth}
\includegraphics[width=\textwidth]{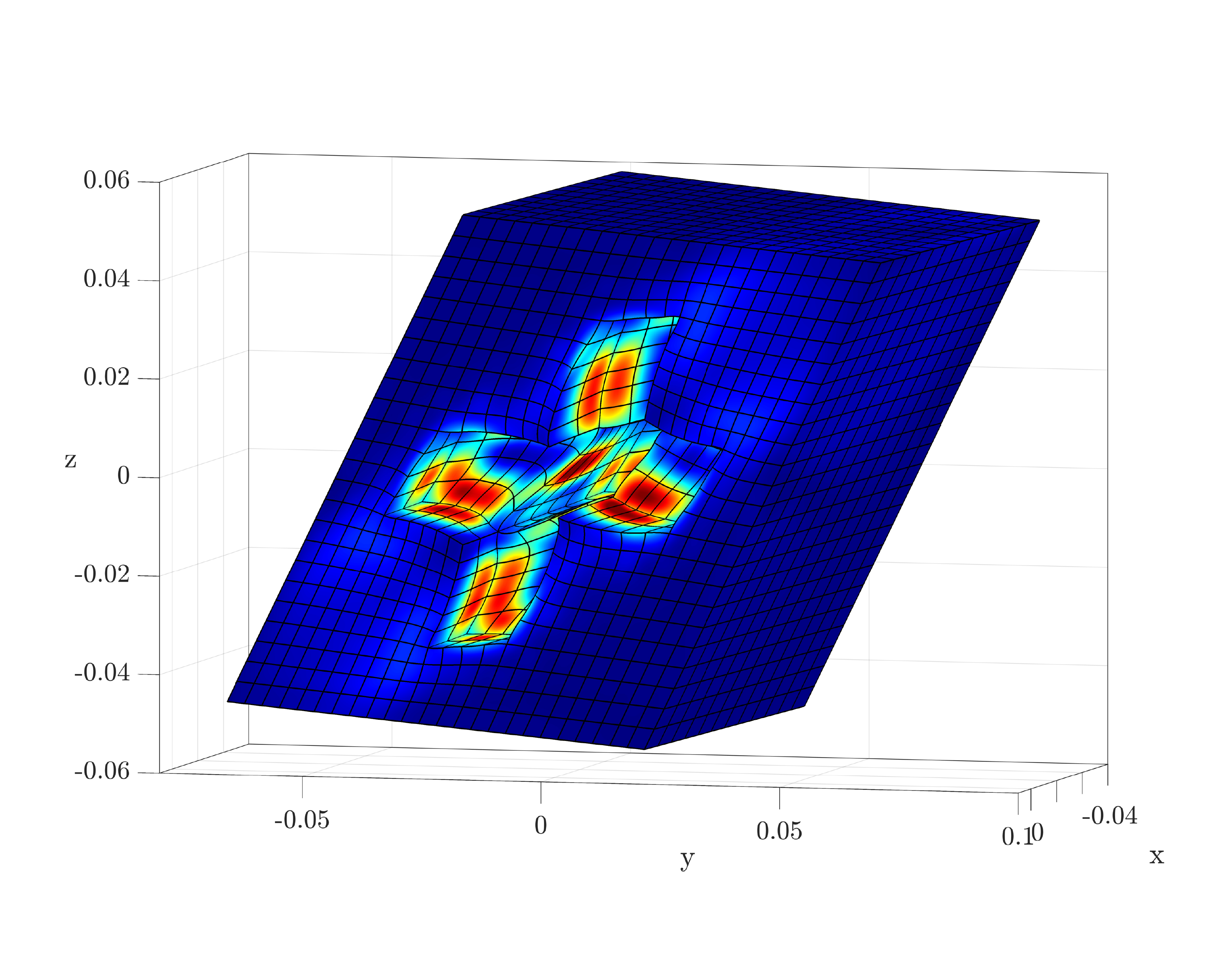}
\end{minipage}
\begin{minipage}{0.49\textwidth}
\includegraphics[width=\textwidth]{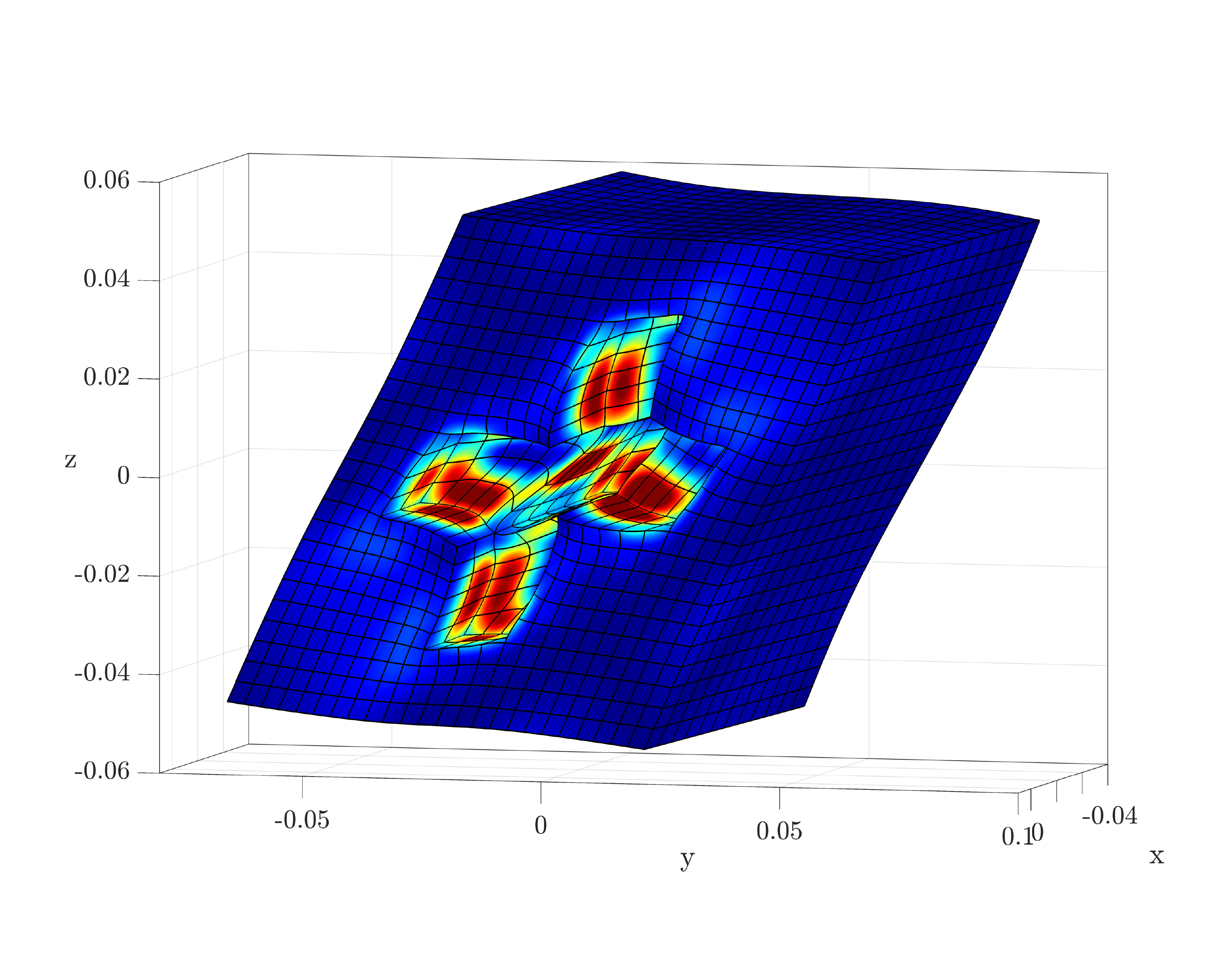}
\end{minipage}
\centering
\psfrag{0}[c][c]{\small\pC}
\psfrag{1}{}
\psfrag{2}[c][c]{\small\pD}
\includegraphics[width=0.5\textwidth]{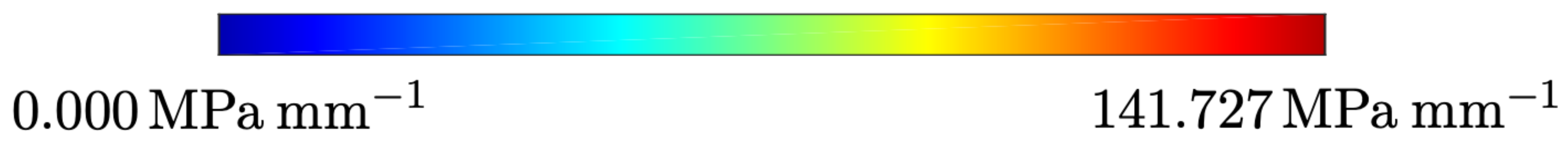}
\caption{\textbf{Second-gradient material with a void.} Von Mises stresses for a half $\RVE$ with $24$ elements in each direction - left to right: Dirichlet and periodic boundaries, top to bottom: von Mises stress and $|| \trimi{P} ||$. Note, that elements within the void are excluded from the plot.}
\label{fig:void_stress}
\end{figure}

\subsection{Cook's membrane}
\begin{figure}[ht]
	\centering	
    \includegraphics[width=0.7\textwidth]{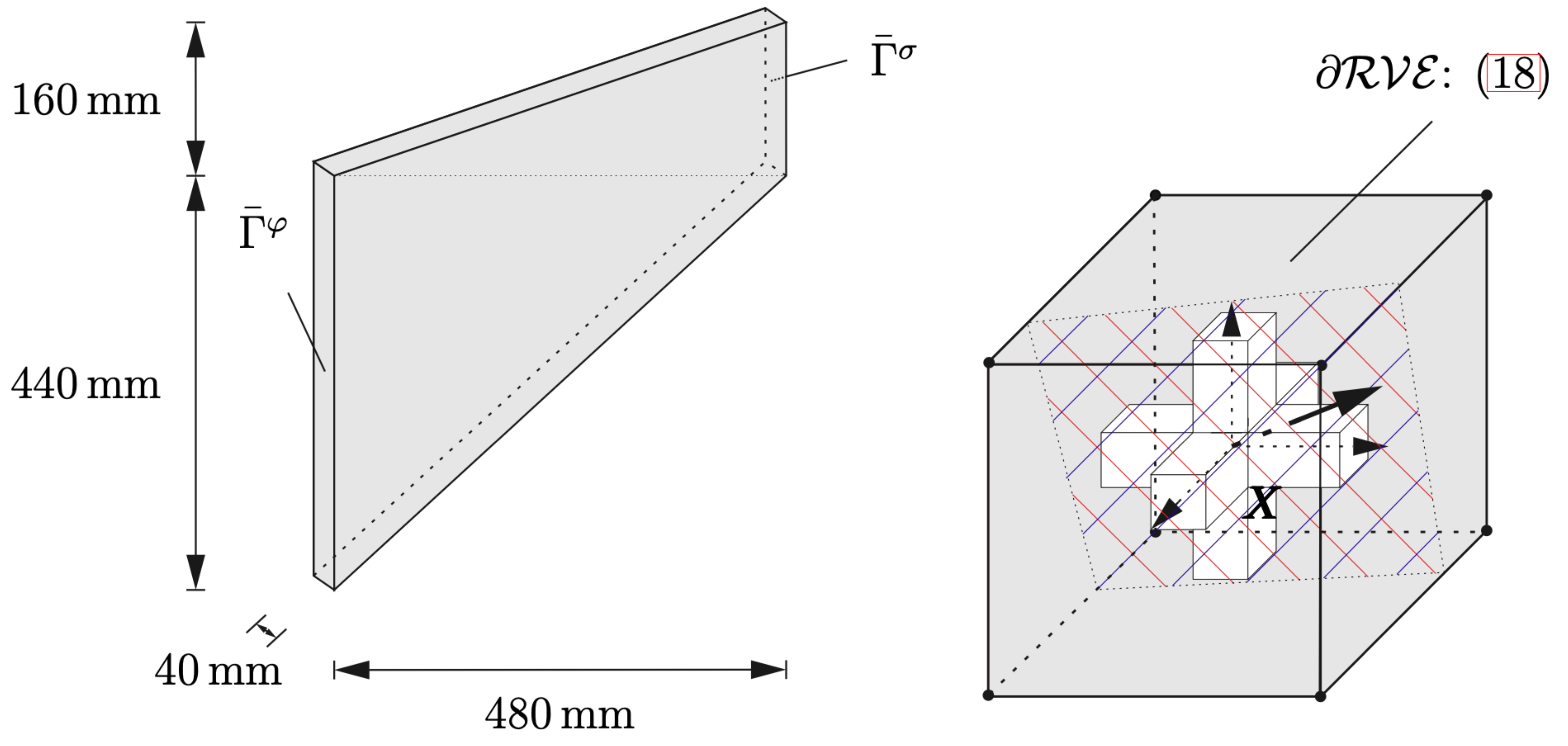}
	\caption{\textbf{Cook's membrane.} Left: Cook's membrane with Dirichlet boundaries $\bar{\Gamma}^\varphi$ on the left side and Neumann boundaries $\bar{\Gamma}^\sigma$ on the right side, right: $\RVE$ of the Cook's membrane with a second gradient material for fiber-reinforced polymers with Dirichlet boundary conditions for the surfaces.}
	\label{fig:Cook}
\end{figure}

In a last example, we examine a Cook's membrane as macroscopic system, see Figure \ref{fig:Cook} left, using again the second-gradient model for the microscopic system inheriting a void as described in Section \ref{sec:void}. All other parameters are given in Table \ref{table:second}.
For the macroscopic system, the Cook's membrane is clamped on the left side, i.e.  $\ma{\varphi} = \vec{0}\,\unit{mm}$ on $\bar{\Gamma}^{\varphi}$. On the right hand side of the Cook's membrane, a constant traction force $\ma{T}_{ext}=\left[0;\,100;\,0\right]\,\unit{N}$  is applied.  We use quadratic B-splines on both scales \MK{with} $27$ Gauss points per element and set up \MK{two mesoscopic systems with $12 \times 12\times 12$ and $24 \times 24\times 24$ elements with in total $2744$ and $17576$ control points, respectively}. Since solving the $\RVE$ for all Gauss points of the macroscopic system in every load increment and Newton iteration requires a high computational effort, we applied a Multigrid-Solution scheme.

\begin{remark}
Multigrid-Solution: For a fast and efficient solution, we construct a series of nested meshes on the macro- and \MK{mesoscale}. Nested meshes are characterized by a linear dependency of the coarse shape functions from those of the fine scale. This can be easily constructed in the context of B-splines and NURBS, if the fine scale is  constructed by a knot insertion technique (see e.g.\ \cite{hughes2005,piegl2010}). This technique provides all necessary topological information for the prolongation matrix $\mathrm{\mathbf{T}}^\mathrm{pro}$. Hence, a first simple algorithm for a fast solution as shown in Box \ref{box:multigrid} can be applied.
\IncMargin{1em}
\begin{center}
\begin{minipage}{0.9\linewidth}
\begin{center}
\begin{algorithm}[H]
\SetAlgorithmName{Box}{}

Construct a coarse scale mesh $M_{F_0}$\\
\For{$i=1:n,\,n:=\text{number of elements}$}
{\vspace{2mm}
Refine the mesh using a knot-insertion to obtain the fine mesh $M_{F_i}$.\\
	Construct prolongation matrix \MK{$\mathrm{\mathbf{T}}_{i,i+1}^\mathrm{pro}$}.\\}
\For{$i=1:n$}
{\vspace{2mm}
Solve the multi-scale problem on $M_{F_i}$.\\
Prolongate the solution \MK{$M_{F_{i+1}} = \mathrm{\mathbf{T}}_{i,i+1}^\mathrm{pro} \circ  M_{F_i}$}.\\}
\caption{Algorithm for fast solution using nested meshes}\label{box:multigrid}
\end{algorithm}
\end{center}
\end{minipage}
\end{center}
Note, that a further decrease of the computational effort can be obtained by using a series of nested meshes $M^\mathrm{RVE}_{F_j}$ on the \MK{mesoscale} as well for each macroscale $M_{F_i}$. \MK{Alternatively, the prolongation on the mesoscale can be circumvented by solving $M_{F_i}$ with $M^\mathrm{RVE}_{F_j}$ and prolongate to $M_{F_i}$ itself but resolved with $M^\mathrm{RVE}_{F_j+1}$.
For the problem at hand we solved  $
\left\{ M_{F_1} | M^\mathrm{RVE}_{F_1} \right\} 
\xrightarrow{\mathrm{\mathbf{T}}_{1,2}^\mathrm{pro} \circ} 
\left\{ M_{F_2} | M^\mathrm{RVE}_{F_1} \right\}  
\xrightarrow{\mathrm{\mathbf{T}}_{2,2}^\mathrm{pro} \circ} 
\left\{ M_{F_2} | M^\mathrm{RVE}_{F_2} \right\} 
\xrightarrow{\mathrm{\mathbf{T}}_{2,3}^\mathrm{pro} \circ} 
\left\{ M_{F_3} | M^\mathrm{RVE}_{F_2} \right\}$. }
\end{remark}

In Figure \ref{fig:cooks_stress} we plotted the von Mises stress of \MK{selected} levels of the Multigrid-Solution and additionally the norm of the second-order stress for the finest resolution of the macroscopic system with a scaled displacement. The second-order stress peaks in the area of clamped left side of the Cook's membrane and matches the expected behaviour.  Furthermore, Table \ref{tab:cooks_convergence} displays the convergence of the macroscopic system in each level of the Multigrid-Solution, indicating the accuracy of the linearization as proposed in Section \ref{sec:linearization}.\ and demonstrating the computational effort.
\renewcommand{\pA}{$-22 \, \unit{MPa}$}
\renewcommand{\pB}{$765 \, \unit{MPa}$}
\renewcommand{\pC}{$0.001 \, \unit{MPa\, mm^{-1}}$}
\renewcommand{\pD}{$0.119 \, \unit{MPa\, mm^{-1}}$}

\begin{figure}[ht!]
\setlength{\medmuskip}{0mu}

\begin{minipage}{0.49\textwidth}
\includegraphics[trim={0cm 0em 0cm 0em},clip,width=\textwidth]{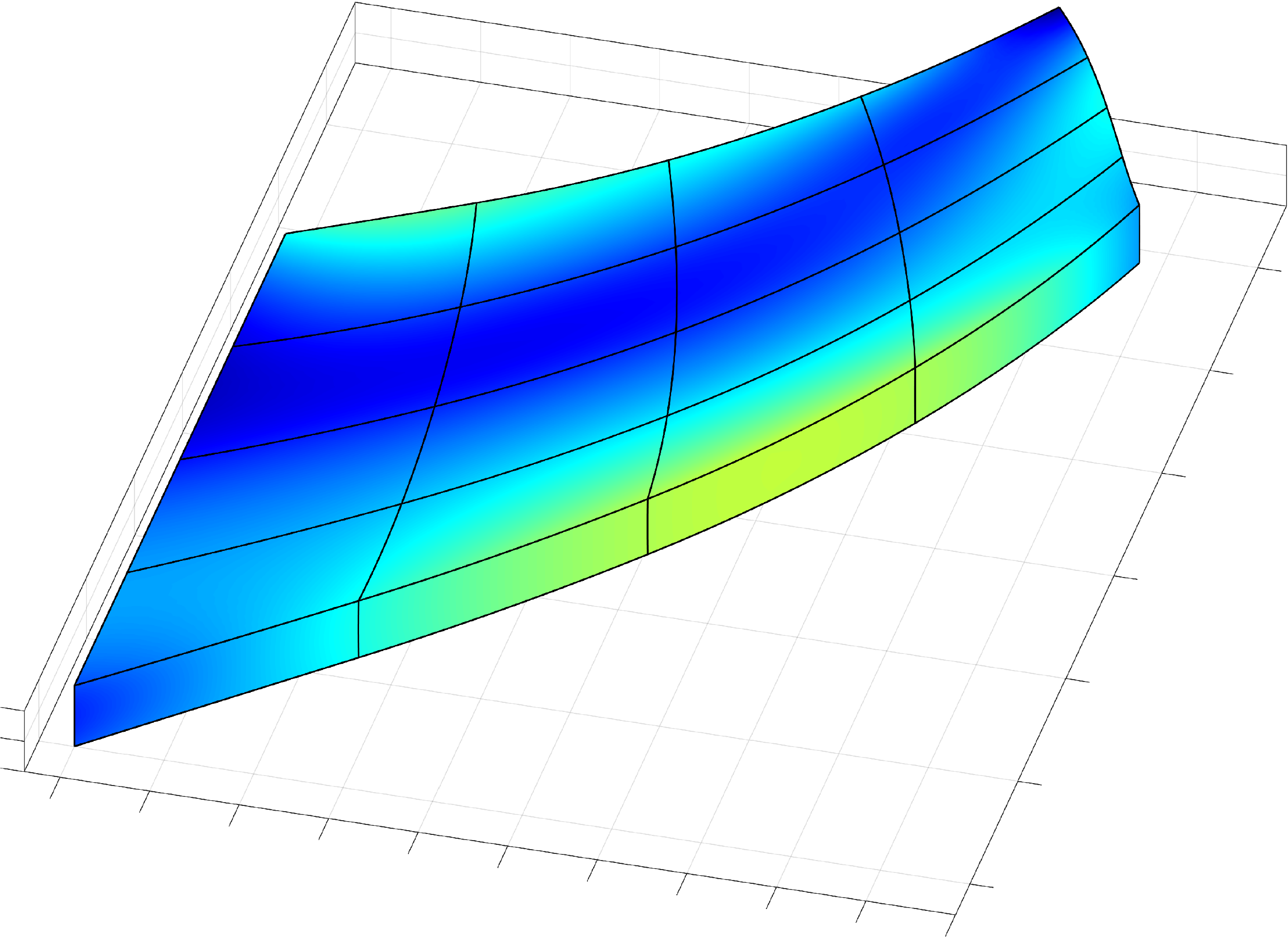}
\end{minipage}\hfill
\begin{minipage}{0.49\textwidth}
\includegraphics[trim={0cm 0em 0cm 0em},clip,width=\textwidth]{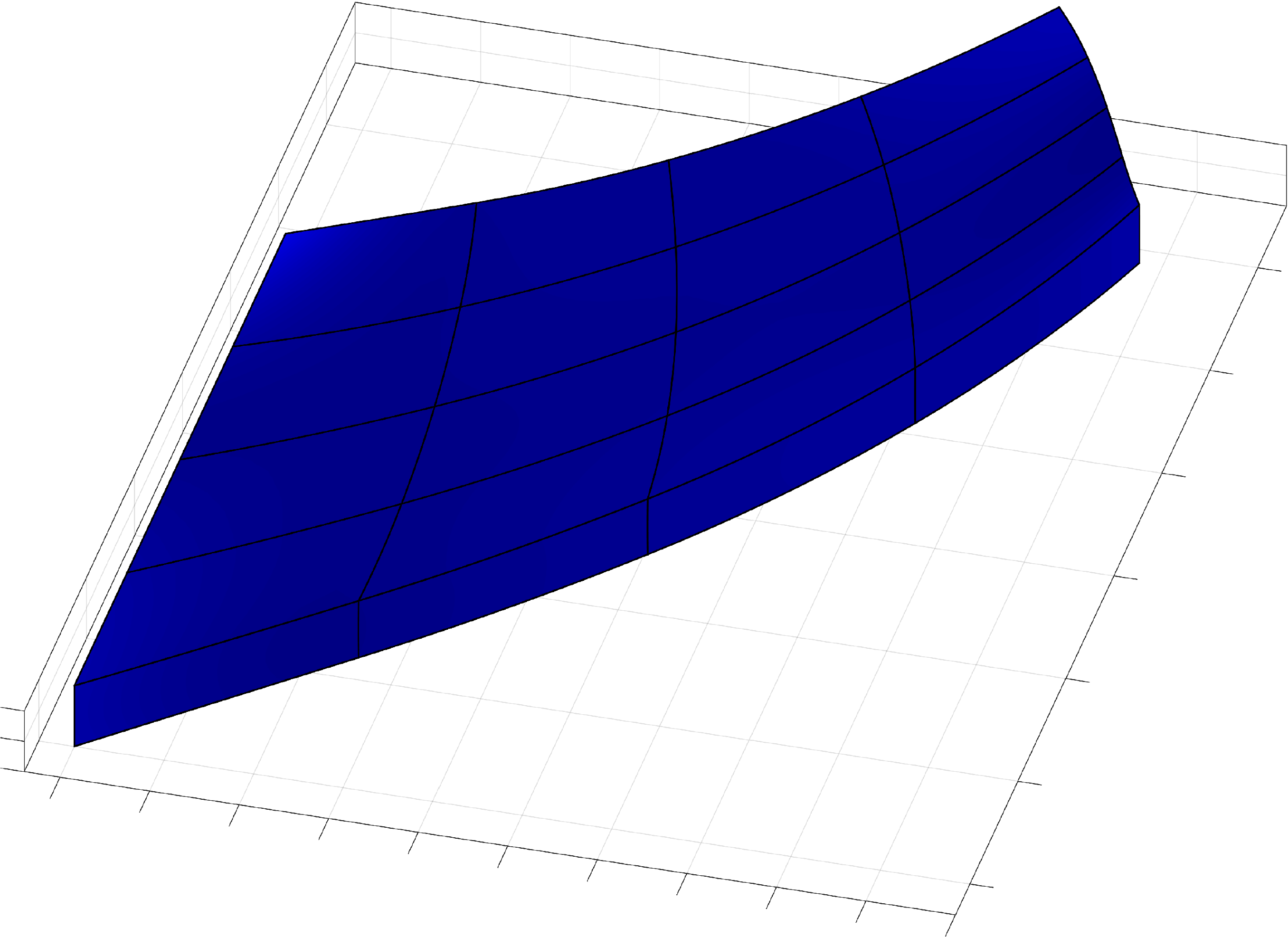}
\end{minipage}\\
\begin{minipage}{0.49\textwidth}
\includegraphics[trim={0cm 0em 0cm 0em},clip,width=\textwidth]{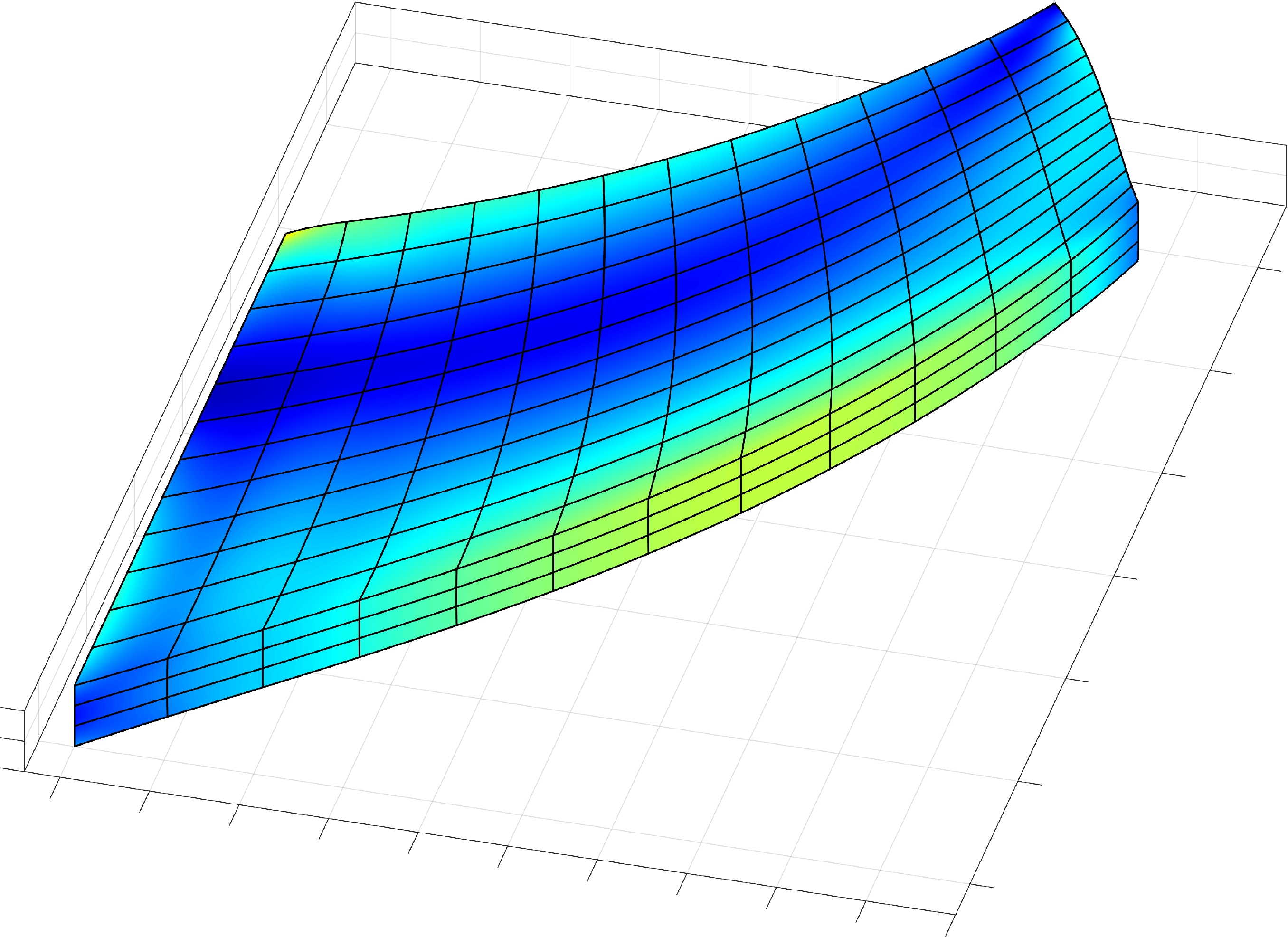}
\end{minipage}\hfill
\begin{minipage}{0.49\textwidth}
\includegraphics[trim={0cm 0em 0cm 0em},clip,width=\textwidth]{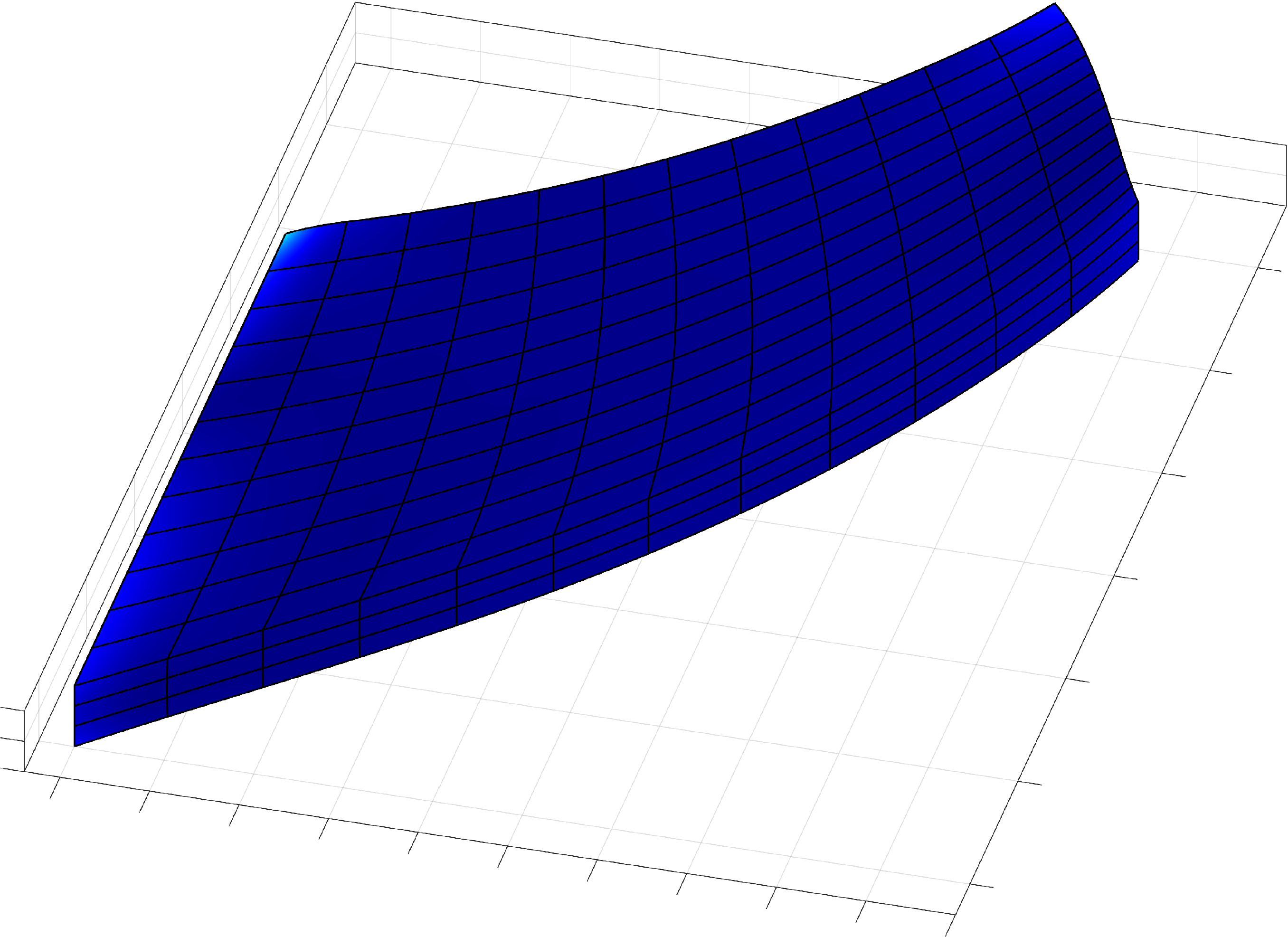}
\end{minipage}\\
\begin{minipage}{0.49\textwidth}
\includegraphics[trim={0cm 0em 0cm 0em},clip,width=\textwidth]{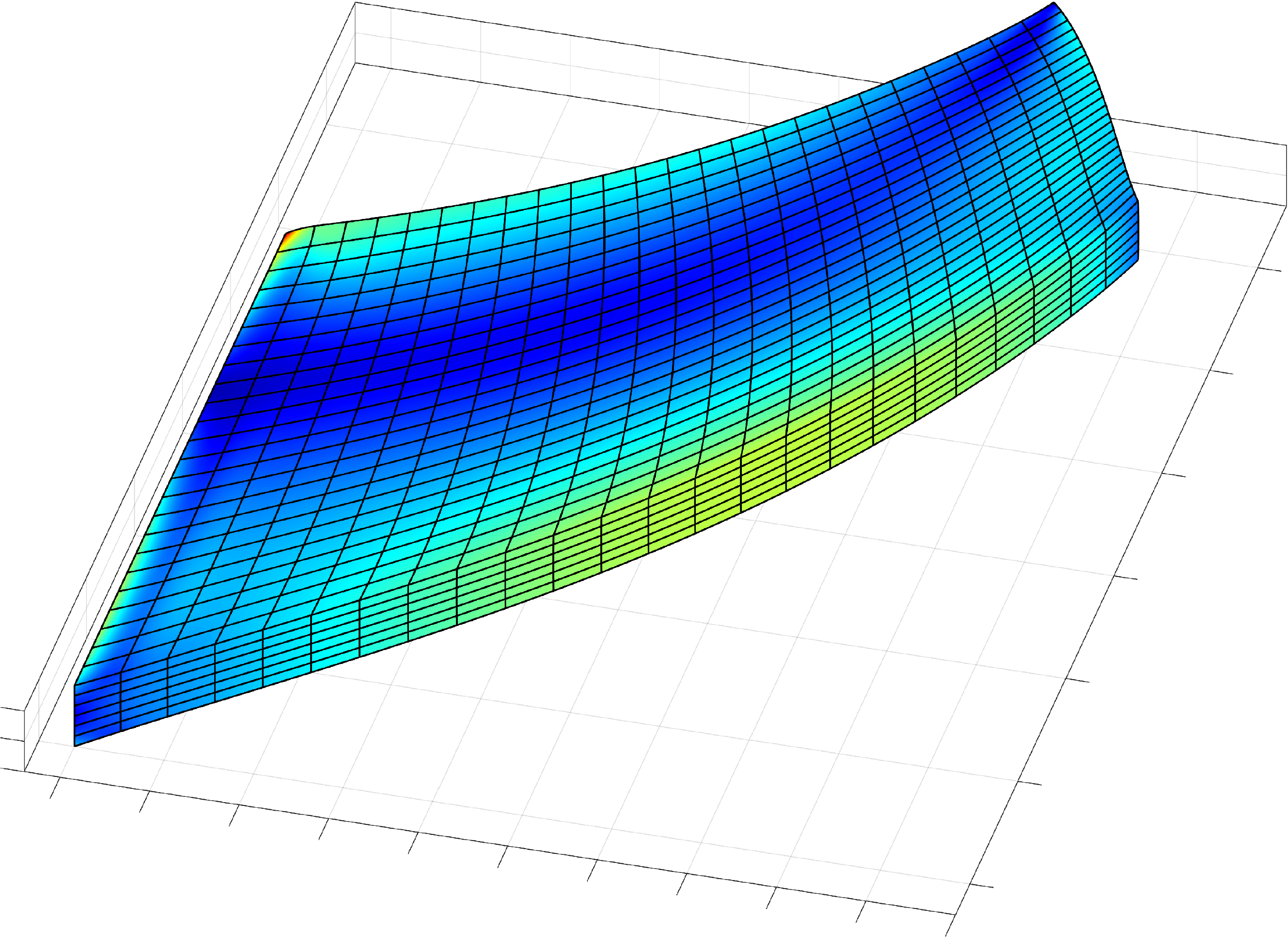}
\end{minipage}\hfill
\begin{minipage}{0.49\textwidth}
\includegraphics[trim={0cm 0em 0cm 0em},clip,width=\textwidth]{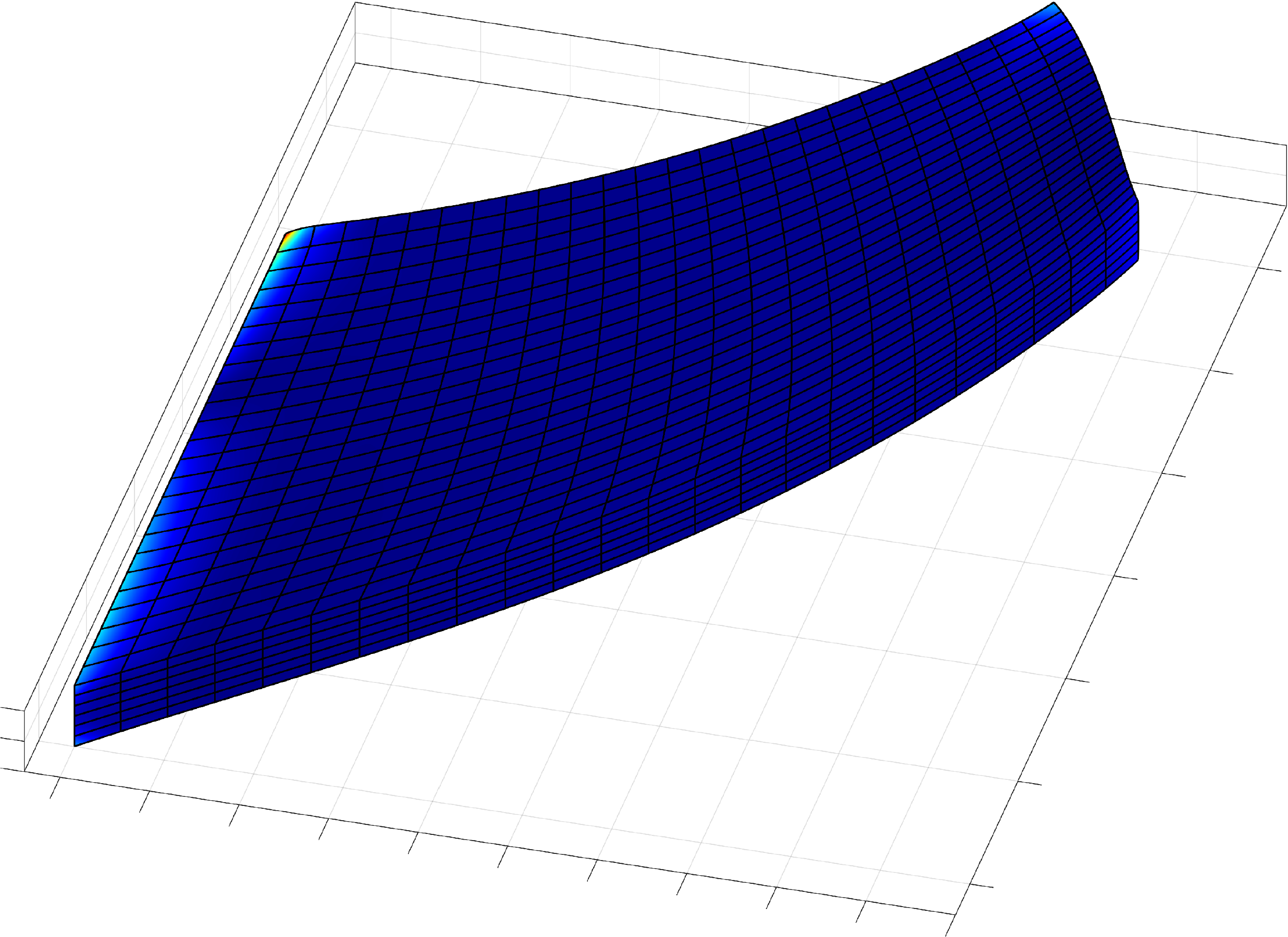}
\end{minipage}\\
\begin{minipage}{0.49\textwidth}
\centering
\psfrag{0}[l][l]{\footnotesize\pA}
\psfrag{1}{}
\psfrag{2}[r][r]{\footnotesize\pB}
\includegraphics[width=0.9\textwidth]{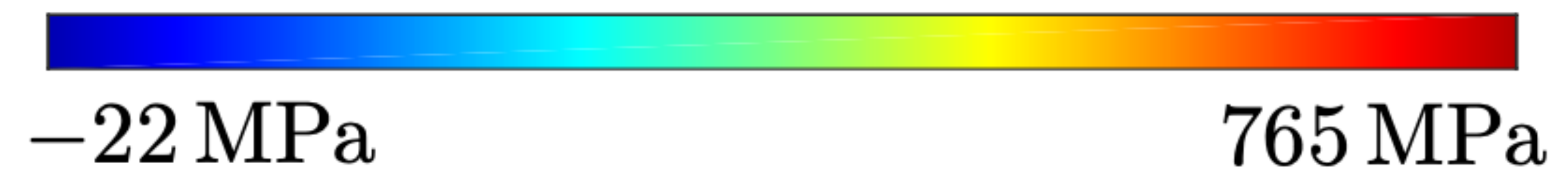}\\\vspace*{1cm}
\end{minipage}\hfill
\begin{minipage}{0.49\textwidth}
\centering
\psfrag{0}[l][l]{\footnotesize\pC}
\psfrag{1}{}
\psfrag{2}[r][r]{\footnotesize\pD}
\includegraphics[width=0.9\textwidth]{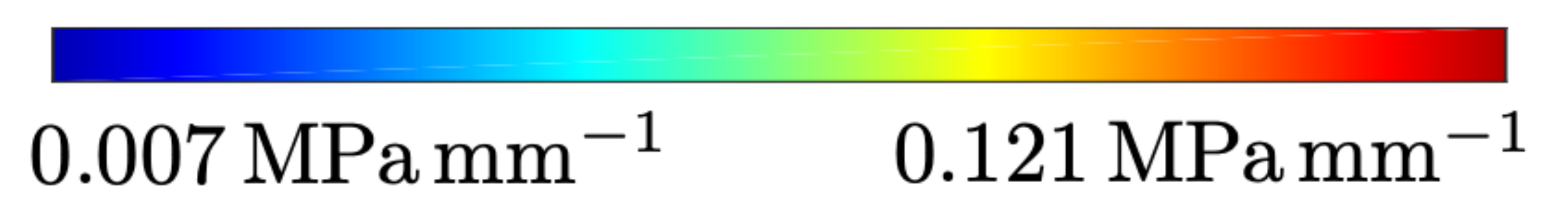}\\\vspace*{1cm}
\end{minipage}

\caption{\textbf{Cook's membrane.} Stresses at different resolutions with a scaled displacement to the factor $5$. \MK{Top to bottom: Level one, three and four according to Table \ref{tab:cooks_convergence}. Left to right: von Mises stress and $\| \trimi{P} \|$.}}
\label{fig:cooks_stress}
\end{figure}

\begin{table}[ht!]
\setlength{\medmuskip}{0mu}
\begin{center}
\small
\renewcommand{\arraystretch}{1}
\MK{
\begin{tabular}[c]{lcccccc}
\toprule
Level & \multicolumn{3}{c}{1} & 2 & 3 & 4   \\
Macro res. & \multicolumn{3}{c}{$4 \times 4\times 1$} & $12 \times 12\times 3$ & $12 \times 12\times 3$ & $ 24 \times 24\times 6$   \\
Micro res. &  \multicolumn{3}{c}{$12 \times 12\times 12$} & $12 \times 12\times 12$ &  $24 \times 24\times 24$ &  $24 \times 24\times 24$  \\
\midrule
Step & 1/10 & $\dots$ & 10/10 & 1/1 & 1/1 & 1/1  \\
\renewcommand{\arraystretch}{.8}\begin{tabular}[t]{@{}c@{}} 
NR-Iterations \end{tabular}
& \renewcommand{\arraystretch}{.8}\begin{tabular}[t]{@{}c@{}}  
  $1.64\mathrm{E}{+}04$ \\ $1.94\mathrm{E}{+}04$ \\ $4.12\mathrm{E}{+}00$ \\ $7.59\mathrm{E}{-}03$ \\ $7.30\mathrm{E}{-}06$ 
  \end{tabular} 
&
& \renewcommand{\arraystretch}{.8}\begin{tabular}[t]{@{}c@{}}  
  $1.63\mathrm{E}{+}04$ \\ $2.37\mathrm{E}{+}04$ \\ $1.17\mathrm{E}{+}00$ \\ $1.66\mathrm{E}{-}04$ \\ $8.93\mathrm{E}{-}07$  
  \end{tabular}   
& \renewcommand{\arraystretch}{.8}\begin{tabular}[t]{@{}c@{}}  
  $3.30\mathrm{E}{+}05$ \\ $2.28\mathrm{E}{+}03$ \\ $8.95\mathrm{E}{-}01$ \\ $7.47\mathrm{E}{-}04$ \\ $8.90\mathrm{E}{-}07$
 \end{tabular} 
& \renewcommand{\arraystretch}{.8}\begin{tabular}[t]{@{}c@{}}  
  $3.96\mathrm{E}{+}02$ \\ $5.54\mathrm{E}{-}01$ \\ $2.03\mathrm{E}{-}04$ \\ $4.66\mathrm{E}{-}07$
  \end{tabular} 
& \renewcommand{\arraystretch}{.8}\begin{tabular}[t]{@{}c@{}}  
  $1.03\mathrm{E}{+}05$ \\ $3.95\mathrm{E}{+}02$ \\ $2.47\mathrm{E}{-}01$ \\ $2.55\mathrm{E}{-}04$ \\ $6.73\mathrm{E}{-}07$
  \end{tabular} \\
\hline
$\sum \RVE$  &  \multicolumn{3}{c}{$21600$} & $58320$ & $46656$ & $466560$ \\
\bottomrule
\end{tabular}
}
\end{center}
\caption{\textbf{Cook's membrane: computational effort and convergence.} Newton-Raphson (NR) convergence utilizing the multigrid solution scheme on the macroscale and total number of solved $\RVE$ per multigrid level. Note, that we conducted an incremental stepping on level one with $10$ steps, whereas the load at higher levels was applied in a single step.}
\label{tab:cooks_convergence}
\end{table}

\section{Conclusions}\label{sec:conclusions}
In this work, we could demonstrate a generalization for the numerical homogenization of higher-order strain gradient materials.  This approach allows to homogenize first- and second-gradient materials on the \MK{mesoscale, containing representative quantities of the microstructure,} towards second- and third-gradient materials on the macroscale. Suitable Dirichlet and periodic boundary conditions have been applied on the \MK{mesoscale} to ensure an energetic consistent formulation, analogously to the Hill-Mandel criterion for first-gradient materials. 

On both scales, the IGA concept using NURBS based shape functions has  proven to be very well suited for these kind of formulations.  Hence, we could implement an IGA$^2$-method and demonstrate the accuracy even for highly anisotropic strain gradient materials on the \MK{mesoscale}.  Eventually, we could derive a generalized framework for a consistent linearization of the macroscale values. The Newton-Raphson iteration for this highly nonlinear problem could be improved by calculating nested meshes on the micro- and the macroscale.  With this framework at hand, novel computational investigations and predictions \MK{of} the constitutive relations of materials with specific microstructures as now widely used in additive manufacturing are feasible. 

\section*{Acknowledgments}
The simulation of the Cook's membrane has been performed on the OMNI Cluster of the University of Siegen.  We gratefully acknowledge the assistance and advice of the HPC support team of ZIMT. M.-A. Keip gratefully acknowledges the financial funding of the German Research Foundation (DFG) within the Collaborative Research Center on Interface-Driven Multi-Field Processes in Porous Media (SFB 1313, Project No. 327154368, Project B01).

\appendix
\section*{Appendix}\label{sec:appendix}
The paper summarizes all necessary equations for a second-gradient micro- and macrocontinuum. In Remark \ref{rem:third_gradient}, we discuss the combination of a second-gradient microcontinuum and a third-gradient macrocontinuum. Therefore, we write the Appendix \ref{app:kinematic} to \ref{app:Deriv_boundary} for the latter one. Omitting the fourth-order tensor $\quarma{F}=\bar{\nabla}^3\ma{\varphi}$ and the triple stress tensor $\quarma{P}$, we end up in the equations for the second-gradient micro- and macrocontinuum mainly used in the paper.

\section{Notation}\label{app:notation}
In the following, we briefly summarize the notation mainly used. The scalar product of two vectors $\vec{a},\,\vec{b}$, two second-order tensors $\vec{A},\vec{B}$, two third-order tensors $\trimi{A},\trimi{B}$ and two fourth-order tensors $\quarmi{A},\quarmi{B}$ is given by\footnote{ Latin indices range in the set $\{1,\,2,\,3\}$. We will make use of the Einstein summation convention on repeated indices.}
\begin{equation}
[\vec{a}\cdot\vec{b}] = a_i\,b_i\,,\quad[\vec{A}:\vec{B}] = A_{ij}\, B_{ij}\,,\quad[\trimi{A}\,\vdots\,\trimi{B}] = A_{ijk}\, B_{ijk}\,,\quad[\quarmi{A}::\quarmi{B}] = A_{ijkl}\, B_{ijkl}
\end{equation}
Other multiplications of two tensors of different order are given in the following way, here for example, for a fourth-order tensor $\quarmi{A}$ with a third-order tensor $\trimi{B}$, second-order tensor $\mi{B}$ and vector $\mi{b}$, respectively
\begin{equation}
[\quarmi{A}\,\vec{b}]_{ijk} = A_{ijkl}\,b_l\,,\quad[\quarmi{A}:\vec{B}]_{ij} = A_{ijkl}\,\, B_{kl}\,,\quad[\quarmi{A}\,\vdots\,\trimi{B}]_{i} = A_{ijkl}\,B_{jkl}\,.
\end{equation}
All other combinations follow analogously. The dyadic product $\otimes$ increases the order of the tensor. For example, a dyadic product of two vectors $\vec{a},\vec{b}$ is given by $\vec{A}=\vec{a}\otimes\vec{b}$ with $A_{ij}=a_i\,b_j$.
Next, we define the macroscopic gradient with respect to the macroscopic reference configuration \(\bar\nabla(\bullet)\) of a vector field \(\ma{a}\) and of a second-order tensor field \(\ma{A}\) as
\begin{equation}\label{macro_grad}
[\bar\nabla\ma{a}]_{iJ}=\frac{\partial[\ma{a}]_i}{\partial[\ma{X}]_J}\quad\text{and}\quad[\bar\nabla\ma{A}]_{iJK}=\frac{\partial[\ma{A}]_{iJ}}{\partial[\ma{X}]_K}.
\end{equation}
For the macroscopic divergence operator it follows
\begin{equation}\label{macro_div}
[\bar{\nabla}\cdot\ma{A}]_{i}=\frac{\partial[\ma{A}]_{iJ}}{\partial[\ma{X}]_J}\quad\text{and}\quad[\ma{\nabla}\cdot\trima{A}]_{iJ}=\frac{\partial[\trima{A}]_{iJK}}{\partial[\ma{X}]_K}.
\end{equation}
The microscopic gradient $\nabla\mi{a}$ and $\nabla\mi{A}$ as well as the divergence operators ${\nabla}\cdot\mi{A}$ and $\nabla\cdot\trimi{A}$ are given analogously to (\ref{macro_grad}) and (\ref{macro_div}) omitting the overlined symbol "$\,\bar{}\,$". Furthermore, the transpose $^{\mathrm{T}i}$ with number $i=1,2,3$ denotes the number of shifted reference magnitudes
\begin{equation}
\begin{aligned}
\left[\mi{A}\right]_{iJ}&=\left[\left[\mi{A}\right]_{Ji}\right]\transp{1}=\left[\left[\mi{A}\right]_{Ji}\right]\transp{}\,,\,&\left[\trimi{A}\right]_{iJK}&=\left[\left[\trimi{A}\right]_{KiJ}\right]\transp{1}\,,\,&\left[\trimi{A}\right]_{iJK}=\left[\left[\trimi{A}\right]_{JKi}\right]\transp{2}\,,\\
\left[\quarmi{A}\right]_{iJKL}\,,&=\left[\left[\quarmi{A}\right]_{KLiJ}\right]\transp{2}\,,&\left[\quarmi{A}\right]_{iJKL}&=\left[\left[\quarmi{A}\right]_{JKLi}\right]\transp{3}\,.
\end{aligned}
\end{equation}
Furthermore, we have to interchange some reference indices with $^{\mathrm{C}{ij}}$, where $i,j$ denote the indices which will be interchanged
\begin{equation}
\begin{aligned}
\left[\trimi{A}\right]_{JKi}&=\left[\left[\trimi{A}\right]_{KJi}\right]\change{12}\,,\quad&\left[\quarmi{A}\right]_{KLiJ}&=\left[\left[\trimi{A}\right]_{LKiJ}\right]\change{12}\,,\\
\left[\quarmi{A}\right]_{JKLi}&=\left[\left[\trimi{A}\right]_{LKJi}\right]\change{13}\,,\quad&\left[\quarmi{A}\right]_{JKLi}&=\left[\left[\trimi{A}\right]_{JLKi}\right]\change{23}\,.\\[2mm]
\end{aligned}
\end{equation}

\section{Macroscopic kinematic}\label{app:kinematic}
The microscopic kinematic for the third-gradient macroscopic continuum is given by
\begin{equation}\label{eq:mi_kinematic_third}
\mi{\varphi}(\mi{X})= \ma{F}\,\mi{X}+\frac{1}{2}\,\trima{F}:(\mi{X}\otimes\mi{X})+\frac{1}{6}\,\quarma{F}\,\vdots\,(\mi{X}\otimes\mi{X}\otimes\mi{X})+\fluc{w}\,.
\end{equation}
The connections between the macroscopic deformations and the averaged microscopic deformations are given by
\begin{equation}\label{eq:mi_ma_F_third}
\begin{aligned}
\frac{1}{V}\,\intRVE\mi{F}\,\d V&=\frac{1}{V}\,\intRVE\,\left(\ma{F}+\trima{F}\,\vec{X}+\frac{1}{2}\,\quarma{F}:(\mi{X}\otimes\mi{X})+\fluc{F}\right)\,\d V\\
&=\ma{F}+\quarma{F}:\frac{1}{V}\,\intRVE\,\frac{1}{2}\,\left(\mi{X}\otimes\mi{X}\right)\,\d V
\end{aligned}
\end{equation}
and
\begin{equation}\label{eq:mi_ma_triF}
\frac{1}{V}\,\intRVE\trimi{F}\,\d V=\frac{1}{V}\,\intRVE\,\left(\trima{F}+\quarma{F}\,\vec{X}+\trifluc{F}\right)\,\d V=\trima{F}
\end{equation}
using $\intRVE\vec{X}\d V=\vec{0}$, which is valid if the coordinate system is in the center of the $\RVE$. Furthermore, $\fluc{F}=\nabla\fluc{w}$ and $\trifluc{F}=\nabla^2\fluc{w}$ are the first and second gradients of the fluctuation field. Since the macroscopic values are exactly the volume averages of the microscopic values and not dependent on the fluctuations, we can write
\begin{equation}\label{restrictions}
\frac{1}{V}\,\intRVE \nabla\fluc{w}\,\d V=\vec{0}\qquad\text{and}\qquad\frac{1}{V}\,\intRVE \nabla^2\fluc{w}\,\d V=\trimi{0}\,.
\end{equation}
Using Gauss's theorem, we can rewrite the volume integrals of \eqref{restrictions} to surface integrals:
\begin{equation}\label{restrictions_dA}
\frac{1}{V}\,\dintRVE \fluc{w}\otimes\mi{N}\,\d A=\vec{0}\qquad\text{and}\qquad\frac{1}{V}\,\dintRVE \nabla\fluc{w}\otimes\mi{N}\,\d A=\trimi{0}\,.
\end{equation}
using the divergence theorem for a unit tensor $\mi{I}$ and a vector $\mi{a}$
\begin{equation}
\begin{aligned}
\dintRVE \mi{a}\cdot(\mi{I}\,\mi{N})\,\d A&=\intRVE \nabla\cdot(\mi{I}\tp\,\mi{a})\,\d V\,,\\
\mi{I}:\dintRVE \mi{a}\otimes\mi{N}\,\d A&=\intRVE \underbrace{\left(\nabla\cdot{\mi{I}}\right)}_{\vec{0}}\cdot\mi{a}\,\d V+\mi{I}:\intRVE\nabla\mi{a}\,\d V\,,\\
\end{aligned}
\end{equation}
or for a tensor $\mi{A}$ and the unit tensor $\mi{I}$
\begin{equation}
\begin{aligned}
\dintRVE \left(\mi{A}\,\mi{I}\right)\,\mi{N}\,\d A&=\intRVE \nabla\cdot(\mi{A}\,\mi{I})\,\d V\,,\\
\dintRVE \mi{A}\otimes\mi{N}\,\d A:\mi{I}&=\intRVE\nabla\mi{A}\,\d V:\mi{I}+\intRVE \mi{A}\,\underbrace{\left(\nabla\cdot{\mi{I}}\right)}_{\vec{0}}\,\d V\,,\\
\end{aligned}
\end{equation}
respectively.

\section{Macroscopic stresses}\label{app:Deriv_stresses}
For the derivation of the macroscopic stresses we use the partial integration and the Gaussian integral theorem for the integral of the left side of the energetic criterion (\ref{eq:Hill_Mandel_third}):
\medskip
\begin{equation}
\begin{aligned}
&\quad\,\intRVE\left(\mi{P}:\nabla\delta\mi{\varphi}+\trimi{P}\,\vdots\,\nabla^2\delta\mi{\varphi}\right)\,\d V\\
&=\intRVE\nabla\cdot\left(\left[\mi{P}-\nabla\cdot\trimi{P}\right]\transp{}\,\delta\mi{\varphi}+\trimi{P}\transp{1}:\nabla\delta\mi{\varphi}\right)\,\d V-\intRVE  \underbrace{\left(\nabla\cdot\left[\mi{P} -\nabla\cdot\trimi{P}\right]\right)}_{=\vec{0}}\cdot \delta\mi{\varphi}\,\d V\\
&=\int\limits_{\partial\RVE}\bigl(\left[\left(\mi{P}-\nabla\cdot\trimi{P}\right)\,\vec{N}\right]\cdot\delta\mi{\varphi}+\left[\trimi{P}\,\vec{N}\right]:\nabla\delta\mi{\varphi}\bigr)\,\d A\,.
\end{aligned}
\end{equation}
The macroscopic stresses are defined in terms of surface integrals since macroscopic values defined by volume integrals could lead to nonphysical results, see Schr\"oder \& Keip \cite{keip2012}. Inserting the variations of the material points $\mi{\varphi}$:
\begin{equation}\label{eq:var_mi_kinematic1}
\delta\mi{\varphi}=\delta\ma{F}\,\mi{X}+\frac{1}{2}\,\delta\trima{F}:(\mi{X}\otimes\mi{X})+\frac{1}{6}\,\delta\quarma{F}\,\vdots\,(\mi{X}\otimes\mi{X}\otimes\mi{X})+\delta\fluc{w}
\end{equation}
and
\begin{equation}\label{eq:var_mi_kinematic2}
\nabla\delta\mi{\varphi}=\delta\ma{F}+\delta\trima{F}\,\mi{X}+\frac{1}{2}\,\delta\quarma{F}:(\mi{X}\otimes\mi{X})+\delta\fluc{F}
\end{equation}
in the last equation, we can split the integral into three parts depending on the variation of the macroscopic deformation gradient $\ma{F}$, the second gradient $\trima{F}$ and the third gradient $\quarma{F}$:
\begin{equation}
\begin{aligned}
&\quad\intRVE\left(\mi{P}:\nabla\delta\mi{\varphi}+\trimi{P}\,\vdots\,\nabla^2\delta\mi{\varphi}\right)\d V\\
=\phantom{+}&\int\limits_{\partial\RVE}\left(\trimi{P}\,\vec{N}+\left[\mi{P}-\nabla\cdot\trimi{P}\right]\,\vec{N}\otimes\mi{X}\right)\,\d A:\delta\ma{F}\\
+&\int\limits_{\partial\RVE}\left(\trimi{P}\,\vec{N}\otimes\mi{X}+\frac{1}{2}\,\left[\mi{P}-\nabla\cdot\trimi{P}\right]\,\vec{N}\otimes\mi{X}\otimes\mi{X}\right)\,\d A\,\vdots\,\delta\trima{F}\\
+&\int\limits_{\partial\RVE}\left(\frac{1}{2}\,\trimi{P}\,\vec{N}\otimes\mi{X}\otimes\mi{X}+\frac{1}{6}\,\left[\mi{P}-\nabla\cdot\trimi{P}\right]\,\vec{N}\otimes\mi{X}\otimes\mi{X}\otimes\mi{X}\right)\,\d A\,::\,\delta\quarma{F}\\
\end{aligned}
\end{equation}
with the restrictions on the boundary, since the macroscopic stresses are not dependent on the fluctuations:
\begin{equation}\label{eq:restr_stresses}
\dintRVE\,(\trimi{P}\,\mi{N}):\nabla\delta\fluc{w}\,\d A=0 \qquad\text{and}\qquad\dintRVE\left[\left(\mi{P}-\nabla\cdot\trimi{P}\right)\,\mi{N}\right]\cdot\delta\fluc{w}\,\d A=0\,.
\end{equation}
The transformation back to volume integrals leads to:
\begin{equation}
\begin{aligned}
&\intRVE\left(\mi{P}:\nabla\delta\mi{\varphi}+\trimi{P}\,\vdots\,\nabla^2\delta\mi{\varphi}\right)\d V\\
=\phantom{+}&\intRVE\left(\nabla\cdot\trimi{P}+\left[\nabla\cdot\left(\mi{X}\otimes\left[\mi{P}-\nabla\cdot\trimi{P}\right]\right)\right]\transp{}\right)\,\d V:\delta\ma{F}\\
+&\intRVE\left(\left[\nabla\cdot\left(\mi{X}\otimes\trimi{P}\right)\right]\transp{1}+\frac{1}{2}\,\left[\nabla\cdot\left(\mi{X}\otimes\mi{X}\otimes\left[\mi{P}-\nabla\cdot\trimi{P}\right]\right)\right]\transp{2}\right)\,\d V\,\vdots\,\delta\trima{F}\\
+&\intRVE\left(\frac{1}{2}\,\left[\nabla\cdot\left(\mi{X}\otimes\mi{X}\otimes\trimi{P}\right)\right]\transp{2}+\frac{1}{6}\,\left[\nabla\cdot\left(\mi{X}\otimes\mi{X}\otimes\mi{X}\otimes\left[\mi{P}-\nabla\cdot\trimi{P}\right]\right)\right]\transp{3}\right)\,\d V::\delta\quarma{F}\\[2mm]
=\phantom{+}&\intRVE\mi{P}\,\d V:\delta\ma{F}+\intRVE\bigl(\trimi{P}+\nabla\cdot\trimi{P}\otimes\mi{X}+\left[\mi{P}-\nabla\cdot\trimi{P}\right]\otimes\mi{X}\bigr)\,\d V\,\vdots\,\delta\trima{F}\\
+& \intRVE\,\bigl(\trimi{P}\otimes\mi{X}+\frac{1}{2}\,\nabla\cdot\trimi{P}\otimes\mi{X}\otimes\mi{X}+\frac{1}{2}\,\left[\mi{P}-\nabla\cdot\trimi{P}\right]\otimes\mi{X}\otimes\mi{X}\bigr)\,\d V::\delta\quarma{F}\\[2mm]
=\phantom{+}&\intRVE\mi{P}\,\d V:\delta\ma{F}+\intRVE\bigl(\trimi{P}+\mi{P}\otimes\mi{X}\bigr)\,\d V\,\vdots\,\delta\trima{F}\\
+& \intRVE\,\bigl(\trimi{P}\otimes\mi{X}+\frac{1}{2}\,\mi{P}\otimes\mi{X}\otimes\mi{X}\bigr)\,\d V::\delta\quarma{F}\,,\\
\end{aligned}
\end{equation}
where we used the following equalities for the divergence of third-, fourth- and fifth-order tensors:
\begin{equation}
\begin{aligned}
\nabla\cdot(\mi{X}\otimes\mi{A})&=\mi{A}\transp{}+\mi{X}\otimes\nabla\cdot\mi{A}\,,\\
\nabla\cdot(\mi{X}\otimes\mathfrak{A})&=\mathfrak{A}\transp{1}+\mi{X}\otimes\nabla\cdot\mathfrak{A}\,,\\[2mm]
\nabla\cdot(\mi{X}\otimes\mi{X}\otimes\mi{A})&=\left(\mi{X}\otimes\mi{A}\transp{}\right)\change{12}+\mi{X}\otimes\mi{A}\transp{}+\mi{X}\otimes\mi{X}\otimes\nabla\cdot\mi{A}\,,\\
\nabla\cdot(\mi{X}\otimes\mi{X}\otimes\trimi{A})&=\left(\mi{X}\otimes\trimi{A}\transp{1}\right)\change{12}+\mi{X}\otimes\trimi{A}\transp{1}+\mi{X}\otimes\mi{X}\otimes\nabla\cdot\trimi{A}\,,\\[2mm]
\nabla\cdot(\mi{X}\otimes\mi{X}\otimes\mi{X}\otimes\mi{A})&=\left(\mi{X}\otimes\mi{X}\otimes\mi{A}\transp{}\right)\change{13}+\left(\mi{X}\otimes\mi{X}\otimes\mi{A}\transp{}\right)\change{23}+\mi{X}\otimes\mi{X}\otimes\mi{A}\transp{}\\
&\quad+\mi{X}\otimes\mi{X}\otimes\mi{X}\otimes\nabla\cdot\mi{A}\,,\\
\end{aligned}
\end{equation}
as well as the strong form of the second-gradient microscopic continuum (\ref{eq:micro_strong}), the symmetry of $\left[\trima{F}\right]_{iJK}$ in $J,K$ and the symmetry of $\left[\quarma{F}\right]_{iJKL}$ in $J,K,L$. 

\section{Mesoscopic boundary conditions}\label{app:Deriv_boundary}
For the third-gradient macroscopic continuum, we can rewrite the energetic criterion \eqref{eq:Hill_Mandel_third} as
\begin{equation}\label{eq:deriv_boundary1_third}
\begin{aligned}
\frac{1}{V}\intRVE\left[\ma{P}-\mi{P}\right]:\left[\delta\ma{F}+\delta\trima{F}\,\vec{X}+\frac{1}{2}\,\delta\quarma{F}:\left(\mi{X}\otimes\mi{X}\right)-\delta\mi{F}\right]\,\d V&\\
+\,\frac{1}{V}\intRVE\left[\trima{P}^{\trimi{P}}-\trimi{P}\right]\,\vdots\,\left[\delta\trima{F}+\delta\quarma{F}\,\vec{X}-\delta\trimi{F}\right]\,\d V&=0\,,
\end{aligned}
\end{equation}
to obtain more information about the boundary conditions. Obviously, the simplest assumption for all points of the microscopic scale, that fulfills the last equation is given by postulating the constraints $\ma{P}:=\mi{P}$ or $\delta\ma{F}+\delta\trima{F}\,\vec{X}+\frac{1}{2}\,\delta\quarma{F}:\left(\mi{X}\otimes\mi{X}\right)=\delta\mi{F}$ and additionally $\trima{P}^\trimi{P}:=\trimi{P}$ or $\delta\trima{F}+\delta\quarma{F}\,\vec{X}=\delta\trimi{F}$, compare Schr\"oder \cite{schroeder2014}.

\subsection{Proof of further representation of energetic criterion}\label{app:rew_Hill_Mandel}
For the derivation of the boundary condition, we have to show, that the energetic criterion (\ref{eq:Hill_Mandel_third}) is equal to (\ref{eq:deriv_boundary1_third}). The first term of (\ref{eq:deriv_boundary1_third}) leads to
\begin{equation}\label{eq:first_term_BC1}
\begin{aligned}
&\frac{1}{V}\intRVE\left[\ma{P}-\mi{P}\right]:\left[\delta\ma{F}+\delta\trima{F}\,\vec{X}+\frac{1}{2}\,\delta\quarma{F}:\left(\mi{X}\otimes\mi{X}\right)-\delta\mi{F}\right]\,\d V\\
=\,&-\ma{P}:\delta\ma{F}-\trima{P}^{\mi{P}}\,\vdots\,\delta\trima{F}-\quarma{P}^{\mi{P}}::\delta\quarma{F}+\frac{1}{V}\intRVE\mi{P}:\delta\mi{F}\,\d V\,,
\end{aligned}
\end{equation}
taking advantage of the fact that the macroscopic quantities are constant over the volume of the $\RVE$ and $\frac{1}{V}\,\intRVE\vec{X}\d V=\vec{0}$, if the coordinate system is in the center of the $\RVE$. Furthermore, the correlations between the microscopic and macroscopic quantities (\ref{eq:mi_ma_F_third}) and (\ref{eq:mi_ma_stresses_third}) are used. For the second term of (\ref{eq:deriv_boundary1_third}) we use additionally (\ref{eq:mi_ma_triF}), which leads to
\begin{equation}\label{eq:sec_term_BC1}
\frac{1}{V}\intRVE\left[\trima{P}^{\trimi{P}}-\trimi{P}\right]\,\vdots\,\left[\delta\trima{F}+\delta\quarma{F}\,\vec{X}-\delta\trimi{F}\right]\,\d V=-\trima{P}^{\trimi{P}}\,\vdots\,\delta\trima{F}-\quarma{P}^{\trimi{P}}::\delta\quarma{F}+\frac{1}{V}\intRVE\trimi{P}\,\vdots\,\delta\trimi{F}\,\d V\,.
\end{equation}

So, we can write for (\ref{eq:deriv_boundary1}) by adding the last two equations
\begin{equation}
\frac{1}{V}\intRVE\mi{P}:\delta\mi{F}\,\d V+\frac{1}{V}\intRVE\trimi{P}\,\vdots\,\delta\trimi{F}\,\d V=\ma{P}:\delta\ma{F}+\left(\trima{P}^{\trimi{P}}+\trima{P}^{\mi{P}}\right)\,\vdots\,\delta\trima{F}+\left(\quarma{P}^{\trimi{P}}+\quarma{P}^{\mi{P}}\right)::\delta\quarma{F}\,,
\end{equation}
which reflects the energetic criterion (\ref{eq:Hill_Mandel_third}).

\subsection{Boundary integral of energetic criterion}\label{app:bc_Hill_Mandel}
Here, the transfer of the volume integrals of (\ref{eq:deriv_boundary1_third}) to boundary integrals of the energetic criterion is explained. Using the partial integration for the first and second term of (\ref{eq:deriv_boundary1_third}), we get for the first term
\begin{equation}\label{eq:first_term_BC2}
\begin{aligned}
&\frac{1}{V}\intRVE\left[\ma{P}-\mi{P}\right]:\left[\delta\ma{F}+\delta\trima{F}\,\vec{X}+\frac{1}{2}\,\delta\quarma{F}:(\mi{X}\otimes\mi{X})-\delta\mi{F}\right]\,\d V\\
=\phantom{+\,}&\frac{1}{V}\intRVE\nabla\cdot\left(\left[\ma{P}-\mi{P}\right]\transp\,\left[\delta\ma{F}\,\mi{X}+\frac{1}{2}\,\delta\trima{F}:(\mi{X}\otimes\mi{X})+\frac{1}{6}\,\delta\quarma{F}\,\vdots\,(\mi{X}\otimes\mi{X}\otimes\mi{X})-\delta\mi{\varphi}\right]\right)\,\d V\\
+\,&\frac{1}{V}\intRVE\nabla\cdot\mi{P}\cdot\left[\delta\ma{F}\,\mi{X}+\frac{1}{2}\,\delta\trima{F}:(\mi{X}\otimes\mi{X})+\frac{1}{6}\,\delta\quarma{F}\,\vdots\,(\mi{X}\otimes\mi{X}\otimes\mi{X})-\delta\mi{\varphi}\right]\,\d V\,,
\end{aligned}
\end{equation}
with $\nabla\cdot\ma{P}=\vec{0}$ and analogously for the second term
\begin{equation}\label{eq:zw_eq}
\begin{aligned}
&\frac{1}{V}\intRVE\left[\trima{P}^{\trimi{P}}-\trimi{P}\right]\,\vdots\,\left[\delta\trima{F}+\delta\quarma{F}\,\mi{X}-\delta\trimi{F}\right]\,\d V\\
=\phantom{+\,}&\frac{1}{V}\intRVE \nabla\cdot\left(\left[\trima{P}^\trimi{P}-\trimi{P}\right]\transp{1}:\left[\delta\trima{F}\,\mi{X}+\frac{1}{2}\,\delta\quarma{F}:(\mi{X}\otimes\mi{X})-\delta\mi{F}\right]\right)\,\d V\\
+\,&\frac{1}{V}\intRVE\nabla\cdot\trimi{P}:\left[\delta\trima{F}\,\mi{X}+\frac{1}{2}\,\delta\quarma{F}:(\mi{X}\otimes\mi{X})-\delta\mi{F}\right]\,\d V\,,\\
\end{aligned}
\end{equation}
with $\nabla\cdot\trima{P}=\vec{0}$. Now, adding the following zero term 
\begin{equation}
\frac{1}{V}\intRVE \left(\nabla\cdot\left(\left[\trima{P}^\trimi{P}-\trimi{P}\right]\transp{1}:\delta\ma{F}\right)+\nabla\cdot\trimi{P}:\delta\ma{F}\right)\,\d V=0\,,
\end{equation}
to the right-hand side of (\ref{eq:zw_eq}) leads only to a change of (\ref{eq:zw_eq}) in the form 
\begin{equation}
\left[\delta\trima{F}\,\mi{X}+\frac{1}{2}\,\delta\quarma{F}:(\mi{X}\otimes\mi{X})-\delta\mi{F}\right]\rightarrow\left[\delta\ma{F}+\delta\trima{F}\,\mi{X}+\frac{1}{2}\,\delta\quarma{F}:(\mi{X}\otimes\mi{X})-\delta\mi{F}\right]\,.
\end{equation}
Using once again a partial integration, the second term of (\ref{eq:deriv_boundary1_third}) is given by
\begin{equation}\label{eq:sec_term_BC2}
\begin{aligned}
&\frac{1}{V}\intRVE\left[\trima{P}^{\trimi{P}}-\trimi{P}\right]\,\vdots\,\left[\delta\trima{F}+\delta\quarma{F}\,\mi{X}-\delta\trimi{F}\right]\,\d V\\
=\phantom{+\,}&\frac{1}{V}\intRVE \nabla\cdot\left(\left[\trima{P}^\trimi{P}-\trimi{P}\right]\transp{1}:\left[\delta\ma{F}+\delta\trima{F}\,\mi{X}+\frac{1}{2}\,\delta\quarma{F}:(\mi{X}\otimes\mi{X})-\delta\mi{F}\right]\right)\,\d V\\
+\,&\frac{1}{V}\intRVE\nabla\cdot\left(\left[\nabla\cdot\trimi{P}\right]\transp\,\left[\delta\ma{F}\,\mi{X}+\frac{1}{2}\,\delta\trima{F}:(\mi{X}\otimes\mi{X})+\frac{1}{6}\,\delta\quarma{F}\,\vdots\,(\mi{X}\otimes\mi{X}\otimes\mi{X})-\delta\mi{\varphi}\right]\right)\,\d V\\
-\,&\frac{1}{V}\intRVE\nabla\cdot\nabla\cdot\trimi{P}\cdot\left[\delta\ma{F}\,\mi{X}+\frac{1}{2}\,\delta\trima{F}:(\mi{X}\otimes\mi{X})+\frac{1}{6}\,\delta\quarma{F}\,\vdots\,(\mi{X}\otimes\mi{X}\otimes\mi{X})-\delta\mi{\varphi}\right]\,\d V\,.
\end{aligned}
\end{equation}
The introduction of the zero term leads to an easy summation of (\ref{eq:first_term_BC2}) and (\ref{eq:sec_term_BC2})
\begin{equation}
\begin{aligned}
&\frac{1}{V}\intRVE\left[\ma{P}-\mi{P}\right]:\left[\delta\ma{F}+\delta\trima{F}\,\vec{X}+\frac{1}{2}\,\delta\quarma{F}:(\mi{X}\otimes\mi{X})-\delta\mi{F}\right]\,\d V\\
+\,&\frac{1}{V}\intRVE\left[\trima{P}^{\trimi{P}}-\trimi{P}\right]\,\vdots\,\left[\delta\trima{F}+\delta\quarma{F}\,\mi{X}-\delta\trimi{F}\right]\,\d V\\
=\phantom{-\,}&\frac{1}{V}\dintRVE\left(\left[\ma{P}-\left(\mi{P}-\nabla\cdot\trimi{P}\right)\right]\,\mi{N}\right)\cdot\left[\delta\ma{F}\,\mi{X}+\frac{1}{2}\,\delta\trima{F}:(\mi{X}\otimes\mi{X})\right.\\
&\hspace*{6cm}\left.+\frac{1}{6}\,\delta\quarma{F}\,\vdots\,(\mi{X}\otimes\mi{X}\otimes\mi{X})-\delta\mi{\varphi}\right]\,\d A\\
+\,&\frac{1}{V}\dintRVE \left(\left[\trima{P}^\trimi{P}-\trimi{P}\right]\,\mi{N}\right):\left[\delta\ma{F}+\delta\trima{F}\,\vec{X}+\frac{1}{2}\,\delta\quarma{F}:(\mi{X}\otimes\mi{X})-\delta\mi{F}\right]\,\d A\\
+\,&\frac{1}{V}\intRVE\underbrace{\nabla\cdot\left(\mi{P}-\nabla\cdot\trimi{P}\right)}_{=\vec{0}}\cdot\left[\delta\ma{F}\,\mi{X}+\frac{1}{2}\,\delta\trima{F}:(\mi{X}\otimes\mi{X})+\frac{1}{6}\,\delta\quarma{F}\,\vdots\,(\mi{X}\otimes\mi{X}\otimes\mi{X})-\delta\mi{\varphi}\right]\,\d V\,,\\
\end{aligned}
\end{equation}
where we make use of the Gaussian integral theorem and the strong form of the microscopic continuum.

\section{Linearization of macroscopic stresses and hyperstresses}\label{app:linearization}
The consistent linearization starts with the incremental macroscopic stresses and hyperstresses  given by the correlated microscopic stresses and hyperstresses, see (\ref{eq:mi_ma_stresses}), which are inserted in the latter equation
\begin{equation}
\Delta\ma{P}=\frac{1}{V}\intRVE\Delta \mi{P}\,\d V\qquad\text{and}\qquad
\Delta\trima{P}=\frac{1}{V}\intRVE\Delta \left(\vec{P}\otimes\mi{X}+\mathfrak{P}\right)\,\d V\,.
\end{equation}
Since the microscopic stresses and hyperstresses depend on $\mi{F}$ and $\trimi{F}$, the chain rule is used
\begin{equation}
\begin{aligned}
\left[\Delta\ma{P}\right]_{iJ}=\phantom{+\,}&\frac{1}{V}\intRVE\left(\left[\quarmi{C}\right]_{iJsT}\,\Delta \left[\mi{F}\right]_{sT}+\left[\fmi{D}\right]_{iJsTU}\,\Delta \left[\trimi{F}\right]_{sTU}\right)\,\d V\,,\\[4mm]
\left[\Delta\trima{P}\right]_{iJK}=\phantom{+\,}&\frac{1}{V}\intRVE \bigl(\left[\quarmi{C}\right]_{iJsT}\,\left[\mi{X}\right]_{K}+\left[\fmi{E}\right]_{iJKsT}\bigr)\,\Delta \left[\mi{F}\right]_{sT}\,\d V\\[2mm]
+\,&\frac{1}{V}\intRVE \bigl(\left[\fmi{D}\right]_{iJsTU}\,\left[\mi{X}\right]_{K}+\left[\smi{G}\right]_{iJKsTU}\bigr)\,\Delta \left[\trimi{F}\right]_{sTU}\,\d V\,.
\end{aligned}
\end{equation}
The derivatives of the stresses with respect to the deformation tensors ($\quarmi{C}$, $\fmi{D}$, $\fmi{E}$, $\smi{G}$) are defined in (\ref{eq:deriv_stresses}). The incremental microscopic deformation measures, see (\ref{eq:deform}), are
\begin{equation}\label{eq:lineariz_kinematic}
\begin{aligned}
\Delta \left[\mi{F}\right]_{sT}&=\Delta\left[\ma{F}\right]_{sT}+\Delta \left[\trima{F}\right]_{sTU}\,\left[\mi{X}\right]_{U}+\Delta\left[\fluc{F}\right]_{sT}\,,\\[2mm]
\Delta \left[\trimi{F}\right]_{sTU}&=\Delta \left[\trima{F}\right]_{sTU}+\Delta \left[\trifluc{F}\right]_{sTU}\,,\\
\end{aligned}
\end{equation}
c.f.\ (\ref{eq:deriv_P}) and (\ref{eq:deriv_tri_P}). 

\section{Approximation of microscopic boundary value problem}\label{app:approx_BVP}
The domain of the representative volume element $\RVE$ is approximated by finite elements $\RVE\approx\RVE\h=\bigcup\limits_{\mathrm{e}=1}^{n}\mathcal{R}\e$ with the number of elements $n$. For the approximation of the microscopic boundary value problem (\ref{eq:micro_equilibrium}) with the incremental deformation tensors of (\ref{eq:incremental_def}), we insert the approximations of (\ref{eq:approx_fluc}) and (\ref{eq:approx_def})
\begin{equation}
\begin{aligned}
\Delta G=\sum\limits_{\mathrm{e}=1}^{n}\,\left[\delta\fluc{q}\right]\eX{A}_{i}\Biggl\{\underbrace{\intRVEe\left[\nabla R\right]^A_{J}\,\left[\quarmi{C}\right]\eh_{iJsT}\,\d V}_{\left[\trimi{L}_1\right]\eX{A}_{isT}}\,\left[\Delta\ma{F}\right]\e_{sT}+\underbrace{\intRVEe\left[\nabla R\right]^A_{J}\,\left[\quarmi{C}\right]\eh_{iJsT}\,\left[\mi{X}\right]\eh_{U}\,\d V}_{\left[\quarmi{M}_1\right]\eX{A}_{isTU}}\,\left[\Delta\trima{F}\right]\e_{sTU}\Biggr.&\\
+\underbrace{\intRVEe\left[\nabla R\right]^A_{J}\,\left[\quarmi{C}\right]\eh_{iJsT}\,\left[\nabla R\right]^B_{T}\,\d V}_{\left[\mi{K}_1\right]\eX{AB}_{is}}\,\left[\Delta\fluc{q}\right]\eX{B}_{s}+\underbrace{\intRVEe\left[\nabla R\right]^A_{J}\,\left[\fmi{D}\right]\eh_{iJsTU}\,\d V}_{\left[\quarmi{M}_2\right]\eX{A}_{isTU}}\,\left[\Delta\trima{F}\right]\e_{sTU}&\\
+\underbrace{\intRVEe\left[\nabla R\right]^A_{J}\,\left[\fmi{D}\right]\eh_{iJsTU}\,\left[\nabla^2 R\right]^B_{TU}\,\d V}_{\left[\mi{K}_2\right]\eX{AB}_{is}}\,\left[\Delta\fluc{q}\right]\eX{B}_{s}
+\underbrace{\intRVEe\left[\nabla^2 R\right]^A_{JK}\,\left[\fmi{E}\right]\eh_{iJKsT}\,\d V}_{\left[\trimi{L}_2\right]\eX{A}_{isT}}\,\left[\Delta\ma{F}\right]\e_{sT}&\\
+\underbrace{\intRVEe\left[\nabla^2 R\right]^A_{JK}\,\left[\fmi{E}\right]\eh_{iJKsT}\,\left[\mi{X}\right]\eh_{U}\,\d V}_{\left[\quarmi{M}_3\right]\eX{A}_{isTU}}\,\left[\Delta\trima{F}\right]\e_{sTU}
+\underbrace{\intRVEe\left[\nabla^2 R\right]^A_{JK}\,\left[\fmi{E}\right]\eh_{iJKsT}\,\left[\nabla R\right]^B_{T}\,\d V}_{\left[\mi{K}_3\right]\eX{AB}_{is}}\,\left[\Delta\fluc{q}\right]\eX{B}_{s}&\\
+\underbrace{\intRVEe\left[\nabla^2 R\right]^A_{JK}\,\left[\smi{G}\right]\eh_{iJKsTU}\,\d V}_{\left[\quarmi{M}_4\right]\eX{A}_{isTU}}\,\left[\Delta\trima{F}\right]\e_{sTU}
\Biggl.+\underbrace{\intRVEe\left[\nabla^2 R\right]^A_{JK}\,\left[\smi{G}\right]\eh_{iJKsTU}\,\left[\nabla^2 R\right]^B_{TU}\,\d V}_{\left[\mi{K}_4\right]\eX{AB}_{is}}\,\left[\Delta\fluc{q}\right]\eX{B}_{s}\Biggr\}\,.&\\
\end{aligned}
\end{equation}
In the end we can write
\begin{equation}
\Delta G=\sum\limits_{\mathrm{e}=1}^{n}\,\left[\delta\fluc{q}\right]\eX{A}_i\,\biggl\{\left[\mi{K}\right]\eX{AB}_{is}\,\left[\Delta\fluc{q}\right]\eX{B}_{s}+\left[\trimi{L}\right]\eX{A}_{isT}\,\left[\Delta\ma{F}\right]\e_{sT}+\left[\quarmi{M}\right]\eX{A}_{isTU}\,\left[\Delta\trima{F}\right]\e_{sTU}\biggr\}
\end{equation}
with
\begin{equation}
\begin{aligned}
\left[\mi{K}\right]\eX{AB}_{is}&=\left[\mi{K}_1\right]\eX{AB}_{is}+\left[\mi{K}_2\right]\eX{AB}_{is}+\left[\mi{K}_3\right]\eX{AB}_{is}+\left[\mi{K}_4\right]\eX{AB}_{is}\,,\\
\left[\trimi{L}\right]\eX{A}_{isT}&=\left[\trimi{L}_1\right]\eX{A}_{isT}+\left[\trimi{L}_2\right]\eX{A}_{isT}\,,\\
\left[\quarmi{M}\right]\eX{A}_{isTU}&=\left[\quarmi{M}_1\right]\eX{A}_{isTU}+\left[\quarmi{M}_2\right]\eX{A}_{isTU}+\left[\quarmi{M}_3\right]\eX{A}_{isTU}+\left[\quarmi{M}_4\right]\eX{A}_{isTU}\,.
\end{aligned}
\end{equation}
After assembling over all elements with $(\bullet)=\bigA\limits_{\mathrm{e}=1}^n(\bullet)\e$, we get in the equilibrium state
\begin{equation}
\Delta G=\left[\delta\fluc{q}\right]^{A}_i\,\biggl\{\left[\mi{K}\right]^{AB}_{is}\,\left[\Delta\fluc{q}\right]^{B}_{s}+\left[\trimi{L}\right]^{A}_{isT}\,\left[\Delta\ma{F}\right]_{sT}+\left[\quarmi{M}\right]^{A}_{isTU}\,\left[\Delta\trima{F}\right]_{sTU}\biggr\}\,.
\end{equation}
Thus, using $\Delta G = 0$, the correlation between the sensitivities and the change of corresponding macroscopic fields is given in the discrete setting as follows
\begin{equation}\label{eq:discr_sensitivities}
\left[\Delta\fluc{q}\right]^{B}_{s}=-\left(\left[\mi{K}\right]^{AB}_{ls}\right)^{-1}\,\left(\left[\trimi{L}\right]^{A}_{lrT}\,\left[\Delta\ma{F}\right]_{rT}+\left[\quarmi{M}\right]^{A}_{lrTU}\,\left[\Delta\trima{F}\right]_{rTU}\right)\,.
\end{equation}

\section{Approximation of macroscopic stresses}\label{app:discr_lin_stresses}
In this section, we derive the discretized macroscopic stresses. We start with the discretization of the linearization of $\ma{P}$:
\begin{equation}
\left[\Delta\ma{P}\right]\h_{iJ}=\left[\quarmi{V}^{\quarmi{C}}\right]\h_{iJsT}\,\left[\Delta\ma{F}\right]_{sT}+\left[\fmi{V}^{\quarmi{C}\fmi{D}}\right]\h_{iJsTU}\,\left[\Delta\trima{F}\right]_{sTU}
+\left[\trimi{N}\right]^{B}_{iJs}\,\left[\Delta\fluc{q}\right]^{B}_{s}\,,
\end{equation}
with the volume-averaged  tensors
\begin{equation}
\begin{aligned}
\left[\quarmi{V}^{\quarmi{C}}\right]\h_{iJsT}&=\frac{1}{V}\intRVEh\left[\quarmi{C}\right]\h_{iJsT}\,\d V\,,\\
\left[\fmi{V}^{\quarmi{C}\fmi{D}}\right]\h_{iJsTU}&=\frac{1}{V}\intRVEh\left(\left[\quarmi{C}\right]\h_{iJsT}\,\left[\mi{X}\right]\h_{U}+\left[\fmi{D}\right]\h_{iJsTU}\right)\d V\,,\\
\left[\trimi{N}\right]^{B}_{iJs}&=\frac{1}{V}\intRVEh\left(\left[\quarmi{C}\right]\h_{iJsT}\,\left[\nabla R\right]^{B}_{T}+\left[\fmi{D}\right]\h_{iJsTU}\,\left[\nabla^2 R\right]^{B}_{TU}\right)\d V\,.
\end{aligned}
\end{equation}
Inserting the discrete sensitivities (\ref{eq:discr_sensitivities}) yields
\begin{equation}
\begin{aligned}
\left[\Delta\ma{P}\right]\h_{iJ}=\phantom{+\,}&\biggl\{\left[\quarmi{V}^{\quarmi{C}}\right]\h_{iJrT}-\left[\trimi{N}\,\right]^{B}_{iJs}\,\left(\left[\mi{K}\right]^{AB}_{ls}\right)^{-1}\,\left[\trimi{L}\right]^{A}_{lrT}\biggr\}\,\left[\Delta\ma{F}\right]_{rT}\\
+\,&\biggl\{\left[\fmi{V}^{\quarmi{C}\fmi{D}}\right]\h_{iJrTU}-\left[\trimi{N}\right]^{B}_{iJs}\,\left(\left[\mi{K}\right]^{AB}_{ls}\right)^{-1}\,\left[\quarmi{M}\right]^{A}_{lrTU}\biggr\}\,\left[\Delta\trima{F}\right]_{rTU}\,.\\
\end{aligned}
\end{equation}
Furthermore, we discretize the linearization of the hyperstresses $\trima{P}$
\begin{equation}
\left[\Delta\trima{P}\right]\h_{iJK}=\left[\fmi{V}^{\quarmi{C}\fmi{E}}\right]\h_{iJKsT}\,\left[\Delta\ma{F}\right]_{sT}+ \left[\smi{V}^{\quarmi{C}\fmi{D}\fmi{E}\smi{G}}\right]\h_{iJKsTU}\,\left[\Delta\trima{F}\right]_{sTU}+\left[\quarmi{N}\right]^{B}_{iJKs}\,\left[\Delta\fluc{q}\right]^{B}_{s}\,,\\
\end{equation}
with the volume-averaged tensors
\begin{equation}
\begin{aligned}
\left[\fmi{V}^{\quarmi{C}\fmi{E}}\right]\h_{iJKsT}=\phantom{+\,}&\frac{1}{V}\intRVEh\bigl(\left[\quarmi{C}\right]\h_{iJsT}\,\left[\mi{X}\right]\h_{K}+\left[\fmi{E}\right]\h_{iJKsT}\bigr)\,\d V\,,\\
\left[\smi{V}^{\quarmi{C}\fmi{D}\fmi{E}\smi{G}}\right]\h_{iJKsTU}=\phantom{+\,}&\frac{1}{V}\intRVEh\bigl(\left[\quarmi{C}\right]\h_{iJsT}\,\left[\mi{X}\right]\h_{K}\,\left[\mi{X}\right]\h_{U}+\left[\fmi{D}\right]\h_{iJsTU}\,\left[\mi{X}\right]\h_{K}\bigr.\\
\bigl.&\hspace*{2cm}+\left[\fmi{E}\right]\h_{iJKsT}\,\left[\mi{X}\right]\h_{U}+\left[\smi{G}\right]\h_{iJKsTU}\bigr)\d V\,,\\
\left[\quarmi{N}\right]^{B}_{iJKs}=\phantom{+\,}&\frac{1}{V}\intRVEh\bigl(\left[\quarmi{C}\right]\h_{iJsT}\,\left[\mi{X}\right]\h_{K}+\left[\fmi{E}\right]\h_{iJKsT}\bigr)\,\left[\nabla R\right]^{B}_{T}\,\d V\\
+\,&\frac{1}{V}\intRVEh\bigl(\left[\fmi{D}\right]\h_{iJsTU}\,\left[\mi{X}\right]\h_{K}+\left[\smi{G}\right]\h_{iJKsTU}\bigr)\,\left[\nabla^2 R\right]^{B}_{TU}\bigr\}\,\d V\,.
\end{aligned}
\end{equation}
Insertion once again of the discrete sensitivities (\ref{eq:discr_sensitivities}) yields
\begin{equation}
\begin{aligned}
\left[\Delta\trima{P}\right]\h_{iJK}=\phantom{+\,}&\biggl\{\left[\fmi{V}^{\quarmi{C}\fmi{E}}\right]\h_{iJKrT}-\left[\quarmi{N}\right]^{B}_{iJKs}\,\left(\left[\mi{K}\right]^{AB}_{ls}\right)^{-1}\,\left[\trimi{L}\right]^{A}_{lrT}\biggr\}
\,\left[\Delta\ma{F}\right]_{rT}\\
 +\,&\biggl\{\left[\smi{V}^{\quarmi{C}\fmi{D}\fmi{E}\smi{G}}\right]\h_{iJKrTU}-\left[\quarmi{N}\right]^{B}_{iJKs}\,\left(\left[\mi{K}\right]^{AB}_{ls}\right)^{-1}\,\left[\quarmi{M}\right]^{A}_{lrTU}\biggr\}\left[\Delta\trima{F}\right]_{rTU}\,.\\
\end{aligned}
\end{equation}

	
\bibliographystyle{plain}
\bibliography{bibliography,literature,IGAsquare}

\end{document}